%% file: arcvix.tex
\documentclass[english,12pt]{article}
\usepackage{lmodern}
\usepackage{amsmath}
\usepackage[T1]{fontenc}
\usepackage{rotating}
\usepackage{epsfig}
\usepackage{eurosym}
\usepackage{pdflscape}
\usepackage{endnotes}
\usepackage{geometry}
\usepackage{graphicx}
\usepackage{epstopdf}
\usepackage{float}
\usepackage{multirow} 
\usepackage{hyperref}
\usepackage{enumitem}
\usepackage{lscape}
\usepackage{tikz}

\makeatletter
\usepackage[para]{threeparttable}
\usepackage{longtable}
\usepackage{caption}

\usepackage{changepage} 
\usepackage{subcaption}
\usepackage{booktabs}
\usepackage{enumitem}
\usepackage{float}
\usepackage[authoryear]{natbib}
\PassOptionsToPackage{normalem}{ulem}
\usepackage{ulem}
\usepackage{pgffor}

\usepackage{setspace}
\usepackage{anyfontsize}

\newlength{\offsetpage}
{\end{adjustwidth}}

\usepackage{babel}
 
\usepackage{rotating}
\usepackage{mdframed}
\usepackage{booktabs}
\usepackage{amsmath}

\usepackage{tabularx}
\usepackage[para]{threeparttable}
\newcommand\startappendixtables{%
    \makeatletter
       \setcounter{table}{0}
   \setcounter{table}{0}\global\def\thetable{A\arabic{table}}%
    \makeatother}
 
 \newcommand\startappendixfigures{%
    \makeatletter
       \setcounter{figure}{0}
   \setcounter{figure}{0}\global\def\thefigure{A\arabic{figure}}%
    \makeatother}

\usepackage[sc]{mathpazo}
\renewenvironment{thebibliography}[1]{
  \begin{oldthebibliography}{#1}
    \setlength{\itemsep}{0.11em}
    \setlength{\parskip}{0em}
}
{
  \end{oldthebibliography}
}
 
\usepackage{xcolor}
\hypersetup{
    colorlinks, 
    linkcolor={black}, 
    citecolor={black}, 
    urlcolor={black}
}

\usepackage{xurl}

\usepackage{dsfont}

\makeatother

\usepackage{babel}

\begin{document}
\title{Pandemic Pressures and Public Health Care: Evidence from England}

\author{Thiemo Fetzer \and Christopher Rauh\thanks{Thiemo Fetzer, University of Warwick, CAGE and CEPR. Email: t.fetzer@warwick.ac.uk. Christopher Rauh: University of Cambridge, Trinity College Cambridge, CEPR. Email: cr542@cam.ac.uk. We would like to thank Dr Karolina Weinmann for helpful comments.}}
%\date{\today}
\date{22 January, 2022}
\maketitle

\begin{center} 
%\emph{Very preliminary and incomplete: do not circulate; do not upload on the web. }
\end{center}
\begin{abstract}
\noindent 
This paper documents that the COVID-19 pandemic induced pressures on the health care system have significant adverse knock-on effects on the accessibility and quality of non-COVID-19 care. We observe persistently worsened performance and longer waiting times in A\&E; drastically limited access to specialist care; notably delayed or inaccessible diagnostic services; acutely undermined access to and quality of cancer care. We find that providers under  COVID-19 pressures experience notably more excess deaths among non-COVID related hospital episodes such as, for example, for treatment of heart attacks. We estimate there to be at least one such non-COVID-19 related excess death among patients being admitted to hospital for non-COVID-19 reasons for every 30 COVID-19 deaths that is caused by the disruption to the quality of care due to COVID-19. In total, this amounts to 4,003 non COVID-19 excess deaths from March 2020 to February 2021. Further, there are at least 32,189 missing cancer patients that should counterfactually have started receiving treatment which suggests continued increased numbers of excess deaths in the future due to delayed access to care in the past. 
\thispagestyle{empty} 
  \smallskip
\end{abstract}

\setstretch{1.5}

\clearpage

\section{Introduction}
Despite the widespread availability of effective and safe COVID-19 vaccinations, the presence of significant pools of unvaccinated groups in society, the inevitable breakthrough infections, and the emergence of new COVID-19 variants imply continued pressures on health care systems around the world (see \citealp{Mahasen2664,deOliveiraAndradem3032}). Not surprisingly, the care for COVID-19 patients is drawing  resources that may have indirect effects on the quantity and quality of care for patients that require medical help for reasons unrelated to COVID-19.  Further, lockdowns and other public-health measures that are taken to slow the spread of COVID-19 may directly affect both the supply as well as the demand for health care services by, for example, discouraging individuals to seek medical advice or, through their impact on the likelihood of falling sick or having an accident (see e.g.\ \citealp{Vandoros2021}). All of these factors combined can lead to  worse public health outcomes for patients that need medical help independent of COVID-19. 

This paper traces out the negative externalities that COVID-19 induced health care  shocks have on non-COVID-19 patients. We study both the arrival of COVID-19 and the initial disruption of the first wave, along with the  effect of ongoing COVID-19 pressures on the health system's performance leveraging a broad array of public administrative data from  the NHS in England lasting until November 2021. Throughout, we study margins of both quantity and quality of health care provision across five domains.  First, we show that the initial wave of the pandemic caused a drastic decline in A\&E attendances that was followed by a sharp increase. While before the pandemic 80\% of A\&E visits were attended within the NHS set goal of 4 hours, this share has collapsed to just two thirds in the most recent months of reporting.  Second, we note drastic impacts on the performance of elective care: since the onset of the pandemic, access to specialist care has been significantly curtailed with waiting lists for referrals increasing by 20\% and waiting times shooting up.  Pre pandemic 97\% of diagnostics took place within the NHS set goal of 6 weeks. The share dropped to  56\% during the first wave and has only increased to 71\% since, implying notable delays in diagnosis and access to treatment. Given that the waiting list tends to be around 1 million entries long, these delays have affected millions of patients over the course of the pandemic.  Third, non-emergency consultant-led treatment  declined dramatically and has only recently approached pre-pandemic levels. We calculate that, in comparison to pre-pandemic treatment, the missing treatments accumulated over the course of the pandemic are 5.6 million. Moreover,  patients referred for treatment have been 9\% less likely to receive treatment within the NHS set goal of 18 weeks. 

Fourth, focusing on the likely longer term repercussions, we show that the disruption caused by the various pandemic waves has had a significant negative impact on cancer care. In total, we estimate there to be 32,189 fewer  cancer patients receiving first treatment following a decision to treat, along with a notable worsening performance in terms of times to first consultation and times of first treatment for urgent cases. For example, while before the pandemic, 91\% of urgent referrals of suspected cancer cases would lead to consultations with a specialist within the NHS set goal of 14 days, this share has been only 86\% since the end of the first wave. Even more importantly, for cancer treatment, the share of patients receiving urgent first treatment after referral within the NHS set goal of 62 days has  declined from  78\%  before the pandemic, to 75\% during and 72\% after the first wave. Alarmingly, the share has been trending downwards: in the most recent months only 66\% of patients received treatment within 62 days of first referral and we document that ongoing COVID-19 pressures further worsen this effect. We estimate that amongst detected urgent cancer cases, 53,068 people had their cancer care delayed past the NHS set goal. This is likely consequential in driving excess mortality in the future as systematic reviews of clinical evidence suggest that even a four weeks delay of cancer treatment is associated with increased mortality for cancer patients (see \citealp{Hanna2020}). For cancer as well as for the other domains, we further show that delays extend way beyond the NHS set goals with previously non-existent large fat tails in delays. Moreover, all previously mentioned aggregate patterns are not driven by few trusts, but instead are found to be statistically significant across outcomes and the health care system.
 
Lastly, we also find robust evidence suggesting that COVID-19 disruptions have had a notable impact increasing excess mortality among patients attending hospital for reasons unrelated to COVID-19. We leverage unique data measuring excess mortality associated with hospital episodes at the provider level. This data is normally used to evaluate the performance of different NHS providers. Underlying the expected mortality is an individual patient level mortality risk estimate that is obtained from a predictive model taking a broad range of patient characteristics into account to predict the probability of death for each admitted patient within 30 days of admission. Importantly, the data exclude any hospital episodes or visits and deaths that involve a COVID-19 diagnosis code such as a positive COVID-19 test result. Since all patients are routinely screened for COVID-19 this implies that virtually the universe of COVID-19 patients is excluded. Further, all deaths that mention COVID-19 on the death certificate are also excluded. Hence, this measure of excess death captures deviations between actually observed deaths and the expected deaths of patients that were admitted to hospital for non COVID-19 reasons \emph{under normal NHS operating circumstances}. Not surprisingly, on average prior to the pandemic, the resulting aggregated measure of excess mortality is centered around zero suggesting that the model has good out-of-sample predictive power. From March onwards, however, there appear systematic deviations in the excess death measure with observed deaths being significantly larger than expected deaths among non COVID-19 patients. This suggests a significant omitted variable: the impact of the pandemic on the care non COVID-19 patients receive. 

We estimate that, for the period from March 2020 to February 2021 alone,  there have been at least 4,003 excess deaths of hospital patients in England that, if it were not pandemic disruptions, would not have been expected to die. This number stands significant in the context of actual and estimated COVID-19 deaths. In the same period, England reported 124,581 deaths which mentioned COVID-19 on the death certificate (which contrasts with other COVID-19 death measures over the same period suggesting  109,250 deaths  occurred within 28 days of a positive COVID-19 test or  102,763 excess deaths estimated by the Office of National Statistics). Thus, this represents a pure excess-death measure capturing excess mortality among patients who sought out medical care for reasons unrelated to COVID-19. This implies that our estimates suggest that  the excess deaths among hospitals episodes of around 4,003 individual cases between March 2020 and February 2021 stand at a non-negligible 3.8\% of all COVID-19 excess deaths or 3.1\% of all deaths mentioning COVID-19 on the death certificate that were captured during that period. 

We document that many of the deteriorations, in particular in the quality of health care provided, are significantly associated with ongoing COVID-19 related pressures experienced across NHS providers. For many performance measures, changes are   economically meaningfully related to 
changes in the number of new COVID-19 admissions that hospitals face.  For instance, ongoing COVID-19 induced hospital pressures significantly increase the waiting times for treatment in A\&E.  We also find that the number of non-COVID-19 excess deaths rises sharply with the number of new COVID-19 admissions. For every doubling of new COVID-19 admissions, there is an additional four non-COVID-19 excess deaths.  

In a final step, we look into potential mechanisms. COVID-19 pressures affect the ability of the NHS to provide care in a multitude of dimensions. Most directly, sharp increases in COVID-19 admissions increase the demand for  care, which naturally, is diverting resources. Yet indirectly, COVID-19 admissions, above and beyond what we account for by controlling for the spread of COVID-19 in the community, are exposing NHS staff to higher infection risks, increasing staff absences. This in turn, is reducing the ability of the NHS to provide care. We document that frontline health care staff -- as opposed to managerial NHS staff -- see notably higher  staff absence rates if they are hit by a significant influx of new COVID-19 patients in need of hospital care. Further, the erratic COVID-19 induced health care demand surges may indirectly affect the quality of care provision through simple physical exhaustion and stress. While these indirect mechanisms cannot be quantified, in the paper we document that the link between staff absence rates and COVID-19 admissions is notably weaker, the higher the share of NHS staff that is fully vaccinated, even after controlling for community transmission and population vaccination rates. 
 
 Measuring the indirect impacts of COVID-19 pressures on non-COVID-19 care is, not surprisingly, difficult. We take advantage of the fact that the NHS had produced measures of excess death based on individual-level patient microdata. This allows us to capture the indirect burden of the pandemic   Measures of excess death are often computed due to a lack of data on COVID-19 deaths. For example, in India, it was recently estimated by numerous studies that the true number of COVID-19 deaths may be actually notably larger \citep{Adam2022,Jha2022}. Yet, these measures of excess death may inadvertently be confounding deaths that arise from disruptions in care for non COVID-19 reasons. In this paper, we can actually quantify the extent of this indirect factor of excess death. In the UK, recorded COVID-19 deaths match very closely with  excess deaths. 
 
The UK is uniquely positioned to enable the study of the impact of the pandemic on wider availability and access of health care. It is one of the few advanced economies that boast a national public health care system -- the National Health Service (NHS) -- implying generally quite good data availability for research while, at the same time, having suffered some of the worlds highest pandemic infection rates and death tolls. Importantly, measuring system-wide disruptions and challenges brought about by the pandemic is generally difficult in many countries due to the often decentralized or fragmented organization of health care systems with a mix of public- and private providers organized across different layers. This is the first paper to show the impact of the arrival and the ongoing pressures imposed by COVID-19 on a countries' health care system as a whole.
Naturally, there is a question on whether the pandemic will have scarring effects on societies.  Within economics, there has been a debate on whether and to what extent there are scarring effects. In this paper we document that, at least based on the data from the health care system, there is likely to be notable scarring effects in terms of worse public health outcomes which are likely to have an economic and health impact for years to come.

Already going into the pandemic, health care systems were deprived of human resources (\citealp{Lasater2021,Clements2008}), and from suffered  losses of life (see \citealp{Bandyopadhyay2020} for a review) and the impact on the physical and mental well being among health care staff (\citealp{Quintana-Domeque2021,Sun2020a}). All these components are likely to further erode the ability of the health care sector to attract and retain human capital.

The ``missing'' cancer patients were already flagged early in the pandemic by charities (\citealp{macmillan2020}), and the medical literature has been discussing how  health care systems are reorganizing to address the issues (e.g.\ \citealp{richards2020impact}) and the trade-offs involved in providing cancer care during the pandemic (e.g.\ \citealp{kutikov2020war}). We are able to provide a lower bound on the likely number of deaths that may have been caused by the deterioration  of care that patients receive in hospital under COVID-19 stress finding that these may easily account for a vast number equivalent to at least 3\% of all officially counted COVID-19 deaths. Our approach contrasts with existing work looking at excess deaths relying on modelling studies of the likely increases,  e.g.\ due to undetected or delayed treatment of cardiovascular diseases (e.g.\ \citealp{Banerjee2021}) or   cancer  (e.g.\ \citealp{lai2020estimated}).  

Much of the economic literature has tried to characterize optimal policies to minimize the lives versus livelihood trade-off (e.g.\ \citealp{bethune2020covid}). Governments have introduced measures curtailing individual freedoms to help put a check on transmission and keeping hospitalization pressures down. Economists have studied the causal impact on the spread of COVID-19 of compulsory face masks (\citealp{Abaluck2021,Mitze2020}); (digital) contact tracing (\citealp{Fetzer2021,Wymant2021}); targeted or untargeted lockdowns or reopenings (e.g.\ \citealp{Fajgelbaum2020,Fetzer2020c}) along with their impacts on  mental health (\citealp{adams2020impact}) as well as its general deterioration during the pandemic (\citealp{etheridge2020gender, proto2021covid}). This paper is among the first to document and quantify the negative externalities that COVID-19 has on the quantity and quality of non-COVID-19 health care.

\section{Data}
We leverage a multitude of data sources to document comprehensively how the COVID-19 induced disruptions affect both access as well as quality of care across NHS Trusts in England.

\subsection{A\&E attendances and emergency admissions}
We measure the performance of the accident and emergency (A\&E) units in the UK in terms of the level of demand as well as the quality of performance as measured by the waiting times to receive treatment using data from the A\&E Attendances and Emergency Admissions dataset. This captures both measures of absolute number of attendances over time for all A\&E types, including Minor Injury Units and Walk-in Centres, and of these, measures of performance capturing e.g.\ the number discharged, admitted or transferred within four hours of arrival.\footnote{This data is available at \url{https://www.england.nhs.uk/statistics/statistical-work-areas/ae-waiting-times-and-activity/}.}  The data is arranged as a monthly panel at the provider level.

\subsection{Referral to treatment and waiting times}
NHS Constitution gives patients a legal right to access services within maximum referral to treatment (RTT) waiting times. Waiting times statistics are collected to ensure that the NHS can be held accountable.\footnote{This data is available at \url{https://www.england.nhs.uk/statistics/statistical-work-areas/rtt-waiting-times/}.} Patients referred for non-emergency consultant-led treatment are on RTT pathways. An RTT pathway is the length of time that a patient waited from referral to start of treatment, or if they have not yet started treatment, the length of time that a patient has waited so far. The incomplete pathway operational standard is the measure of patients' constitutional right to start treatment within 18 weeks.\footnote{An RTT pathway ends with the start of first treatment; the start of active monitoring of a condition initiated by the patient or care professional; a decision not to treat a condition; if a patient declined an offered treatment; or if a patient died before treatment.} Each pathway relates to an individual referral rather than an individual patient, so if a patient was waiting for multiple treatments they may be included in the figures more than once. 
Incomplete pathways, often referred to as waiting list times, are the waiting times for patients waiting to start treatment, as at the end of each month. The incomplete waiting time standard was introduced in 2012 and states that the time waited must be 18 weeks or less for at least 92\% of patients on incomplete pathways.

These data are available at the provider level. We leverage the data pertaining to the NHS Trusts that carry out the vast majority of treatments. The data is arranged as a panel at the provider by treatment function by the pathway status by month providing measures of the total number of patients on waiting lists, along with a breakdown capturing how long individuals have been waiting for treatment.\footnote{The treatment functions are Cardiology, Rheumatology, General Surgery, Urology, Trauma \& Orthopaedics, Ear, Nose \& Throat (ENT), Ophthalmology, General Medicine, Thoracic Medicine, Gynaecology, Gastroenterology, Dermatology, Geriatric Medicine, Oral Surgery, Neurosurgery, Neurology, Cardiothoracic Surgery, Plastic Surgery, and a composite Other category.} 

\subsection{Diagnostics waiting times and activity}
The NHS collects monthly data on waiting times and activity for 15 key diagnostic tests and procedures.  This data collection effort is intended to monitor activity and identify bottlenecks in diagnostic services recognizing that early diagnosis is central to improving health outcomes.  The waiting list data is a ``snap shot'' of the waiting list on the last day of the month in question. The activity data is the actual number of procedures carried out during the month in question. Delayed diagnostics can significantly lengthen patient waiting times to start treatment. Diagnostic tests refers to set tests or procedures used to identify and monitor a person's disease or condition, allowing a medical diagnosis to be made. This contrasts with actual therapeutic procedures that aim to actually treat a persons condition.\footnote{This data is available at \url{https://www.england.nhs.uk/statistics/statistical-work-areas/diagnostics-waiting-times-and-activity/monthly-diagnostics-waiting-times-and-activity/}. The following diagnostic tests are considered: MRI, CT, Ultrasound, barium enema, dexa scans, audiology assessments, echocardiography, electrophysiology, peripheral neuropathy, sleep studies, urodynamics, colonoscopy, flexi sigmoidoscopy, cystoscopy, gastroscopy.} The data is arranged at the provider by diagnostic test by time level. %and is available from 2017 to October 2021. 

\subsection{Cancer referrals, treatment, and waiting times}
The NHS collects data and sets targets for the performance of cancer care services. These are arranged at three levels.  Following an urgent referral for suspected cancer, at least 93\% of patients should be seen by a specialist within two weeks. A second target involves first treatment:  for all cancer treatment types, at least 96\% of patients should start a first treatment for a new primary cancer within one month (31 days) of the decision to treat. The overarching target is that at least 85\% of patients should start a first treatment for cancer within two months (62 days) of an urgent referral.  
This allows a construction of measures both on the extensive margin capturing numbers of individuals referred to a specialist; number of treatment decisions taken and number of actual treatments commenced. Similarly, it allows for measures on the intensive margin capturing care quality measured by the time it takes for individuals to be seen by a specialist and to commence treatment.\footnote{This data is available at \url{https://www.england.nhs.uk/statistics/statistical-work-areas/cancer-waiting-times/monthly-prov-cwt/}.}
This data is arranged across a range of datasets at the provider by (suspected) cancer type by care setting (admitted or not-admitted) and by time.

\subsection{Measuring non-COVID-19 excess mortality}
The Summary Hospital-level Mortality Indicator (SHMI) reports on mortality at the NHS trust level across England and is produced as an official monthly statistic by NHS Digital. The SHMI includes deaths which occurred in hospital or within 30 days of discharge and is calculated using Hospital Episode Statistics (HES) data linked to Office for National Statistics (ONS) death registrations data.\footnote{The data is available on \url{https://digital.nhs.uk/data-and-information/publications/statistical/shmi/}.}  The SHMI is the ratio between the actual number of patients who die following hospitalisation at the trust level and the number that would be expected to die.  The expected probability of an individual patients death is estimated from a statistical model based on the characteristics of the patients. These characteristics include the condition the patient is in hospital for, other underlying conditions the patient suffers from, age, gender, method and month of admission to hospital, and birthweight (for perinatal diagnosis groups).  For each admission a risk of death score is computed for which then the cumulative expected deaths can be computed and contrasted with the observed number of deaths that occur while patients were in hospital or within 30 days of discharge. Crucially, the SHMI data remove any activity or death that is related to COVID-19. Specifically, if any hospital episode within a provider spell have a COVID-19 diagnosis code recorded (such as, for example, if a patient tests positively for COVID-19), then the spell is excluded from the analysis. Since all admitted patients are routinely tested for COVID-19 this implies that virtually all hospital episodes under consideration exclude COVID-19 patients. Moreover, for all deaths included in the SHMI, if COVID-19 is recorded anywhere on the death certificate, then the death and the spell it is linked to are also excluded from the SHMI. This ensures that we focus exclusively on deaths and in particular, excess deaths in care settings that are not directly attributable to COVID-19, but may still be driven by COVID-19, due to its impact on the quality of care that can be provided. 

 The data is reported as twelve month rolling cumulative totals, that is, for example, the monthly publication of March 2020 includes the cumulative total number of hospital episodes or ``spells'', the number of observed deaths or the number of expected deaths over the twelve month window ranging from April 2019 to March 2020 inclusive. That is, for every reporting month $t$, the measures we capture the twelve month cumulative totals, that is, $\sum_{\tau = t-12}^{t} \text{Obs}_{p,\tau} $, $\sum_{\tau = t-12}^{t} \text{Exp}_{p,\tau} $ and $\sum_{\tau = t-12}^{t} \text{Spells}_{p,\tau} $. 

We can compute the number of excess deaths in a twelve month rolling window as reported in month $t$ as

$$\sum_{\tau = t-12}^{t} \text{Excess}_{p,\tau}  = \sum_{\tau = t-12}^{t} \text{Obs}_{p,\tau}  - \sum_{\tau = t-12}^{t} \text{Exp}_{p,\tau} $$

Naturally, the above measure can be considered to be the residual of a regression that is the result of having aggregated the individual predicted mortality risks $h(x_{i,p,t})$ of  observation $i$ that is captured in a set of features $x$ about the individual $i$. If this model was unbiased, we would expect that the expected value of this measure $\mathop{\mathbb{E}} (\sum_{\tau = t-12}^{t} \text{Excess}_{p,\tau} | h(x)) = 0$.  Naturally, if there was an omitted variable $z_{i,p,t}$ either at the individual, provider- or time level that affects the number of observed deaths in a way that the statistical model to generate the expected deaths measure has not taken into account for -- i.e. if there is an \emph{omitted variable} - in the risk model, we would expect the above condition to be violated, i.e. that there is indeed some structure in the residuals.  We document that up to February 2020, there is no structure in the residuals with the average excess deaths across providers and over time to hover close to zero. Yet, from March 2020 onwards, the pattern suggests that there is indeed an important omitted variable in the risk model that results in a notable divergence between the observed number of deaths and the expected number of deaths.  In section \ref{sec:excessdeath} we document that measures of COVID-19 pressures at the provider level   $p$ capture this pattern quite well.

\subsection{NHS sickness absence and vaccination rates}
We also study staff absence rates as a potential mechanism along with staff vaccination rates. Detailed breakdown of staff absence rates by staff group types which broadly distinguishes doctors, nurses, management and other support staff is available as a monthly measure.\footnote{These data are available on \url{https://digital.nhs.uk/data-and-information/publications/statistical/nhs-sickness-absence-rates}.} We also construct a (cross sectional) measure of the share of staff fully vaccinated. This data was broken down by NHS Trust level from October 2021 onwards and we use this first reporting month as a cross sectional characteristic to explore heterogenous treatment effects. We also construct wider population vaccination rates and case numbers in catchment areas of NHS Trusts described in more detail next.

\subsection{Measuring provider-level exposure to COVID-19}
We construct a range of measures to a specific providers exposure to COVID-19. We observe three measures directly at the health care provider level: the number of new hospital admissions who tested positive for COVID-19 in the 14 days prior to hospital admissions or who during their stay in hospital inpatients were diagnosed with COVID-19 after admission. The number of cases in hospital measured as the number of people currently in hospital with confirmed COVID-19 through a positive PCR test for COVID-19 in the past 14 days. The number of COVID-19 patients in beds which can deliver mechanical ventilation.

 We also construct a measure of the number of cases within the community across catchment areas of NHS providers. NHS trusts are not defined spatially explicitly, but rather, can serve multiple regions. Yet, most NHS Trusts are spatially quite concentrated. To allocate NHS trusts and providers to specific locations and to merge in additional data, we leverage an analysis of individual-level micro data from the Hospital Episodes Statistics dataset which breaks down all hospital visits to an NHS provider location by the location of residence of the patients at the granular middle layer super output area (MSOA) which have, on average, a population of 8,000 residents.\footnote{This data is available on \url{https://app.powerbi.com/view?r=eyJrIjoiODZmNGQ0YzItZDAwZi00MzFiLWE4NzAtMzVmNTUwMThmMTVlIiwidCI6ImVlNGUxNDk5LTRhMzUtNGIyZS1hZDQ3LTVmM2NmOWRlODY2NiIsImMiOjh9}.} We allocate MSOA's to NHS trusts on the basis of a first-past-the-post basis -- that is, an MSOA is counted towards the catchment area of an NHS trust if that trust handles the most hospital episodes across all NHS trusts that serve residents from this MSOA. As illustrated in Appendix Figure \ref{fig:spatialallocation} there is, not surprisingly, ample spatial clustering implicit in this. Having this mapping of MSOA's that are spatially explicit to NHS trusts (which may operate out of several sites within an area) allows us to construct measures of the cumulative community exposure to COVID-19 as COVID-19 case figures along with vaccination rates are provided at the MSOA level. 

We next describe the empirical analysis that we carry out.

\section{Empirical analysis}
Most of the datasets we leverage here are monthly panel datasets allowing us to study the evolution of key measures of health care system within providers and over time.  We carry out two main sets of exercises: first, studying the evolution of NHS performance across a broad range of metrics within providers over time, contrasting the time before- and after March 2020 and second, the performance of NHS providers since March 2020 and to what extent COVID-19 pressures continues to affect the ongoing operations during the pandemic.  The former allows us to quantify the pandemic-induced backlogs and care quality concerns that arise on the extensive margin, while the latter allows us to study how ongoing COVID-19 pressures affect the quality of care on a recurrent basis -- allowing us to further shed light on the underlying mechanisms behind the shock.

\subsection{Before and after the arrival of COVID-19}
We begin by documenting the impact that the arrival of COVID-19 had, in particular, the initial wave in March 2020, on care provision contrasting both quantity as well as quality of care before and after the arrival of the pandemic over time. We separate three distinct regimes: i)  before the pandemic, ii) during the  first wave, and iii) in all subsequent waves. Contrasting across these three regimes is helpful, as naturally, the pre-pandemic performance and implied trends allow us to construct a simple counterfactual evolution of how NHS performance may have evolved over time, had it not been for the arrival of the pandemic. 

Specifically, we estimate simple models of the form

$$ y_{p, t} = \alpha_p + \nu_t + \epsilon_{p} $$

for a measure of quantity or quality of NHS care provided in provider $p$ during month $t$. The provider fixed effect, $ \alpha_p $, captures time-invariant provider specific idiosyncratic level differences in both quality and quantity of care, while the time fixed effects $\nu_t$ capture the distinct time variation in performance common across providers. The arrival of the pandemic constitutes a common shock that affected the health care system as a whole, while the subsequent waves were handled notably differently with much more specific interventions and decisions to maintain service quality taken at the individual provider level.

The above allows us to estimate counterfactual time-paths and evaluate how the estimated performance $\hat{\nu_t}$ compares with such counterfactual time-paths. We distinguish three time periods: the period prior to the pandemic up to March 2020; the period of the first lockdown constituting major disruption of the health care system from March to June 2020; and the period since July 2020. Contrasting the $\hat{\nu_t}$ across the three regimes allows us to estimate the gaps in health outcomes and the extent to which such gaps may or may not be closed.

The analysis will highlight that, while the outsets of the pandemic may have been a system-wide common shock, the ongoing pandemic may exert different pressures across providers. While the arrival of the pandemic marks an unexpected shock, the initial reaction to the shock may not be representative of the ongoing pressures that COVID-19 poses on hospitals in the foreseeable future. Hence, we focus explicitly on exploiting cross-provider variation across NHS trusts in the \emph{within pandemic} exercise.
 
\subsection{Within pandemic}
The second set of exercises focuses on the period since March 2020, documenting how, within the pandemic, the idiosyncratic variation in COVID-19 cases affecting health care providers differentially, impact the quantity and quality of care provided. That is, we specifically document how ongoing pandemic pressures affect the provision of care. 

For that purpose we estimate specifications of the form

$$ y_{p,  t} = \alpha_p + \nu_t  + \beta \times \text{COVID-19}_{p,t} + \xi \times X_{p,t} + \epsilon_{p} $$

Compared with the previous exercise, this in essence studies to what extent we can attribute the variation around the common time fixed effects $\nu_t$ since March 2020 can be attributed to providers being differentially affected by the pandemic since March 2020.  Here, $\text{COVID-19}_{p,t}$ captures a provider-specific COVID-19 exposure measure, such as, the number of COVID-19 cases within the provider's typical area of operation; the number of COVID-19 hospital cases; the number of admitted COVID-19 patients or the number of patients on mechanical ventilation beds. Throughout the results presented in the main body include additionally a control capturing the  log number of COVID-19 cases detected across MSOAs that make up the main catchment area of a different provider. This is naturally a very demanding specification as it puts specific focus on pressures faced by providers in form of hospital admissions of said COVID-19 cases. Depending on the outcome data of interest, we estimate more demanding specifications that, for example, capture provider and function area specific fixed effects, capturing, e.g.\ different demand levels for various health care services such as cardiology or dermatology services.

\paragraph{Identification assumption} We treat the variation in COVID-19 pressures hitting different NHS providers as exogenous. Given the notable heterogeneity both in intensity and timing of COVID-19 infections across the country this is not an unreasonable assumption to make. Throughout the exercises presented in the main paper we control for the extent of community transmission of COVID-19 within catchment areas of different providers. This implies that we put specific focus on the shocks to COVID-19 cases hitting different providers at different points in time which, in turn, is a function of the underlying demographic makeup of the population in the catchment area. All results are robust, and most even larger, when excluding the measure of community transmission as  control variable as shown in Appendix Tables \ref{table:ae_statistics-pressures-nocomm}--\ref{table:excess-deaths-nhsvaxx-nocomm}. 

Treating COVID-19 pressures at the provider level as an exogenous source of variation would be violated, if, for example COVID-19 patients were avoiding specific providers that are under significant distress already -- that is, if patients sick with COVID-19 purposefully avoid going to a hospital that has notable new numbers of COVID-19 patients. Since the extent to which providers are overwhelmed in local areas is not public knowledge and given that COVID-19 patients can deteriorate quickly, it is unlikely that such strategic behavior would affect which hospital to go to. 
Similarly, our identification strategy is relying on the assumption that there is no reverse causality, i.e.\ that COVID-19 patients do no strategically avoid hospitals due changes in non-COVID-19 related care which we use as dependent variable. Again this sort of sophisticated behavior is unlikely due to lagged publication of the data and the urgency of the situation.

\subsection{Non-COVID-19 excess mortality}
As indicated, the data from \cite{SHMI2021} provide us with an estimate of the expected mortality of hospital admissions for each diagnostic based on a range of patient characteristics. Importantly, this excludes all COVID-19 related deaths. We study  whether with the start of the pandemic, the structure of excess deaths is different compared to before the pandemic started, and further, to what extent month-on-month variation in COVID-19 pressures is affecting the excess deaths.  As the data is reported at the monthly level but as twelve month cumulative rolling totals this dampens the month-on-month variation. We carry out two complementariy exercises that document however, that this is not an issue.\footnote{A monthly rather than twelve month rolling sum of the excess mortality data was requested by the researchers via email and via a Freedom of Information request -- all communication relating to this FOI request can be tracked here \url{https://www.whatdotheyknow.com/request/shmi_data_by_provider_at_monthly}.} 
The reported data in a given reporting month $t$ provides the cumulative totals of the observed- and expected deaths $\sum_{\tau = t-12}^{t} \text{Obs}_{p,\tau}$ and $ \sum_{\tau = t-12}^{t} \text{Exp}_{p,\tau}$. This implies we can compute the month-on-month changes as
\begin{align}
\Delta \text{Excess}_{p,t} = & [\sum_{\tau = t-12}^{t} \text{Obs}_{p,\tau}  -  \sum_{\tau = t-12}^{t} \text{Exp}_{p,\tau}] -  [\sum_{\tau = t-13}^{t-1} \text{Obs}_{p,\tau} -   \sum_{\tau = t-13}^{t-1} \text{Exp}_{p,\tau}]  \\
  = & [ \text{Obs}_{p,t} -  \text{Exp}_{p,t} ]- [ \text{Obs}_{p,t-13} -  \text{Exp}_{p,t-13}].  \nonumber 
 \end{align}
This implies we can capture the number of excess deaths in a given month $t$, rather than the twelve month rolling window in the above expression. If we denote the genuine monthly excess number of deaths as $Excess_{p,t} =  \text{Obs}_{p,t} -  \text{Exp}_{p,t} $, we can exploit month-on-month variation in COVID-19 pressures at the hospital level by estimating variations of the below specification:
\begin{align}
Excess_{p,t} =  \alpha_i + \nu_p + \gamma_t + \beta \times \text{COVID-19}_{p,t}  + \xi \times \mathbf{X}_{p,t} + \nu_{i,p,t}. 
\end{align}
Crucially, given the above transformation, the vector of additional control variables $\mathbf{X}_{p,t}$  should include $[ \text{Obs}_{p,t-13} -  \text{Exp}_{p,t-13}] $ as control variable.

Alternatively, we also estimate alternative specifications that do not transform the data in the above fashion. Given the reporting in twelve month cumulative totals this implies we need to measure the COVID-19 pressures not month-on-month but similarly compute cumulative totals over a time window. For example, we can estimate the impact of COVID-19 pressures in the last $\xi$ month relative to the reporting month $t$ on the log difference in observed- vis-a-vis expected number deaths cumulatively in the last twelve months as in
$$ \log{\sum_{\tau = t-12}^{t}  \text{Obs}_{p,\tau}} - \log{\sum_{\tau = t-12}^{t} \text{Exp}_{p,\tau}} =  \nu_p + \gamma_t + \beta \times \sum_{\tau = t-12}^{\xi}\text{COVID-19}_{p,\tau} + \nu_{p,t}. $$

We explore a range of variations of the above to document the robustness. 

We next present some descriptive evidence across the outcomes.

\section{Quantity and quality of care before and after the arrival of COVID-19}
We begin by presenting some descriptive statistics that highlight how the pandemic affected both the quantity as well as the quality of care being provided by the NHS system across a broad range of outcome measures.

\subsection{A\&E attendance and waiting times}
\begin{landscape}
\begin{figure}[h!]
\caption{Measuring quantity and quality of health care across NHS over time  \label{fig:quantity-qual}}
\begin{center}
$
\begin{array}{llll}
\multicolumn{2}{c}{\text{\emph{Panel A:} Measures of quantity }} & \multicolumn{2}{c}{\text{\emph{Panel B:} Measures of quality }}.  \\  \\
\multicolumn{1}{c}{\text{ A\&E attendance }} & \multicolumn{1}{c}{\text{Completed RT referral}}  & \multicolumn{1}{c}{\text{ A\&E < 4h }} & \multicolumn{1}{c}{\text{Completed RT referral < 18 weeks}} \\  

    \includegraphics[width=.26\columnwidth]{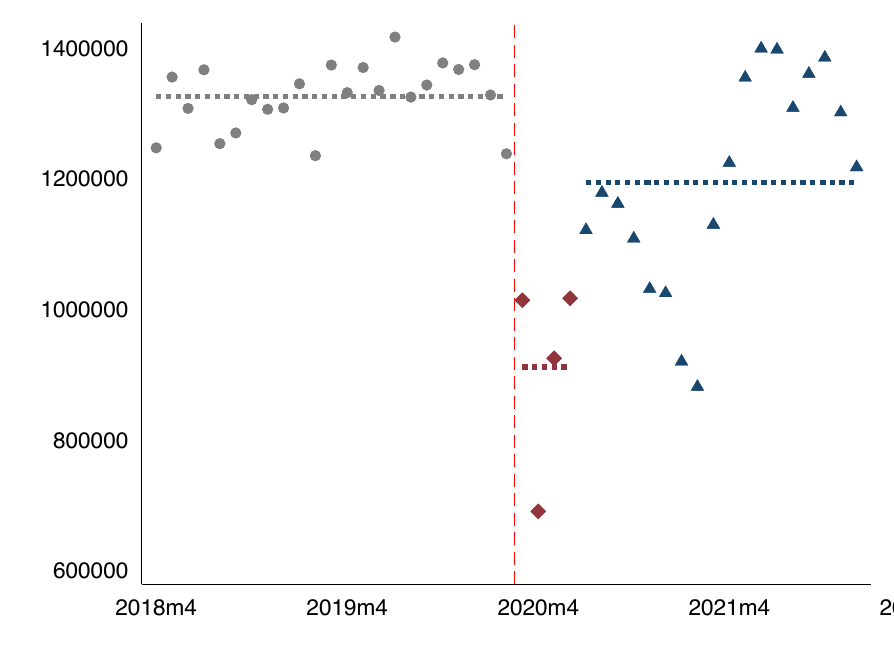}  &     \includegraphics[width=0.26\columnwidth]{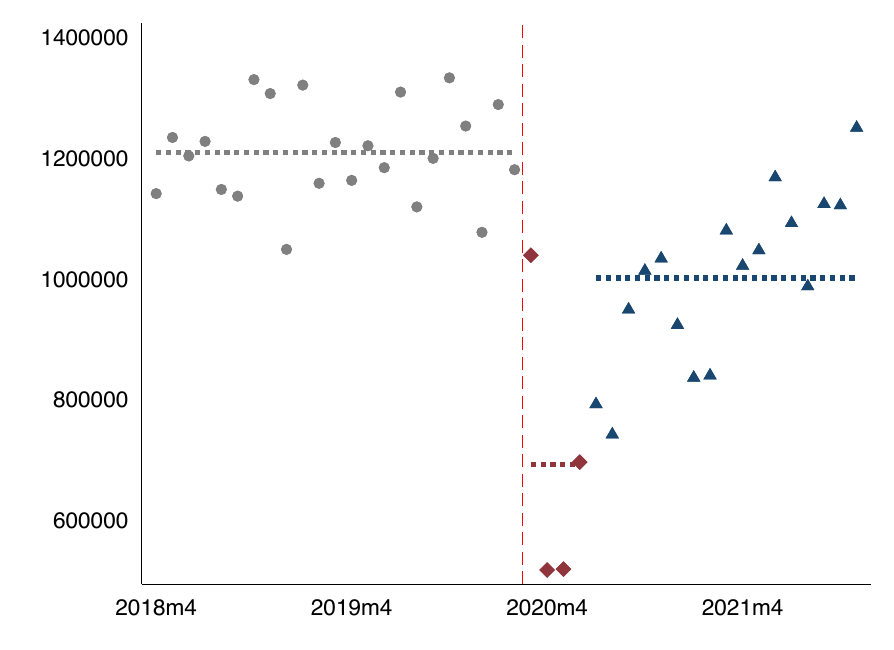} &     \includegraphics[width=.26\columnwidth]{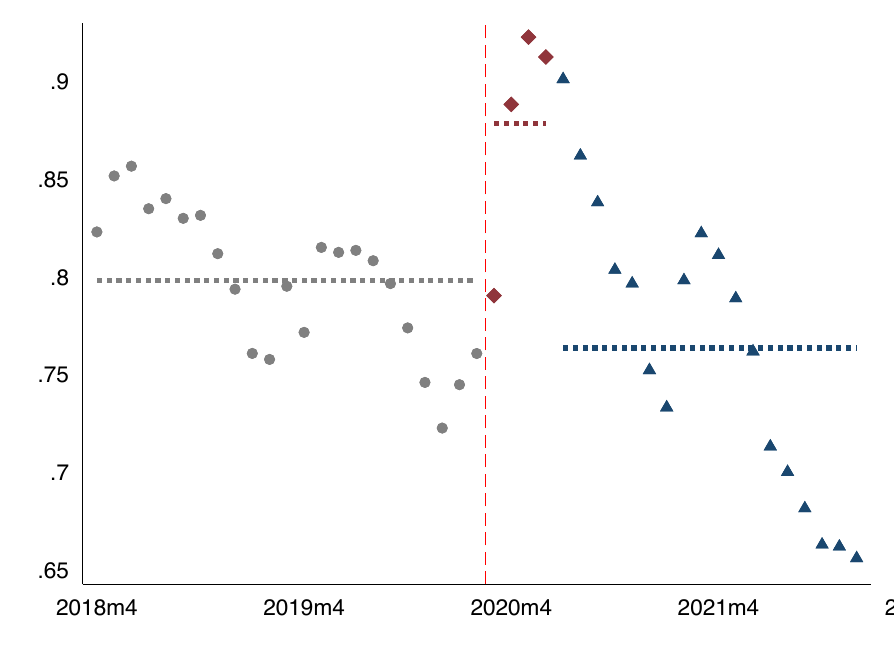}  &     \includegraphics[width=0.26\columnwidth]{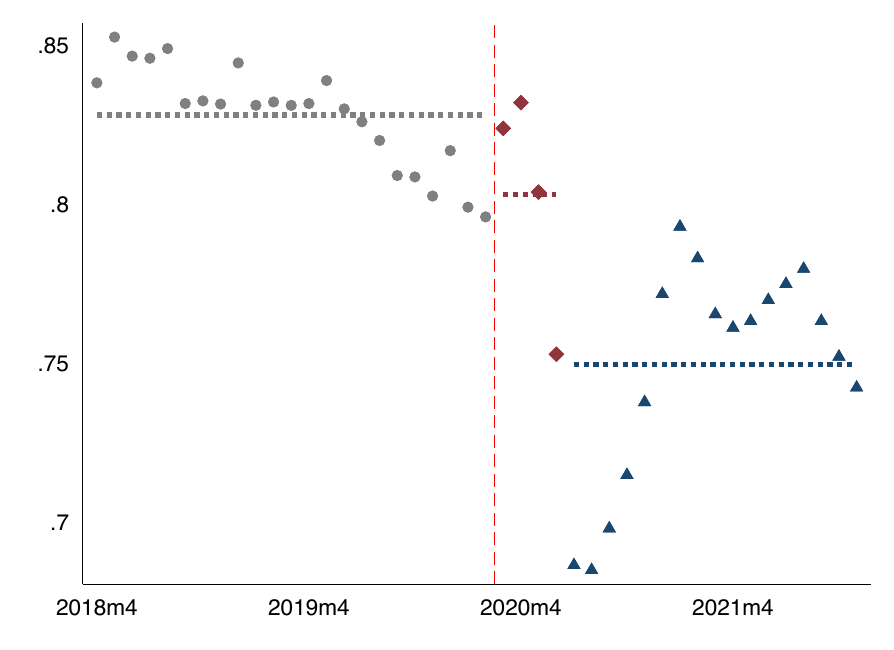} \\
    \\
\multicolumn{1}{c}{\text{Diagnostic waiting list }} & \multicolumn{1}{c}{\text{Cancer cases treated}}  & \multicolumn{1}{c}{\text{Diagnostic < 6 weeks }} & \multicolumn{1}{c}{\text{Cancer treatment < 62 days}} \\\  
 \includegraphics[width=0.26\columnwidth]{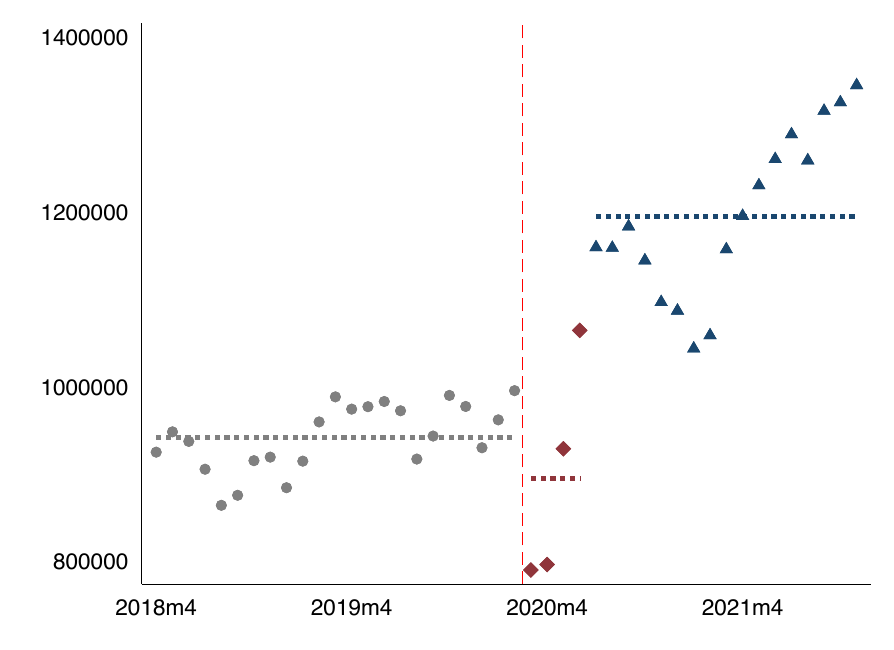} & \includegraphics[width=0.26\columnwidth]{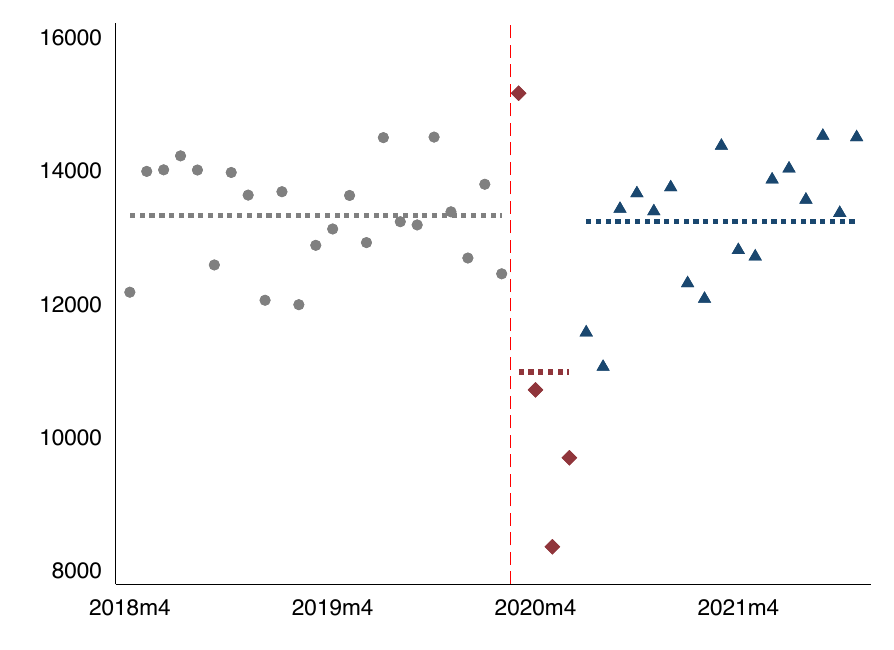} &  \includegraphics[width=0.26\columnwidth]{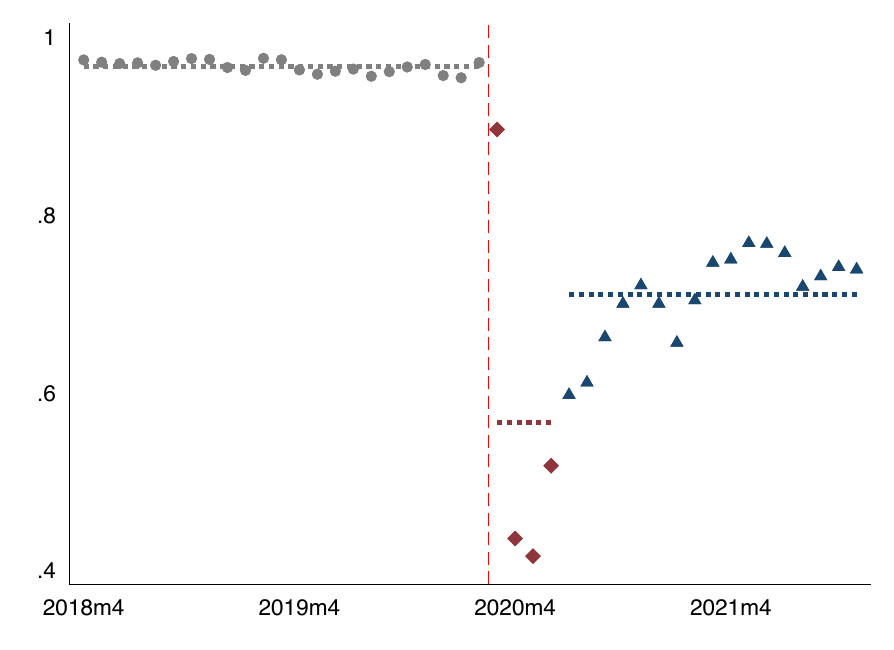} & \includegraphics[width=0.26\columnwidth]{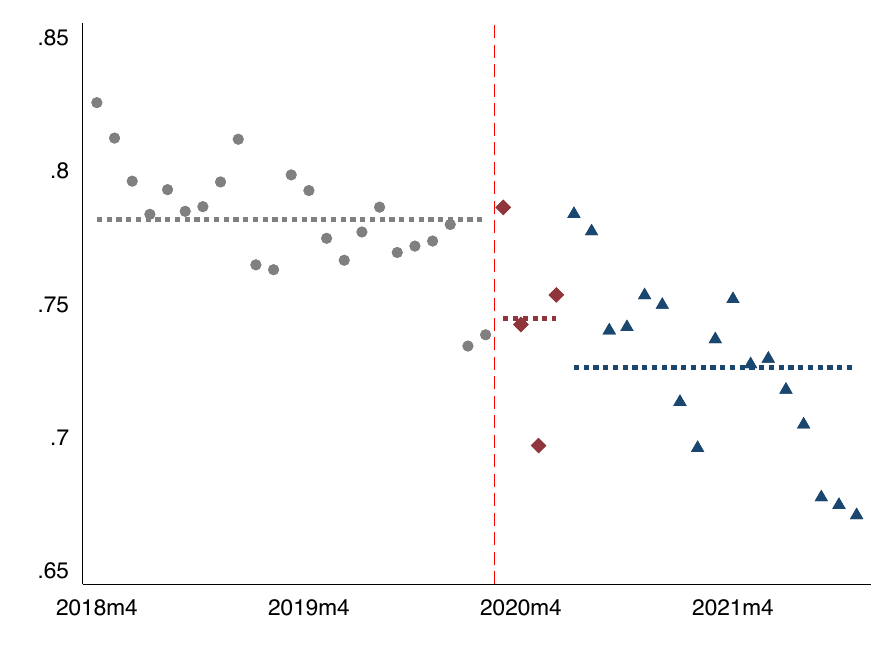}   \\

    \end{array}$
\end{center}
\scriptsize{\textbf{Notes:} Figures present aggregate measures across a broad range of metrics indicative of extent as well as timeliness or accessibility of health care across NHS. The pre-pandemic mean is represented by the gray dotted line, during the first wave by red dotted line, and after the first wave by the blue dotted line. Left panel A studies quantity metrics  clockwise capturing the number of A\&E attendances; the number of completed specialist referrals; the number of diagnostic tests performed; the number of cancer cases that received first treatment. Panel B captures measures indicative of quality or accessibility clockwise measuring the share of A\&E attendances seeing a doctor within 4 hours; the share of completed specialist referrals that had their referral completed within 18 weeks; the share of patients waiting less than 6 weeks for a diagnostic test; the share of cancer patients that have received first treatment within 62 days.}

\end{figure}
\end{landscape}

\paragraph{Quantity} In the top left corner of panel A of Figure \ref{fig:quantity-qual} we see the absolute number of visits to A\&E departments by month. The pre-pandemic months are marked as gray circles, the observations during the first wave by red diamonds, and after the first wave by blue triangles. The averages during these three periods are added as horizontal lines. We see that before the pandemic the absolute number is relatively stable, oscillating around a mean of 1.3 million visits per month. We see that  during the first wave, visits drop sharply to  an average of 0.9 million visits. This drop can likely be attributed to both   people being less exposed to other infectious diseases and risky activities during the lockdown, as well as a demand effect driven by people fearing potential exposure to COVID-19, and therefore avoiding a visit to A\&E. After the first wave, we see strong fluctuations trending upwards towards the original pre-pandemic levels.

\begin{landscape}
\begin{figure}[h!]
\caption{Measuring distribution of quantity and quality of health care across NHS trusts over time \label{fig:quantity-qual-timefe}}
\begin{center}
$
\begin{array}{llll}
\multicolumn{2}{c}{\text{\emph{Panel A:} Measures of quantity }} & \multicolumn{2}{c}{\text{\emph{Panel B:} Measures of quality }}.  \\  \\
\multicolumn{1}{c}{\text{ A\&E attendance }} & \multicolumn{1}{c}{\text{Completed RT referral}}  & \multicolumn{1}{c}{\text{ A\&E < 4h }} & \multicolumn{1}{c}{\text{Completed RT referral < 18 weeks}} \\  

     \includegraphics[width=.26\columnwidth]{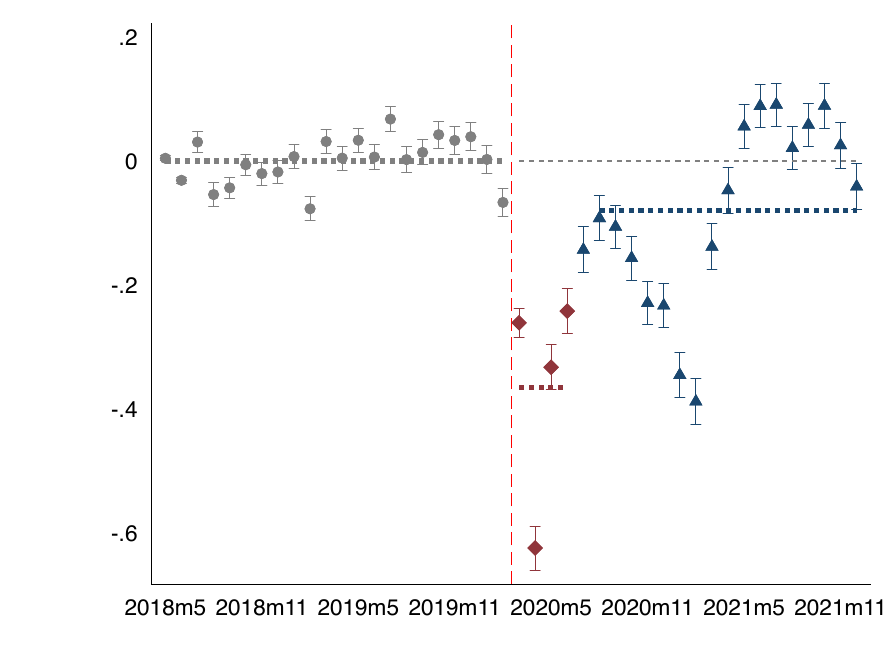}  &     \includegraphics[width=0.26\columnwidth]{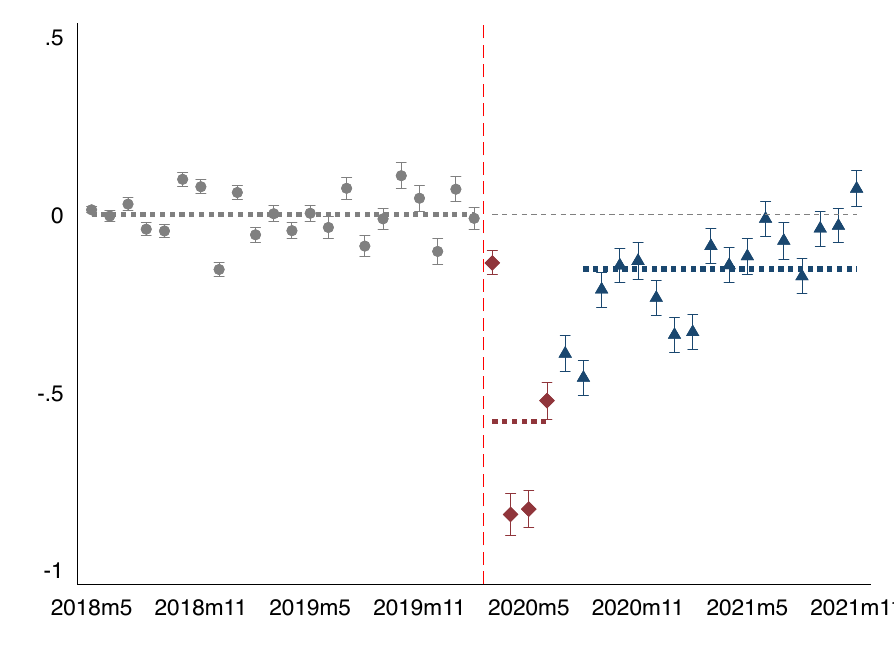}    & 
 \includegraphics[width=.26\columnwidth]{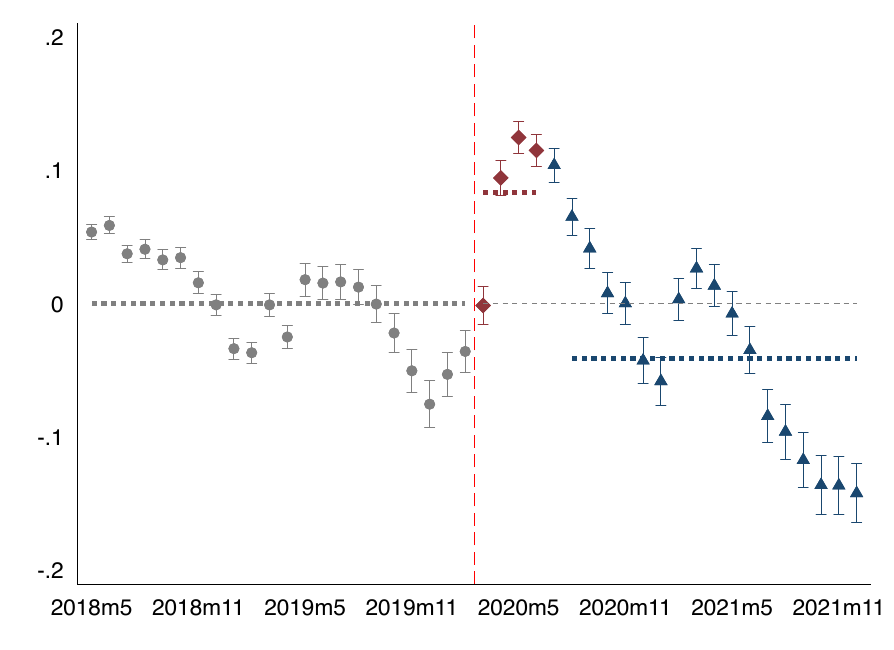}  &     \includegraphics[width=0.26\columnwidth]{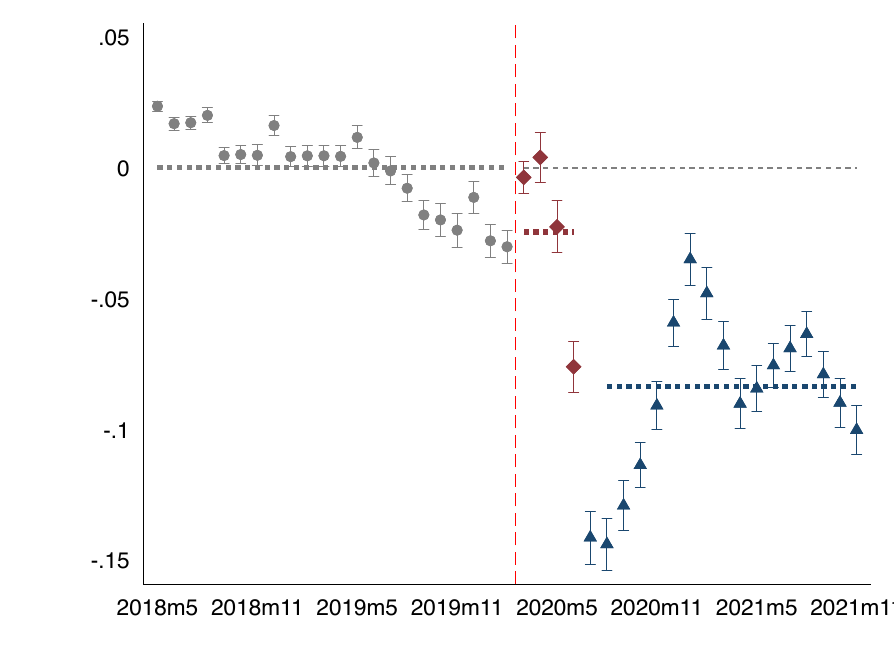}    \\
 
\multicolumn{1}{c}{\text{Diagnostic waiting list }} & \multicolumn{1}{c}{\text{Cancer cases treated}}  & \multicolumn{1}{c}{\text{Diagnostic < 6 weeks }} & \multicolumn{1}{c}{\text{Cancer treatment < 62 days}} \\\  
 \includegraphics[width=0.26\columnwidth]{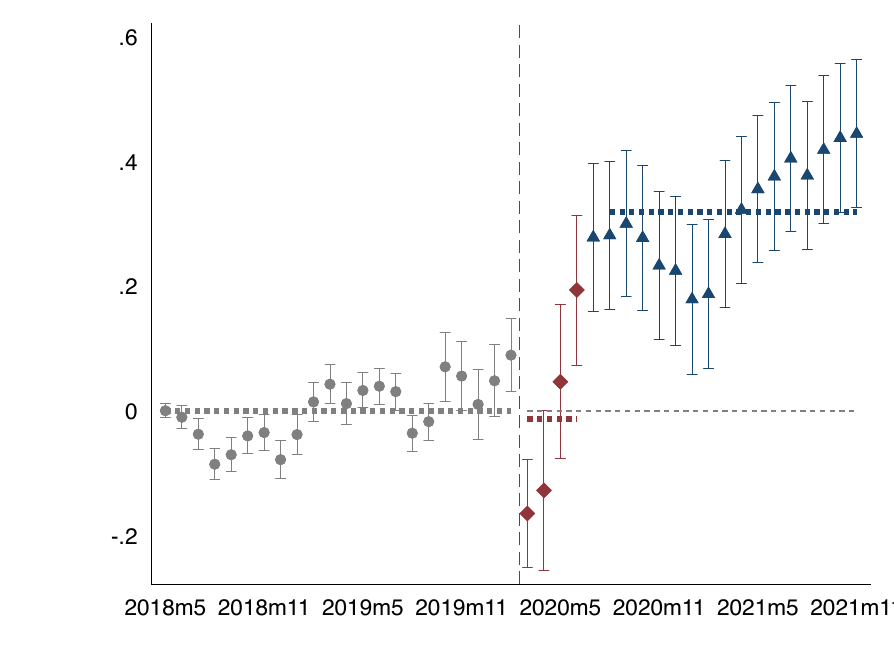} & \includegraphics[width=0.26\columnwidth]{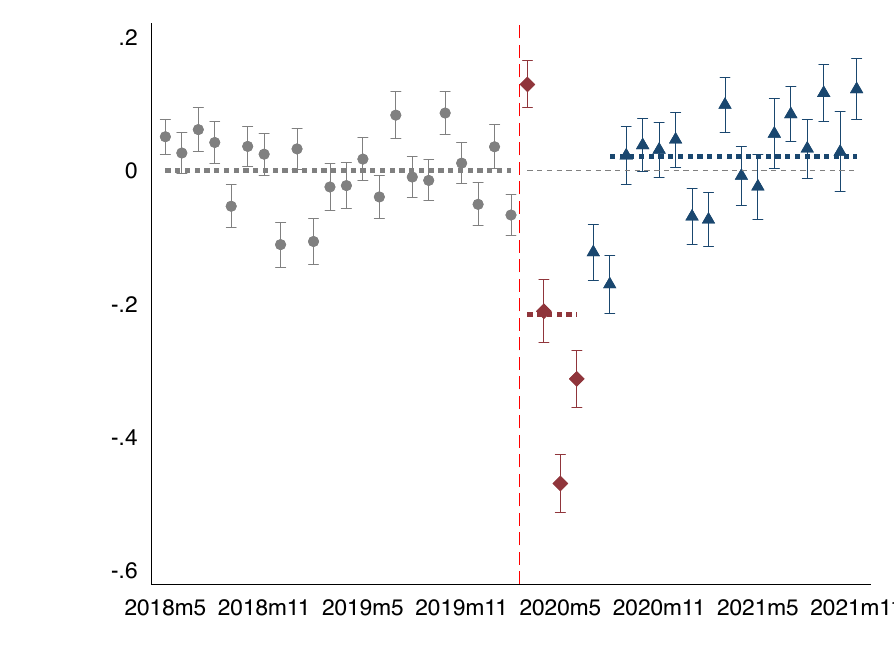}   &  
 \includegraphics[width=0.26\columnwidth]{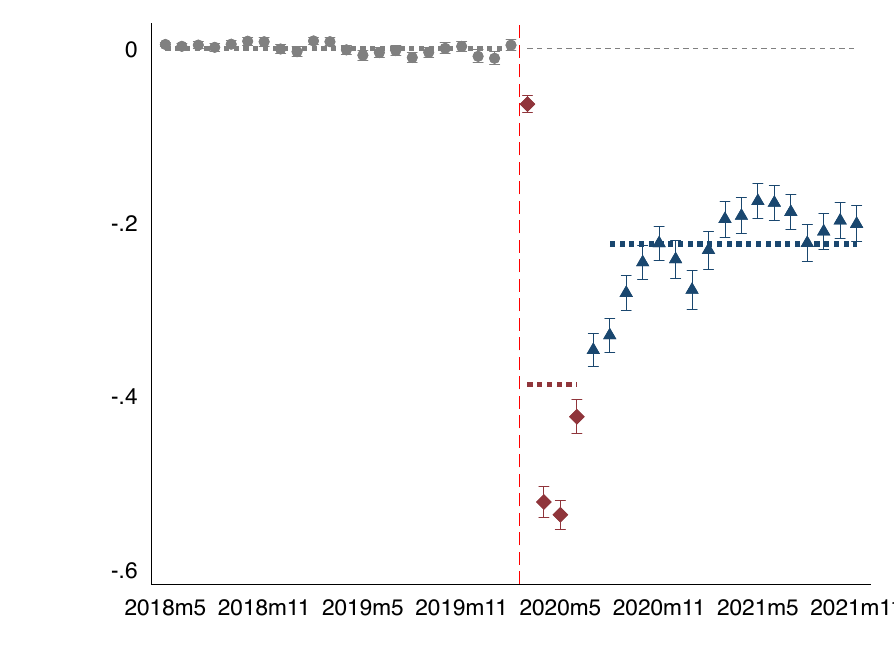} & \includegraphics[width=0.26\columnwidth]{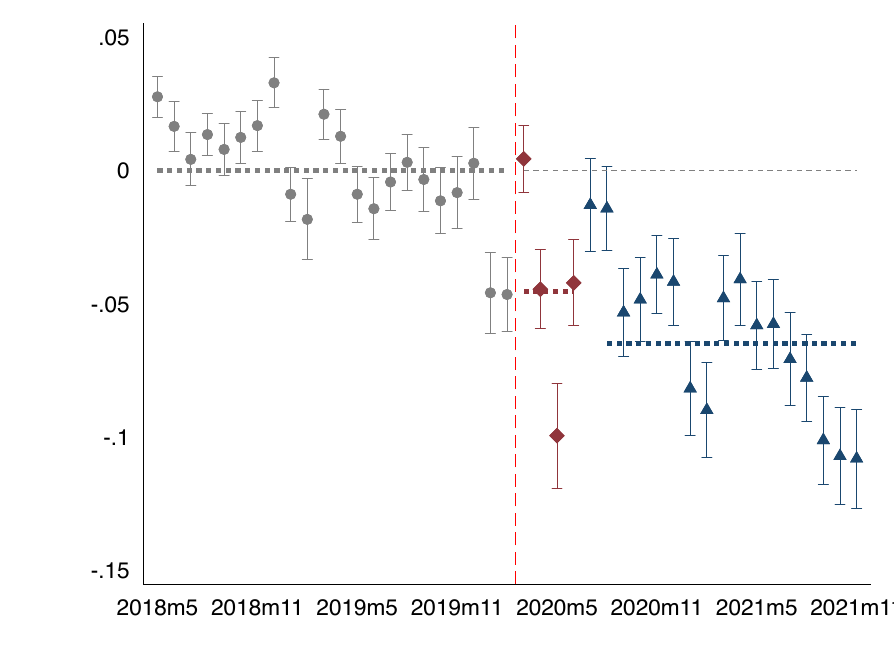}  \\

    \end{array}$
\end{center}
\scriptsize{\textbf{Notes:} Figures plot out estimated time effects and 90\% confidence intervals capturing both the time average as well as the distribution of that time average across a broad range of metrics indicative of extent as well as timeliness or accessibility of health care across different NHS providers. All regressions include provider fixed effects centering the pre COVID-19 arrival data around zero. The pre-pandemic mean of coefficients is represented by the gray dotted line, during the first wave by red dotted line, and after the first wave by the blue dotted line. Left panel A studies quantity metrics  clockwise capturing the number of A\&E attendances; the number of completed specialist referrals; the number of diagnostic tests performed; the number of cancer cases that received first treatment. Panel B captures measures indicative of quality or accessibility clockwise measuring the share of A\&E attendances seeing a doctor within 4 hours; the share of completed specialist referrals that had their referral completed within 18 weeks; the share of patients waiting less than 6 weeks for a diagnostic test; the share of cancer patients that have received first treatment within 62 days.}

\end{figure}
\end{landscape}

\paragraph{Quality}  In the top left corner of panel B of Figure \ref{fig:quantity-qual} we show that the share of visits to A\&E whose waiting times are less than the NHS set goal of 4 hours. Pre pandemic, the share fluctuates around 80\% with seasonal variation. While the share increases with onset of the pandemic to an average of almost 90\% during the first wave, the share begins to drop post pandemic, and is around 65\% for the last months in the dataset. Looking at the aggregates of the three defined periods in Table  \ref{table:summary-statistic-disruption}, we see in column (5) that after the first wave the share of those waiting less than the NHS set goal of  4h is  3.8 percentage points (pp) lower  than before the pandemic.\footnote{Following a similar logic as panel B of Figure  \ref{fig:quantity-qual}, we also look at the increase in the fat tail of waiting times in Appendix Figures  \ref{fig:quality_long} and \ref{fig:quality_long_time_fe}. Moreover, we show the distributions of qualities in the different time intervals in Appendix Figure \ref{fig:distrib_quality}. Pre-pandemic, close to no patients wait more than 12 hours when visiting A\&E. However, in the mean time the share is rising steadily to 1\%. While only 2.5\% wait more than 36 weeks for a referral to a specialist before the pandemic, this share skyrockets to 15\% after the first wave, before stabilizing at just below 10\%. For diagnostics, the fat tail of long waits above 12 weeks is nearly non-existent before the pandemic but peaks at about one third during the first wave, and still remains fairly flat above 10\% since. The share of those waiting more than 104 days for cancer treatment after an urgent referral is close to 6\% before the pandemic, but almost doubles at one point during the first wave and is close to 9\% over the last months in the dataset. In Figure \ref{fig:quantity-qual-timefe} we  look at the time fixed effects when running the regressions at the trust level. The patterns are confirmed and the confidence intervals suggest that the increases in high waiting times is not driven by few trusts.}

\subsection{Referrals to specialist treatment}

\paragraph{Quantity} In the top right corner of panel A of Figure \ref{fig:quantity-qual} we see the absolute number of completed referrals to specialists. Before the pandemic, on average about 1.2 million referrals to specialists are completed per month. During the first wave this number drops dramatically  by almost half to around 0.7 million completed referrals. Following the  first wave the number of completed referrals rebounds to pre-pandemic levels.

\paragraph{Quality} In the top right corner of panel B of Figure \ref{fig:quantity-qual} we see the share of  referrals that are completed within the NHS set goal of 18 weeks, which trended downwards before the pandemic hitting 80\% before the outbreak. While the initial levels during the onset remain at comparable levels, the share drops to two-thirds after the first wave, then shoots up again yo almost 80\% before  trending downwards to around three-quarters in the last months of the dataset. The aggregates in  Table  \ref{table:summary-statistic-disruption} indicate that the share of completed referrals within the NHS set goal of 18 weeks is 2.4pp and 7.7pp lower during and after first wave compared to before the pandemic. While during the first wave the share increases by 4.8pp for the admitted and decreases by 4.9pp for the non-admitted patients, after the first wave the drop is similar across both types of patients.

\subsection{Diagnostic waiting list}
\paragraph{Quantity} In the bottom left corner of panel A of Figure \ref{fig:quantity-qual} we see the absolute number patients on the diagnostic waiting list. Before the pandemic the average length of this waiting list is below one million people. When the pandemic hits, the list shortens briefly during first two months of the first wave and then increases to almost 1.4 million people in the last month of the dataset.

\paragraph{Quality} In the bottom left corner of panel B of Figure \ref{fig:quantity-qual} we see the share of patients on the waiting list who have been waiting less than the NHS set goal of 6 weeks. While pre pandemic the great majority of the  list is waiting less than 6 weeks, with only very moderate fluctuations across months, during the first wave the  share of the list  waiting less than 6 weeks drops to nearly half. After the first wave this share increases but stabilizes below 80\%. The aggregates in  Table  \ref{table:summary-statistic-disruption} suggest that instead of 97\% of the waiting less being within the NHS set goal before the pandemic, the shares drops to 56\% during and 71\% after the first wave.

\subsection{Cancer treatment and detection}

\paragraph{Quantity} In the bottom right corner of panel A of Figure \ref{fig:quantity-qual} we present the total number of urgent suspected cancer patients receiving their first treatment. We see that before and after the first wave the average levels of treatments is very similar at  13,326 treated cases per month. However, during the first wave many urgent treatments are displaced dropping by an average of  2,352 cases per month. In panel A of Appendix Figure  \ref{fig:quantity-qual-cancer} we further show that the absolute number of consultations with specialists following urgent referrals fluctuates around an average of 189,983 per month before the pandemic, and drops below 100,000 at the height of the first wave, before increasing above pre-pandemic levels after the first wave. A similar pattern can be observed for treatments following decisions to treat.\footnote{The difference between referrals to first treatment, with a goal of 62 days, and decisions to treat leading to first treatment, with a goal of 31 days, is that the referrals to first treatment measure includes only urgent cases, whereas decisions to treat leading to first treatment include all suspected cancer types.} During the first wave the monthly average of urgent referrals being seen by a specialist dropped  by  60,772 cases per month as can be seen in Table  \ref{table:summary-statistic-disruption}.

Given that cancer is a serious negative shock and an outcome which is unlikely to be strongly affected by short-term behavioral responses to lockdowns, we look into the heterogeneity across different types of cancer. In Panel A of Appendix Figure \ref{fig:suspected-type-cancer-quantity} we show  the absolute number of cancer referrals that received their first treatment for breast, lower gastrointestinal, lung, skin, urological, and other cancers  drop drastically during the first wave. However, only for lung and urological cancer levels remain below pre-pandemic levels after the first first wave. In panel B of Appendix Figure \ref{fig:suspected-type-cancer-quantity} we show the number of urgent referrals leading to a consultation with a specialist. Even for suspected children's cancer, a disease affecting a demographic group which in general is only mildly affected by COVID19,  the number of urgent referrals leading to a consultation with a specialist drops from an average above 800 cases per month before the pandemic to less than half at the peak of the first wave, despite the arrival rate of this disease likely being constant.

\paragraph{Quality} In the bottom right corner of panel B of Figure \ref{fig:quantity-qual} it becomes clear that waiting times to first treatment has increased over the course of the pandemic. For the last observations only two-thirds of cases receive their first treatment within the NHS set goal of 62 days.  Panel B of Appendix Figure  \ref{fig:quantity-qual-cancer} shows that the share of referrals being seen by a specialist within 14 days remains close to pre-pandemic levels of above 90\% during the first wave, but  deteriorates  to almost 75\% after the first wave. Similarly, the share of treatments taking place within 31 days after the decision to treat drops down to below 94\% compared to an average above 96\% before the pandemic. In panel A of Appendix Figure \ref{fig:suspected-type-cancer-quality} we show for each   type of cancer that the quality, measured in terms of receiving treatment within less than 62 days after an urgent referral, suffers not only during the first wave, but that the negative impact persists, and in some cases even deteriorates further in more recent months.

\subsection{Across trust dispersion}

\begin{figure}[h!]
\caption{Performance targets of referrals and treatment of suspected cancers by type of cancer  \label{fig:suspected-type-cancer-quality-timefe}}
\begin{center}
$
\begin{array}{l}
\multicolumn{1}{l}{\text{\emph{Panel A}:   Referrals  to treatment < 62 days}} \\
  \includegraphics[width=1\columnwidth]{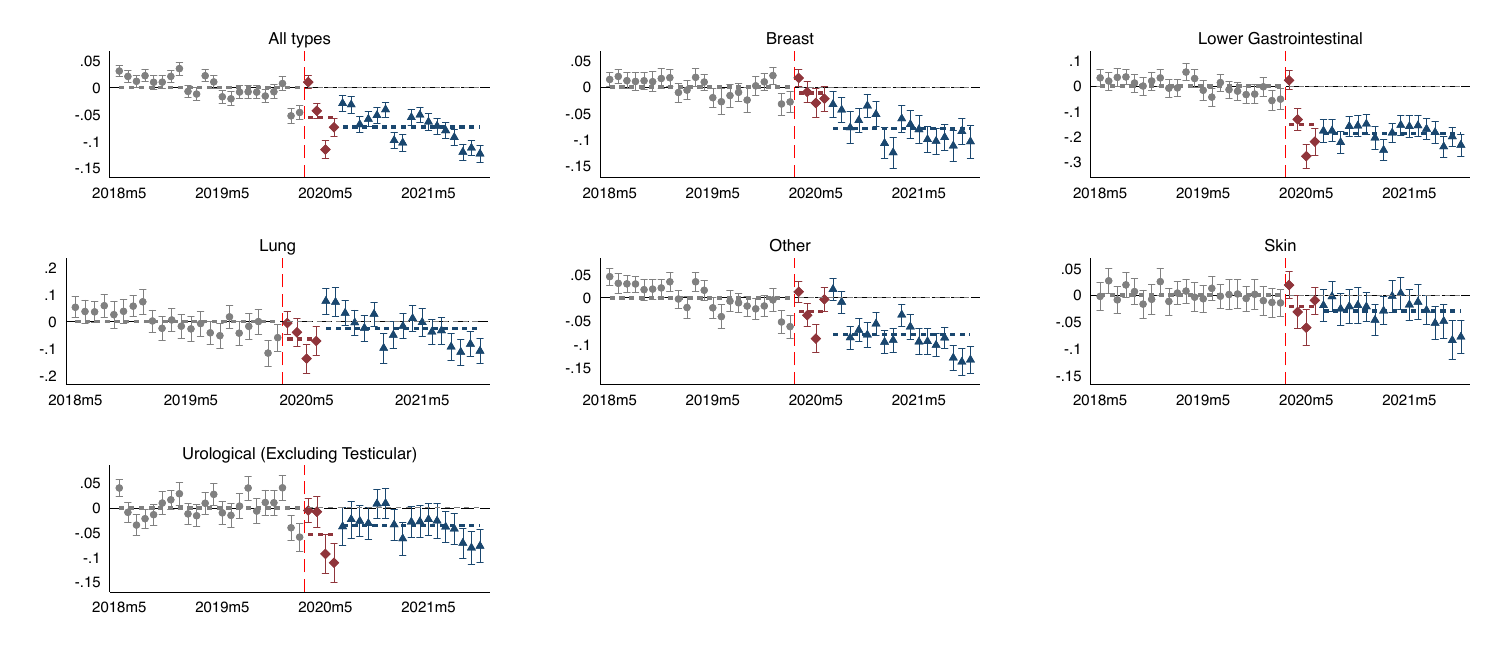}  \\
  \\
    \multicolumn{1}{l}{\text{ \emph{Panel B}:  Referrals  to consultation < 14 days}} \\
  \includegraphics[width=1\columnwidth]{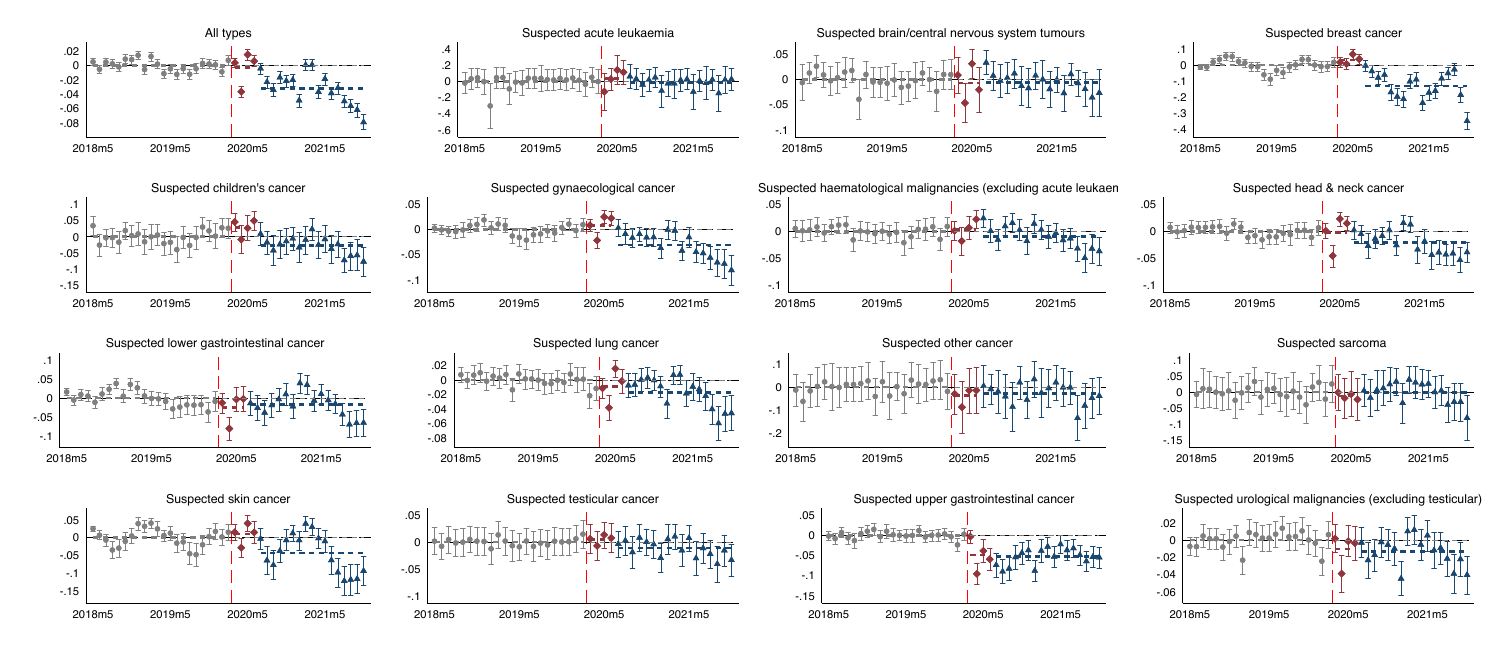}  \\

  %&   \includegraphics[width=0.55\columnwidth]{zoom-in-allocation-msoa-hes.pdf} \\  
    \end{array}$
\end{center}
\scriptsize{\textbf{Notes:} Figures plot out estimated time effects and 90\% confidence intervals capturing both the time average as well as the distribution of that time average across  treatment and detection of different types of cancer  indicative of timeliness or accessibility of health care across different NHS providers. The pre-pandemic mean is represented by the gray dotted line, during the first wave by red dotted line, and after the first wave by the blue dotted line.  Panel A captures measures indicative of quality or accessibility from left to right measuring the share of urgent referrals leading to first treatment within 62 days and panel B shows urgent cancer referrals seen by a specialist within 14 days .}

\end{figure}

For each of the previously discussed   outcomes we also run regressions at the trust level, while including trust and time fixed effects, and clustering the standard errors at the trust level. In Figure \ref{fig:quantity-qual-timefe} we plot the coefficients for each month with the corresponding 90\% confidence interval. For the sake of interpretation we center the average of the pre-pandemic coefficients around zero such that coefficients during the pandemic can be interpreted as deviations from the mean in normal times. Both the patterns for quantity (panel A) and quality (panel B) are confirmed.\footnote{In Appendix Figures   \ref{fig:suspected-type-cancer-quantity-timefe} and \ref{fig:suspected-type-cancer-quality-timefe} we show the same coefficients for difference types of cancer.}

\subsection{Overall non-COVID-19 excess mortality}\label{sec:excessdeath}
\begin{figure}[h!]
\caption{Estimated non-COVID-19 excess deaths \label{fig:non-covid19-excess-deaths}}
\begin{center}
$
\begin{array}{l}
     \includegraphics[width=1\columnwidth]{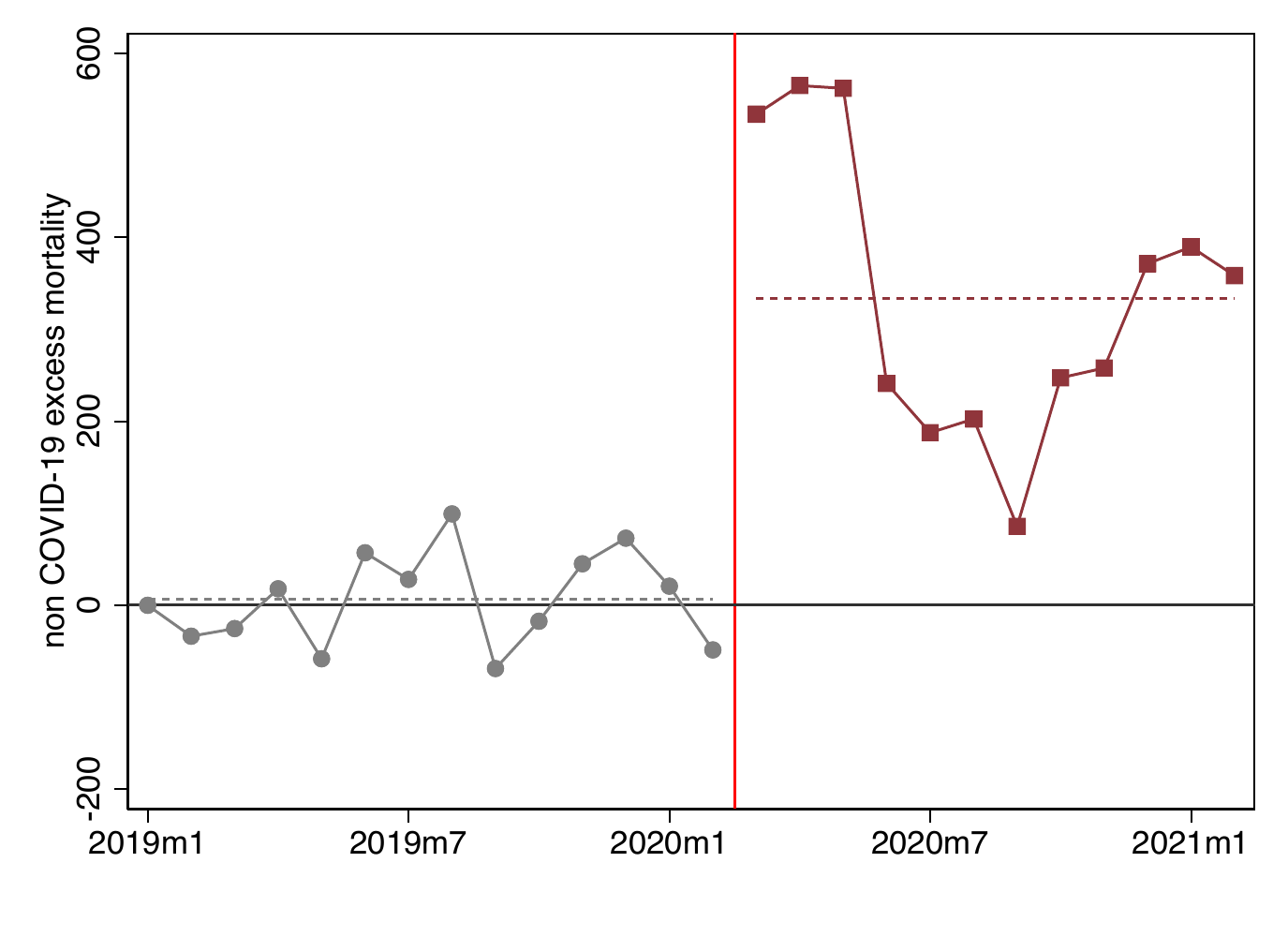}   

    \end{array}$
\end{center}
\scriptsize{\textbf{Notes:} Figures plot the evolution of the estimated non-COVID-19 excess mortality across all providers covered with in the SHMI dataset. We observe that up until February 2020 the aggregate number of excess deaths hovered around zero. From March 2020 onwards excess deaths as measured in the SHMI data shoot up. The cumulative total of non-COVID-19 excess deaths stands at 4,003. }

\end{figure}

 Figure \ref{fig:non-covid19-excess-deaths} provides a picture capturing the evolution of this estimate excess death up until February 2021 summed up across NHS trusts. Reassuringly, we observe that the estimate of excess death is close to zero up until February 2020, before shooting up. In total, we estimate that there are 4,003 excess deaths that are not related to COVID-19 from March 2020 to February 2021. These are cases that, under normal circumstances, considering the patients characteristics, should \emph{not have died} within 30 days of their hospital episode.

\section{Within pandemic pressures on non-COVID-19 care resulting from COVID-19 admissions}
In the previous section we saw that both quantity and quality across a range of treatments and services deteriorated drastically with the arrival and continuation of the pandemic. The results, however, also suggest that there is significant heterogeneity across NHS trusts and across time. We aim to understand that heterogeneity a bit better. While the arrival of the pandemic marks an unexpected shock, the initial reaction to the shock may not be representative of the ongoing pressures that COVID-19 poses on hospitals in the foreseeable future. 

For each of the outcomes, we regress measures of quantity and quality discussed previously on three  different measures of COVID-19 pressures at the provider level: the logs of i) new COVID-19 admissions, the number of COVID-19 cases in  hospital, and the number of COVID-19 cases on ventilators.  We control for an exhaustive set of fixed effects adapted to the unit of analysis and listed at the bottom of the regression results in Tables  \ref{table:ae_statistics-pressures}-\ref{table:cancer-pressures}, while clustering standard errors at the provider level.

\subsection{A\&E attendance and waiting times}
In  columns (1) and (2) of  Table \ref{table:ae_statistics-pressures}  we do not see much of a relationship between COVID-19 pressures and A\&E visits or admissions. However,  in column (3) - (5), we find that COVID-19 related pressures exhibit a positive and significant relationship with increases in the share of patients waiting more than 4 hours, 4-12 hours, and more than 12 hours. For instance, in column (3) of panel A, a 1\% increase in new COVID-19 admissions is associated with an  increase of 0.03\% in the share of visitors to A\&E waiting more than the NHS set goal of 4 hours. While an increase of 0.03\% might not sound dramatic, one has to bear in mind  that COVID-19 admissions fluctuate wildly during the observed period.  Increases of 100\% are not uncommon as can be seen in Appendix Figure \ref{fig:month_on_month_distrib_new_admissions} which shows the distribution of month on month changes in log(1+new COVID-19 admissions) at the trust level. The standard deviation of changes is 0.65 and  13.5\% of observations see swings of more than 100 log points.

\subsection{Referrals to specialist treatment}
In  Table \ref{table:rtt-pressures} we look at three different groups of outcomes for referrals: i)  total referrals, ii) length of the waiting list, and iii) waiting times. When outcomes are related to multiple months of actions, we  capture cumulative pressures by   summing  pressures over the corresponding number of months. In column (1) we see that new referrals tend to decrease when new COVID-19 admissions or cases in the hospital increase. However, when inspecting columns (2) and (3) it becomes clear that this effect is predominantly driven by completed referrals amongst admitted patients who see a drop of 0.09\% when new COVID-19 admissions increase by 1\%.

In column (3) we see a weakly positive relationship between the length of the waiting list and pressures. However, in column (5) we find that the aggregate waiting list exhibits a strong significant relationship with pressures, for instance, increasing by 0.03\% with a 1\% increase in COVID-19 cases on ventilators.

Concerning waiting times the evidence is stark. The share of referrals waiting more than the NHS set goal of 4 weeks in column (7), 8 weeks in column (8), or 12 weeks in column (9) is highly significant for each of the proxies for COVID-19 related pressures.

\subsection{Diagnostic waiting list}

In column (1) of  Table \ref{table:diagnostic-pressures}  we find no significant evidence about the relationship between COVID-19 pressures and the quantity of diagnostic activities. A similar conclusion can be drawn from the impacts on average waiting times or the share waiting more than NHS set goal of 8 weeks. However, when we look at scans using computerized tomography (CT), which are specialized  scarce machines, we detect a strong significant increase in waiting times and the share of referrals waiting for a scan when new COVID-19 admissions increases. This can likely be explained by the reliance on CT scans to gauge the extent of damage to the lung exerted by COVID-19 and resulting congestions in the health care system.

\subsection{Cancer treatment and detection}

In Table \ref{table:cancer-pressures} we look at how COVID-19 pressures relate to cancer consultations and treatments. In columns (1) and (4) we show that neither the share of urgent referrals seen by a specialist  nor their waiting times seem to be systematically related to COVID-19 pressures. However, for  treatments  we see both a significant reduction in those following the decision to treat as well as those following referrals  when COVID-19 pressures increase. For instance, as can be seen in  column (3), a 1\% increase in new admissions is associated with a 0.02\% drop in those receiving urgent first treatment.  Moreover, in columns (5) and (6) we show that waiting times for treatments following the decision to treat and referrals increase substantially with COVID-19 pressures.

\subsection{Non-COVID-19 excess mortality}
We next turn to study the impact of ongoing COVID-19 pressures on non COVID-19 hospital excess mortality. The analysis of aggregate figures suggests that with the onset of COVID-19, there has been a quite persistent increase in non COVID-19 related hospital excess mortality. We next explore to what extent there is cross-provider variation in this excess mortality and to what extent it can be linked to COVID-19 induced pressures. 

The results from this analysis are presented in Table \ref{table:excess-deaths-main}. Across the three panels we capture different measures of COVID-19 pressures with the average monthly number of new daily hospital admissions as main measure being presented in Panel A.  The dependent variable measures the month on month change in the number of excess deaths, which, as was indicated in section \ref{sec:excessdeath}, approximates the monthly number of excess death quite accurately as long as we control for the base effects  $\text{Obs}_{p,t-13} -  \text{Exp}_{p,t-13}$.   Throughout Panel A, the results are quite stable suggesting that increased COVID-19 admission numbers at the provider level translates into some notable structure in the estimate of the excess deaths. The point estimates suggest that a 10\% increase number in average daily new hospital admissions due to COVID-19 translates into 0.41 additional excess deaths. This result remains robust across different specifications, including, adding linear-time trends at the provider level in column (6).

\begin{figure}[h!]
\caption{Impact of COVID-19 pressures on non-COVID-19 excess mortality: effect across different deciles of the COVID-pressure intensity \label{fig:coefplot-shmi-nonlinear}}
\begin{center}
$
\begin{array}{ll}
%\text{\emph{Panel A}: log(Total \# of cancer patients treated)   }  
  \includegraphics[width=1\columnwidth]{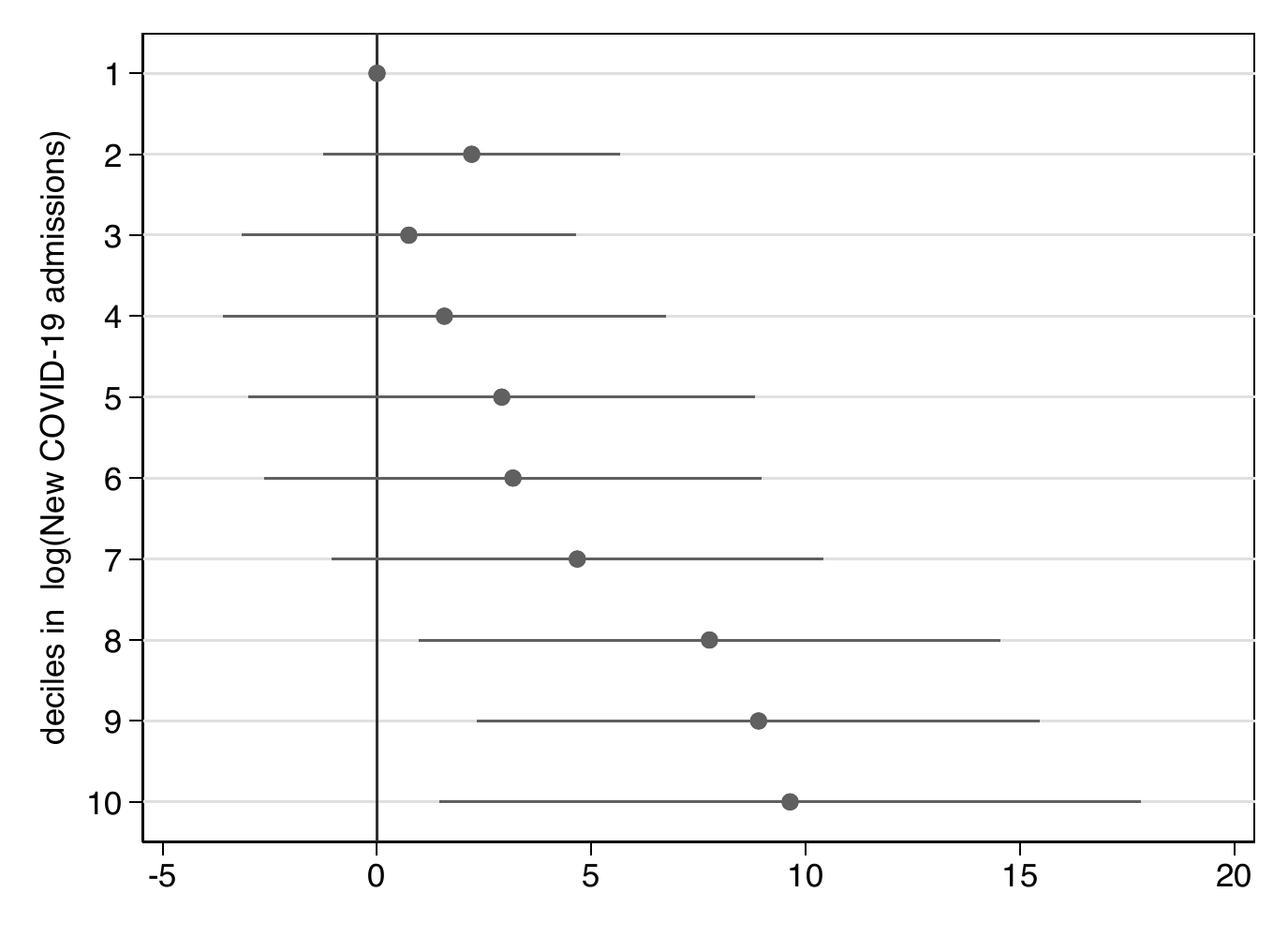} \\
     \end{array}$
\end{center}
\scriptsize{\textbf{Notes:} Figure presents heterogenous treatment effects capturing the impact of COVID-19 pressures on excess mortality. The dependent variable measures the month-on-month changes in excess mortality to proxy month-specific excess mortality. All regressions control for provider fixed effects , time fixed effects and provider-specific linear trends along with as well as month-on-month changes in number of spells.  90\% confidence intervals obtained from clustering standard errors at the provider level are indicated.}

\end{figure}

This suggests that, in particular, pressures resulting from increases in new COVID-19 hospital admissions and occupation of beds with mechanical ventilation is associated with significant worsening survival chances for patients that get admitted to hospital for non COVID-19 reasons. Figure \ref{fig:coefplot-shmi-nonlinear} suggests that the effects are strongest in hospitals that experience, based on its empirical distribution, relatively large shocks of new COVID-19 patients being admitted. Further, Appendix Tables \ref{table:did-excess-mortality}, 
\ref{table:did-logobserved-mortality} and \ref{table:did-deathrate-mortality} find very similar results when adapting the empirical design to study the excess mortality data as reported by NHS digital as twelve months rolling cumulative totols, for which, we then construct a rolling cumulative measure of the COVID-19 pressures to match that data structure.

\begin{figure}[h!]
\caption{Impact of COVID-19 pressures on diagnosis specific excess mortality \label{fig:coefplot-shmi-diagnosis}}
\begin{center}
$
\begin{array}{ll}
%\text{\emph{Panel A}: log(Total \# of cancer patients treated)   }  
  \includegraphics[width=1\columnwidth]{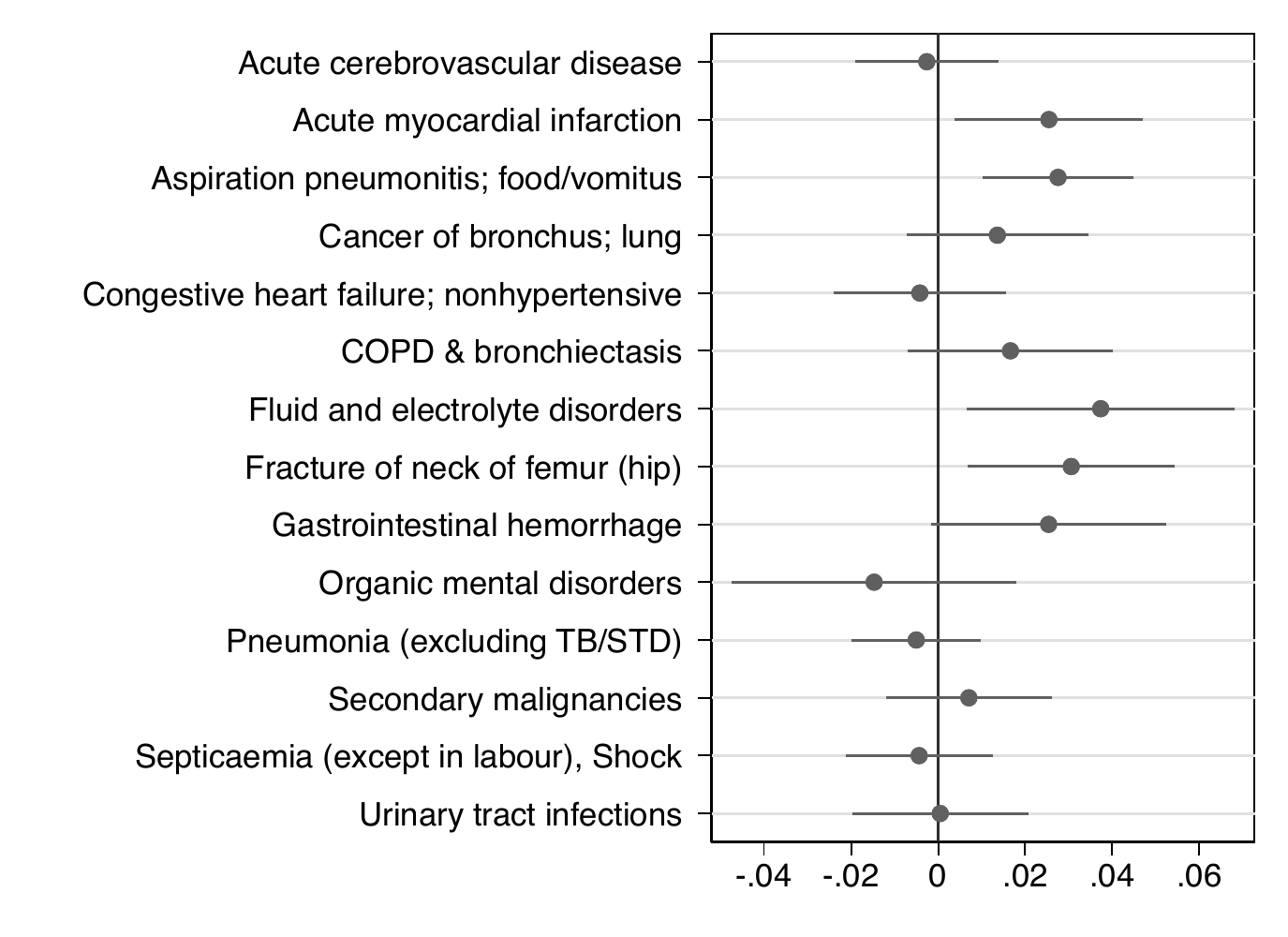} \\
     \end{array}$
\end{center}
\scriptsize{\textbf{Notes:} Figure presents heterogenous treatment effects capturing the impact of COVID-19 pressures on diagnosis specific excess mortality. The estimating equation explores variation in the log differences in observed minus expected deaths for hospital episodes and diagnosis for which data is available for the whole sample period and for diagnosis that are consistently included in the data across at least 100 of the 127 NHS providers for which the data is constructed. All regressions control for provider by diagnosis fixed effects as well as diagnosis by time fixed effects and control for the diagnosis-specific relationship between log(spells) and excess deaths. 90\% confidence intervals obtained from clustering standard errors at the provider level are indicated.}

\end{figure}

 The diagnosis-specific data is too sparse to allow us to estimate the preferred specification exploiting month-on-month changes.\footnote{This is driven by the noisiness of month-on-month changes  due to the fact that confidentiality protection implies many missing observations once the excess death is broken down by diagnosis group.} Hence, we work with the cumulative twelve month rolling window design to study to what extent COVID-19 pressures in the last three months affect the cumulative twelve month rolling sum of excess death by diagnostic group. Figure \ref{fig:coefplot-shmi-diagnosis} presents results from estimating such a heterogenous effects version focusing on 15 of the 142 diagnosis codes for which the vast majority of NHS providers have a near complete record of observed and expected deaths, along with the number of spells. The results suggest that the increase in mortality at the provider level is driven, to a significant extent, by urgent care needs, such as heart attacks (acute myocardial infraction). 
This suggests that COVID-19 induced hospital pressures are causing a notable increase in non COVID-19 related excess deaths due to, likely, a worse quality of care. We next explore this in some detail.

\section{Mechanisms}
Naturally there is a broad range of mechanisms that may be at play. In this section we focus on two datasets to explore the underlying mechanisms. We focus in particular on the cancer care as well as the excess death dataset as they provide hard outcome measures capturing performance measures that are end-to-end. 

\subsection{Staff absence rates}
We begin by studying staff absence rates. Using monthly data at the provider by staff group level measuring the number of full-time equivalent (FTE) days that are available in a given month by provider and staff group, along with the number of FTE days that staff are absent, we construct staff absence rates for each staff group $s$ by provider $p$ in month $t$, $s_{i,p,t}$.

We estimate

$$s_{i,p,t} =   \alpha_i + \nu_p + \gamma_t + \beta \times \text{COVID-19}_{p,t} + \epsilon_{p,t}. $$

The results are presented in Table \ref{table:staff absences}. The results suggest that COVID-19 pressures across providers are associated with significantly higher staff absence rates, even after controlling for the vaccination rate in the broader population. Column (1) suggests that an increase in average new COVID-19 daily admissions per month is associated with an increase in staff absence rates by 0.5 percentage points. This effect is notably carried by nurses, the most important staff group in terms of size, and much less so by managers and doctors. This highlights that a likely mechanism driving the differential absence rates  between different types of staff is the likely more direct and ongoing exposure to patients that nurses have with patients vis-a-vis NHS managers.

We would expect that vaccination should weaken this relationship notably. We carry out this exercise in Table \ref{table:staff absences-vaccination} adding an interaction term capturing two different measures of NHS staff vaccination take up at the provider level. This data is only available from October 2021 onwards at which point staff vaccination rates had been quite high, averaging at around 91\% across providers. Yet, there remains residual variation with the vaccination rate ranging from 82\% to 96\%.  The results suggest that vaccination rates at the NHS level significantly reduce the link between COVID-19 admissions and staff absence rates. This effect is only to be seen robustly for measures of new COVID-19 admissions and hospital cases, rather than the number of patients on mechanical ventilation beds, which is not surprising, given that the latter is varying less on a month-on-month basis as patients may be on mechanical ventilation beds for a prolonged period.

\subsection{Excess mortality} 
We next explore to what extent the excess mortality results we documented before appear to weaker among providers with higher vaccination uptake. The previous exercise would suggest that if a part of the increase in excess mortality may be due to staff absence rates increasing in COVID-19 pressures, this could have a moderating effect on non-COVID-19 excess mortality. 

To do so, we estimate similar specifications as above, interacting the COVID-19 pressures with the NHS vaccination uptake cross-sectional measure. That is, we estimate

\begin{align}
Excess_{p,t}  & = &  \alpha_i + \nu_p + \gamma_t + \beta_1 \times \text{NHS vaccination}_{p}  \times \text{COVID-19}_{p,t}    \\
& & + \beta_2 \times \text{COVID-19}_{p,t} + \xi \times \mathbf{X}_{p,t} + \epsilon_{p,t}. \nonumber 
\end{align}

The results pertaining to this analysis are presented in  Table \ref{table:excess-deaths-nhsvaxx}.  As before, we note that hospital providers that see a large influx of COVID-19 patients see a notable increase in non-COVID-19 excess death. This effect, however, is notably weaker for providers that have a higher staff vaccination uptake. Providers with high vaccination take up have one fourth less  non-COVID-19 excess deaths for any given increase in COVID-19 pressures. 

This suggests that staff absence rates may be an important mechanisms, but by far, is not the only mechanism that drives the unobserved variation in the worsening of the care that patients receive in times of COVID-19 stress.

\section{Conclusion}

The COVID-19 pandemic has put drastic pressures on health care systems across the world. In this paper, we look into the knock-on effects on the quantity and quantity of non-COVID19 related health care provision. Further, we study whether these pressures are systematically related to excess mortality and investigate potential mechanisms. 

While vaccinations reduce the probability that hospitalized patients die from COVID-19, our findings show that COVID-19 hospitalizations still have an indirect knock-on effect for other health outcomes. Moreover, absence rates amongst staff have meaningful negative implications for the provision of health care and excess mortality. Therefore, any decision to allow a widescale spread of infections has not only to focus on COVID-19 mortality, but also has to factor in potential spillover and congestion effects.

The fact that vaccination of NHS staff has a positive impact reducing staff absence rates could be important for the heated debate on the  vaccine mandate for NHS health care workers expected to be coming into effect on 1 April 2022 in England.\footnote{The necessity to have received two doses by 1 April 2022 would require a first dose of vaccine by 3 February 2022 given the policy of requiring an 8 week interval between first and second dose in  England.} However, the results  also could lead way to the interpretation  that NHS trusts are inefficiently understaffed  in terms of critical staffs, such as nurses. We leave this important open question to future research.

\newpage
\bibliography{coronabib}
\bibliographystyle{aer}

\clearpage

\section*{Figures and tables}

%%%%%%%%%%%%%%%%%%%%%%%%%%%%%%
%%%%%%%%%%%%%%%%%%%%%%%%%%%%%%

%%%%%%%%%%%%%%%%%%%%%%%%%%%%%%
%%%%%%%%%%%%%%%%%%%%%%%%%%%%%%

%%%%%%%%%%%%%%%%%%%%%%%%%%%%%%
%%%%%%%%%%%%%%%%%%%%%%%%%%%%%%

%%%%%%%%%%%%%%%%%%%%%%%%%%%%%%
\begin{figure}[h!]
\caption{Effect of COVID-19 admissions on staff absence rates across NHS providers by staff type \label{fig:coefplot-stafftype-absence}}
\begin{center}
$
\begin{array}{l}
     \includegraphics[width=1\columnwidth]{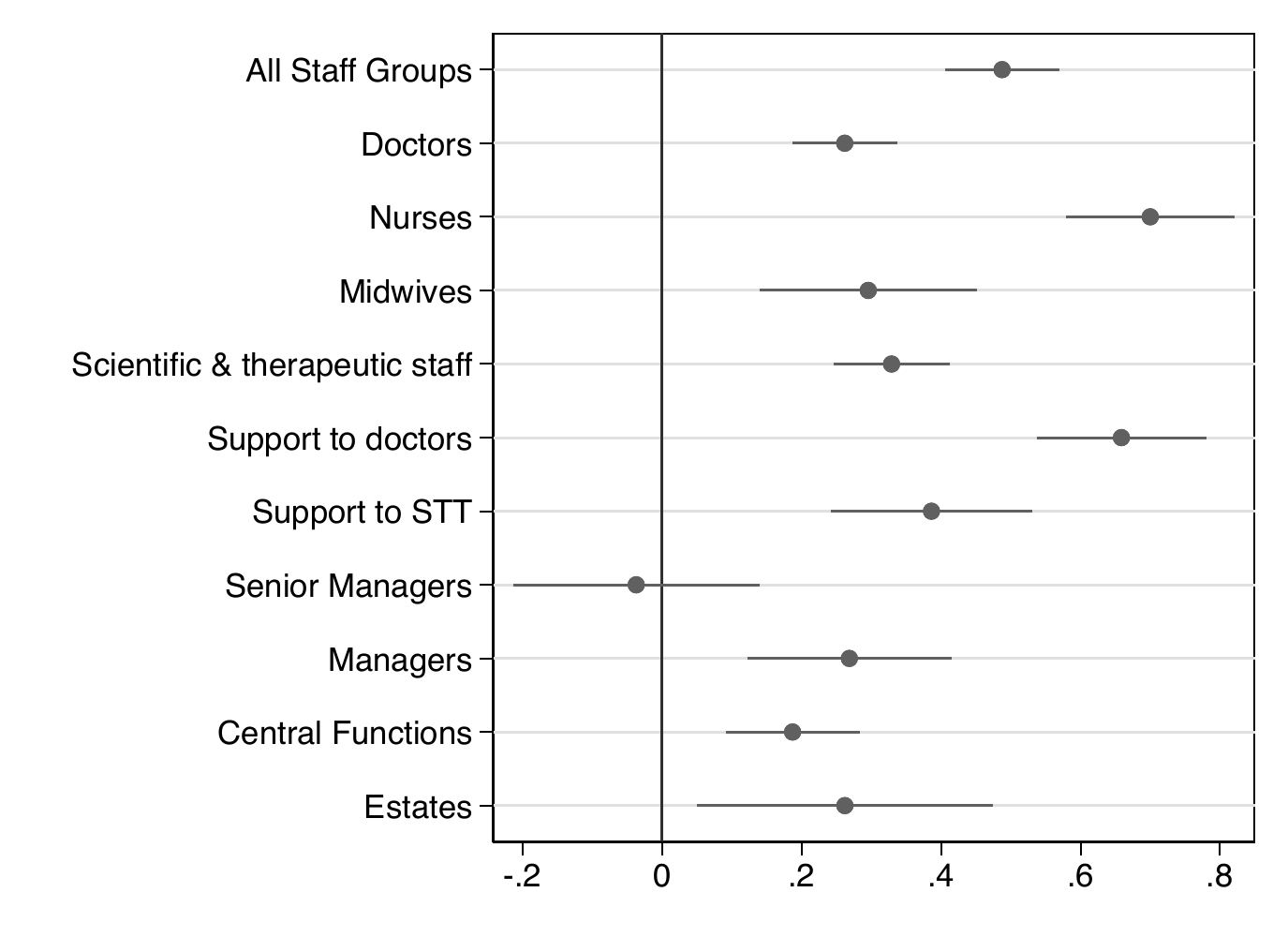}   

    \end{array}$
\end{center}
\scriptsize{\textbf{Notes:} Figures plot the effect of COVID-19 hospital admissions on staff absence rates by staff type. All regressions are estimated after removing provider fixed effects and time fixed effects. 90\% confidence intervals obtained from clustering standard errors at the provider level are indicated. }
\end{figure}
%%%%%%%%%%%%%%%%%%%%%%%%%%%%%%

\clearpage
\newpage

%%%%%%%%%%%%%%%%%%%%%%%%%%%%%%
 \begin{table}[h]
\centering
\scalebox{0.625}{
  \begin{threeparttable}
  \caption{Monthly averages of quantity and quality of health care before and during pandemic  \label{table:summary-statistic-disruption}}
\begin{tabular}{lccccc}
  \hline

\addlinespace
	&\multicolumn{1}{c}{(1)}   &\multicolumn{1}{c}{(2)}   &\multicolumn{1}{c}{(3)}   &\multicolumn{1}{c}{(4)}   &\multicolumn{1}{c}{(5)}  \\
\hline
   
     & \multicolumn{5}{c}{\emph{   Quantity  measures }}        \\
              \cmidrule(lr){2-4}        \cmidrule(lr){5-6} 

      & \multicolumn{1}{c}{Pre pandemic  }  &\multicolumn{1}{c}{ First wave} &\multicolumn{1}{c}{Post first wave}  & \multicolumn{1}{c}{$\Delta$ First wave - Pre}  & \multicolumn{1}{c}{$\Delta$ Post - Pre} \\
  \addlinespace
  
%  A \&E Visits \\
% 
%  A \&E Hospital Admissions \\
% 
%  Referral to Treatment   \\
%  
%  Diagnostics   \\
%  
%  Cancer Referrals   \\
%  Cancer First Treatment    \\

\input{quantity_outbreak.tex}\\

\addlinespace

     & \multicolumn{5}{c}{\emph{   Quality  measures }}          \\
              \cmidrule(lr){2-4}        \cmidrule(lr){5-6} 

       & \multicolumn{1}{c}{Pre pandemic  }  &\multicolumn{1}{c}{ First wave} &\multicolumn{1}{c}{Post first wave}  & \multicolumn{1}{c}{$\Delta$ First wave - Pre}  & \multicolumn{1}{c}{$\Delta$ Post - Pre} \\

%  
%  A \&E Visits \\
% 
%  A \&E Hospital Admissions \\
% 
%  Referral to Treatment   \\
%  
%  Diagnostics   \\
%  
%  Cancer Referrals   \\
%  Cancer First Treatment    \\
  
  \input{quality_outbreak.tex}\\

\addlinespace
\addlinespace
   \hline
   \end{tabular}

    \begin{tablenotes} {\footnotesize
    \item Notes:  Table presents monthly averages of key outcome measures studied capturing various margins of the health care systems absolute performance in terms of quantity as well as quality of health care services being produced. The monthly averages are taken during time windows before the pandemic; during the first wave from March to June 2020; and since July 2020. `Cancer from decision to treat to treatment' includes all cases requiring treatment, while `cancer from referral to treatment' includes only urgent suspected cases. The quality measures are attendance within 4 hours for `A\&E attendance', less than 6 weeks for `diagnostic waiting list', less than 18 weeks for `referral to treatment', less than 14 days for `cancer referral to specialist consultation',  less than 62 days for `cancer from referral to treatment', and less than 31 days for `cancer from decision to treat to treatment'.} \end{tablenotes}
      \end{threeparttable}
      
 }

\end{table}
 %%%%%%%%%%%%%%%%%%%%%%%%%%%%%%

%%%%%%%%%%%%%%%%%%%%%%%%%%%%%%
 \begin{table}[h!]
\centering{
\scalebox{0.725}{
  \begin{threeparttable}
  \caption{Impact of COVID-19 health care system pressures on A\&E activity and performance \label{table:ae_statistics-pressures}}
\begin{tabular}{lccccc}
  \hline

%   & \multicolumn{6}{c}{\emph{DV:  }}  \\
% \cmidrule(lr){2-7}     
\addlinespace
 &\multicolumn{1}{c}{(1)}   &\multicolumn{1}{c}{(2)}   &\multicolumn{1}{c}{(3)}   &\multicolumn{1}{c}{(4)}   &\multicolumn{1}{c}{(5)}  \\

  & \multicolumn{2}{c}{\emph{log(A\&E visits$_t$) }} & \multicolumn{3}{c}{\emph{Patients waiting to be treated or admitted for}}     \\
   \cmidrule(lr){2-3}        \cmidrule(lr){4-6} 
      & \multicolumn{1}{c}{All }  &\multicolumn{1}{c}{ Resulting admissions} &\multicolumn{1}{c}{$>$ 4h } &\multicolumn{1}{c}{  4 - 12 h } &\multicolumn{1}{c}{$ >$ 12 h } \\ 
   \hline
  \addlinespace
\multicolumn{5}{l}{\emph{Panel A}:    } \\
\input{fragment-ae_statisticslognew_cases-new_admissions.tex} \\
\addlinespace
\addlinespace
\multicolumn{5}{l}{\emph{Panel B}:  }\\
\input{fragment-ae_statisticslognew_cases-hospital_cases.tex} \\
\addlinespace
\addlinespace
\multicolumn{5}{l}{\emph{Panel C}:   } \\
\input{fragment-ae_statisticslognew_cases-occupied_mv_beds.tex} \\
\addlinespace
\addlinespace

Provider   FE &X & X & X &X & X\\
Time FE &X & X & X &X & X\\ 
Community transmission &X & X & X &X & X\\ 

   \hline
   \end{tabular}

    \begin{tablenotes} {\footnotesize
    \item Notes:  Regressions present results at the NHS provider level documenting the relationship between different measures of COVID-19 pressures at the provider level across panels on the performance of the A\&E departments and waiting times in a given month. This does not distinguish between COVID-19 and non-COVID-19 related A\&E visits. Columns (1) and (2) measure the quantity of A\&E visits and resulting admissions respectively, while columns (3)-(5) capture the impact that pressures COVID pressures have on waiting times for A\&E visits.  Community Transmission indicates that the regressions control for the log of the number of COVID-19 cases within the catchment areas of NHS providers. Standard errors are clustered at the provider level with stars indicating *** p$<$ 0.01, ** p$<$ 0.05, * p$<$ 0.1.} \end{tablenotes}
      \end{threeparttable}
      
 }
  }    
  
\end{table}
 %%%%%%%%%%%%%%%%%%%%%%%%%%%%%%

%%%%%%%%%%%%%%%%%%%%%%%%%%%%%%
\begin{landscape}
 \begin{table}[h!]
\centering{
\scalebox{0.7}{
  \begin{threeparttable}
  \caption{Impact of COVID-19 health care system pressures on specialist referral \label{table:rtt-pressures}}
\begin{tabular}{lcccccccccc}
  \hline

%   & \multicolumn{6}{c}{\emph{DV:  }}  \\
% \cmidrule(lr){2-7}     
\addlinespace
 &\multicolumn{1}{c}{(1)}   &\multicolumn{1}{c}{(2)}   &\multicolumn{1}{c}{(3)}   &\multicolumn{1}{c}{(4)}   &\multicolumn{1}{c}{(5)} &\multicolumn{1}{c}{(6)} &\multicolumn{1}{c}{(7)} &\multicolumn{1}{c}{(8)} &\multicolumn{1}{c}{(9)} \\

  & \multicolumn{3}{c}{\emph{log(Referrals$_t$) }} & \multicolumn{3}{c}{\emph{log(Waiting list$_t$) }}   & \multicolumn{3}{c}{\emph{Share waiting$_t$ }}     \\
   \cmidrule(lr){2-4}        \cmidrule(lr){5-7}  \cmidrule(lr){8-10}

   & \multicolumn{1}{c}{New}  &\multicolumn{2}{c}{ Completed } &\multicolumn{1}{c}{ Length } &\multicolumn{1}{c}{ Aggregate wait}  &\multicolumn{1}{c}{ Avg.\ wait} &\multicolumn{1}{c}{ $>$ 4 weeks } &\multicolumn{1}{c}{ $> 8$ weeks}  &\multicolumn{1}{c}{ $> 12$ weeks} \\
      & \multicolumn{1}{c}{ }  &\multicolumn{1}{c}{ Admitted } &\multicolumn{1}{c}{Non-admitted} &\multicolumn{1}{c}{   } &\multicolumn{1}{c}{  }  &\multicolumn{1}{c}{  } &\multicolumn{1}{c}{  } &\multicolumn{1}{c}{}  &\multicolumn{1}{c}{ } \\ 
   \hline
  \addlinespace
\multicolumn{5}{l}{\emph{Panel A}:    } \\
\input{fragment-rttlognew_cases-new_admissions.tex}\\ 
\addlinespace
 
\multicolumn{5}{l}{\emph{Panel B}:  }\\
\input{fragment-rttlognew_cases-hospital_cases.tex} \\
\addlinespace
 \multicolumn{5}{l}{\emph{Panel C}:   } \\
\input{fragment-rttlognew_cases-occupied_mv_beds.tex} \\
\addlinespace

Provider x Treatment function FE &X & X & X &X & X & X&X & X & X\\
Treatment function x Time FE &  X& X &  X &  X & X & X &X & X & X\\ 
Community transmission &X & X & X &X & X & X&X & X & X\\
   \hline
   \end{tabular}

    \begin{tablenotes} {\footnotesize
    \item Notes:  Regressions present results at the NHS provider level documenting the relationship between different measures of COVID-19 pressures at the provider level across panels on the accessibility and quality of referrals to specialist treatment. This does not distinguish between COVID-19 and non-COVID-19 related referrals. Columns (1) - (3) focus on measures of output capturing new referrals to specialists and completion of referral pathways. Columns (4) - (6) study broad characteristics of the  waiting list for non-completed  specialist referrals. Columns (7)-(9) provide breakdown of stock of average wait on waiting list for non-completed referrals.   Community Transmission indicates that the regressions control for the log of the number of COVID-19 cases within the catchment areas of NHS providers. Standard errors are clustered at the provider level with stars indicating *** p$<$ 0.01, ** p$<$ 0.05, * p$<$ 0.1.} \end{tablenotes}
      \end{threeparttable}
      
 }
  }    
  
\end{table}
\end{landscape}
 %%%%%%%%%%%%%%%%%%%%%%%%%%%%%%

%%%%%%%%%%%%%%%%%%%%%%%%%%%%%%
 \begin{table}[h!]
\centering{
\scalebox{0.725}{
  \begin{threeparttable}
  \caption{Impact of COVID-19 health care system pressures on diagnostic  activity and performance \label{table:diagnostic-pressures}}
\begin{tabular}{lccccc}
  \hline

%   & \multicolumn{6}{c}{\emph{DV:  }}  \\
% \cmidrule(lr){2-7}     
\addlinespace
 &\multicolumn{1}{c}{(1)}   &\multicolumn{1}{c}{(2)}   &\multicolumn{1}{c}{(3)}   &\multicolumn{1}{c}{(4)}   &\multicolumn{1}{c}{(5)}  \\

  & \multicolumn{1}{c}{\emph{log(Diagnostic activity$_t$) }} & \multicolumn{2}{c}{\emph{log(Average wait$_t$) }}   & \multicolumn{2}{c}{\emph{Share waiting$_t$  > 8 weeks}}     \\
   \cmidrule(lr){2-2}        \cmidrule(lr){3-4}   \cmidrule(lr){5-6}

      & \multicolumn{1}{c}{ }  &\multicolumn{1}{c}{ All } &\multicolumn{1}{c}{CT} &\multicolumn{1}{c}{  All } &\multicolumn{1}{c}{ CT } \\ 
   \hline
  \addlinespace
\multicolumn{5}{l}{\emph{Panel A}:    } \\
\input{fragment-diagnosticlognew_cases-new_admissions.tex} \\
\addlinespace
\addlinespace
\multicolumn{5}{l}{\emph{Panel B}:  }\\
\input{fragment-diagnosticlognew_cases-hospital_cases.tex} \\
\addlinespace
\addlinespace
\multicolumn{5}{l}{\emph{Panel C}:   } \\
\input{fragment-diagnosticlognew_cases-occupied_mv_beds.tex} \\
\addlinespace
\addlinespace

Provider x Diagnostic  FE &X & X & X &X & X\\
Diagnostic x Time FE &X & X & X &X & X\\ 
Community transmission &X & X & X &X & X\\ 

   \hline
   \end{tabular}

    \begin{tablenotes} {\footnotesize
    \item Notes:  Regressions present results at the NHS provider level documenting the relationship between different measures of COVID-19 pressures at the provider level across panels on the diagnostic performance and diagnostic waiting times. This does not distinguish between COVID-19 and non-COVID-19 related diagnostic activity. Columns (1) measures total diagnostic activity across 15 diagnostic functions performed. Columns (2) - (3) study average waiting times for all diagnostic activity (column 2) and CT diagnostic (column 3). Columns (4)-(5) study as dependent variable the share of individuals waiting more than 6 weeks across all diagnostic activity (column 4) and specifically for CT diagnostic (column 5). Community Transmission indicates that the regressions control for the log of the number of COVID-19 cases within the catchment areas of NHS providers. Standard errors are clustered at the provider level with stars indicating *** p$<$ 0.01, ** p$<$ 0.05, * p$<$ 0.1.} \end{tablenotes}
      \end{threeparttable}
      
 }
  }    
  
\end{table}
 %%%%%%%%%%%%%%%%%%%%%%%%%%%%%%

%%%%%%%%%%%%%%%%%%%%%%%%%%%%%%
\begin{landscape}
 \begin{table}[h!]
\centering{
\scalebox{0.75}{
  \begin{threeparttable}
  \caption{Impact of COVID-19 health care system pressures on cancer treatment pathways and performance \label{table:cancer-pressures}}
\begin{tabular}{lcccccc}
  \hline

%   & \multicolumn{6}{c}{\emph{DV:  }}  \\
% \cmidrule(lr){2-7}     
\addlinespace
 &\multicolumn{1}{c}{(1)}   &\multicolumn{1}{c}{(2)}   &\multicolumn{1}{c}{(3)}   &\multicolumn{1}{c}{(4)}   &\multicolumn{1}{c}{(5)} &\multicolumn{1}{c}{(6)}  \\

  & \multicolumn{3}{c}{\emph{log(cases$_t$) with }} & \multicolumn{3}{c}{\emph{\% with time taken to }}      \\
     \cmidrule(lr){2-4}        \cmidrule(lr){5-7} 

   &\multicolumn{1}{c}{ }   &\multicolumn{2}{c}{treatment}     &\multicolumn{1}{c}{referral}   & \multicolumn{2}{c}{treatment}   \\ 
   
   &\multicolumn{1}{c}{referrals}   &\multicolumn{1}{c}{decision}   &\multicolumn{1}{c}{start}   &\multicolumn{1}{c}{seen $>$ 14 days}   &\multicolumn{1}{c}{decision $>$ 31 days} &\multicolumn{1}{c}{start $>$ 62 days}  \\

  %    & \multicolumn{1}{c}{ }  &\multicolumn{1}{c}{ All } &\multicolumn{1}{c}{CT} &\multicolumn{1}{c}{  All } &\multicolumn{1}{c}{ CT } \\ 
   \hline
  \addlinespace
\multicolumn{5}{l}{\emph{Panel A}:    } \\
\input{fragment-cancerlognew_cases-new_admissions.tex} \\
\addlinespace
\addlinespace
\multicolumn{5}{l}{\emph{Panel B}:  }\\
\input{fragment-cancerlognew_cases-hospital_cases.tex} \\
\addlinespace
\addlinespace
\multicolumn{5}{l}{\emph{Panel C}:   } \\
\input{fragment-cancerlognew_cases-occupied_mv_beds.tex} \\
\addlinespace
\addlinespace

Provider x  Care setting x Cancer  FE &X & X & X &X & X  & X\\
Cancer x Care setting x Time FE &X & X & X &X & X  & X\\ 
Community transmission &X & X & X &X & X  & X\\ 

   \hline
   \end{tabular}

    \begin{tablenotes} {\footnotesize
    \item Notes:  Regressions present results at the NHS provider level documenting the relationship between different measures of COVID-19 pressures at the provider level across panels on the diagnostic performance and diagnostic waiting times. This does not distinguish between COVID-19 and non-COVID-19 related diagnostic activity. Columns (1) measures total diagnostic activity across 15 diagnostic functions performed. Columns (2) - (3) study average waiting times for all diagnostic activity (column 2) and CT diagnostic (column 3). Columns (4)-(5) study as dependent variable the share of individuals waiting more than 6 weeks across all diagnostic activity (column 4) and specifically for CT diagnostic (column 5). Community Transmission indicates that the regressions control for the log of the number of COVID-19 cases within the catchment areas of NHS providers. Standard errors are clustered at the provider level with stars indicating *** p$<$ 0.01, ** p$<$ 0.05, * p$<$ 0.1.} \end{tablenotes}
      \end{threeparttable}
      
 }
  }    
  
\end{table}
\end{landscape}

 %%%%%%%%%%%%%%%%%%%%%%%%%%%%%%

%%%%%%%%%%%%%%%%%%%%%%%%%%%%%%
 \begin{table}[h!]
\centering{
\scalebox{0.75}{
  \begin{threeparttable}
  \caption{Impact of COVID-19 health care system pressures on non-COVID-19 excess deaths \label{table:excess-deaths-main}}
\begin{tabular}{lcccccc}
  \hline

%   & \multicolumn{6}{c}{\emph{DV:  }}  \\
% \cmidrule(lr){2-7}     
\addlinespace
 &\multicolumn{1}{c}{(1)}   &\multicolumn{1}{c}{(2)}   &\multicolumn{1}{c}{(3)}   &\multicolumn{1}{c}{(4)}   &\multicolumn{1}{c}{(5)}  \\

%  & \multicolumn{3}{c}{\emph{log(cases$_t$) with }} & \multicolumn{3}{c}{\emph{\% with time taken to }}      \\
%     \cmidrule(lr){2-4}        \cmidrule(lr){5-7} 

%   &\multicolumn{1}{c}{ }   &\multicolumn{2}{c}{treatment}     &\multicolumn{1}{c}{referral}   & \multicolumn{2}{c}{treatment}   \\ 
   
 %  &\multicolumn{1}{c}{referrals}   &\multicolumn{1}{c}{decision}   &\multicolumn{1}{c}{start}   &\multicolumn{1}{c}{seen $>$ 14 days}   &\multicolumn{1}{c}{decision $>$ 31 days} &\multicolumn{1}{c}{start $>$ 62 days}  \\
   \hline
  \addlinespace
\multicolumn{5}{l}{\emph{Panel A}:    } \\
\input{fragment-shmilognew_cases-dchexcess-new_admissions.tex} \\
\addlinespace
\addlinespace
\multicolumn{5}{l}{\emph{Panel B}:  }\\
\input{fragment-shmilognew_cases-dchexcess-hospital_cases.tex} \\
\addlinespace
\addlinespace
\multicolumn{5}{l}{\emph{Panel C}:   } \\
\input{fragment-shmilognew_cases-dchexcess-occupied_mv_beds.tex} \\
\addlinespace
\addlinespace
\addlinespace

Provider  FE &X & X & X &X & X \\
Time FE &X & X & X &X & X\\ 
Community transmission &X & X & X &X & X \\
$\Delta \text{Spells}_{p,t}$   & & X & X &X & X\\ 
$ \text{Excess deaths}_{p,t-13} $   & &  & X &  &  \\ 
$ \text{Obs}_{p,t-13}$ and  $\text{Exp}_{p,t-13}$    & &  &  &X & X\\ 
Provider specific linear time trend  & &  &  & & X\\ 

   \hline
   \end{tabular}

    \begin{tablenotes} {\footnotesize
    \item Notes:  Regressions present results at the NHS provider level documenting the relationship between different measures of COVID-19 pressures at the provider level and overall excess deaths reported in a given month. The excess death measure captures month-on-month changes in excess death constructed from the twelve month cumulative windows. Across columns subsequently more control variables are added that aim to capture the potential confounding effect that base effects could have on the estimates. Community Transmission indicates that the regressions control for the log of the number of COVID-19 cases within the catchment areas of NHS providers. Standard errors are clustered at the provider level with stars indicating *** p$<$ 0.01, ** p$<$ 0.05, * p$<$ 0.1.} \end{tablenotes}
      \end{threeparttable}
      
 }
  }    
  
\end{table}
 %%%%%%%%%%%%%%%%%%%%%%%%%%%%%%

%%%%%%%%%%%%%%%%%%%%%%%%%%%%%%
 \begin{table}[h!]
\centering{
\scalebox{0.725}{
  \begin{threeparttable}
  \caption{Impact of COVID-19 pressures on staff absence rates\label{table:staff absences}}
\begin{tabular}{lccccccccc}
  \hline

%   & \multicolumn{6}{c}{\emph{DV:  }}  \\
% \cmidrule(lr){2-7}     
\addlinespace
 &\multicolumn{1}{c}{(1)}   &\multicolumn{1}{c}{(2)}   &\multicolumn{1}{c}{(3)}   &\multicolumn{1}{c}{(4)}   &\multicolumn{1}{c}{(5)}  &\multicolumn{1}{c}{(6)} \\ %&\multicolumn{1}{c}{(7)}   &\multicolumn{1}{c}{(8)}  &\multicolumn{1}{c}{(9)}  \\

%  & \multicolumn{6}{c}{\emph{NHS staff vaccination uptake measure}} \\
%     \cmidrule(lr){2-7} 
  & \multicolumn{3}{c}{\emph{All staff groups}} & \multicolumn{3}{c}{\emph{By staff group}}  \\% & \multicolumn{3}{c}{\emph{Above 75\% percentile}}      \\
    \cmidrule(lr){2-4}        \cmidrule(lr){5-7}  % \cmidrule(lr){8-10} 

%   &\multicolumn{1}{c}{ }   &\multicolumn{2}{c}{treatment}     &\multicolumn{1}{c}{referral}   & \multicolumn{2}{c}{treatment}   \\ 
   
   &\multicolumn{1}{c}{ }   &\multicolumn{1}{c}{ }   &\multicolumn{1}{c}{}   &\multicolumn{1}{c}{Nurses}   &\multicolumn{1}{c}{Doctors} &\multicolumn{1}{c}{Managers}  \\
 
  \addlinespace
Panel A: \\
\input{fragment-absencelognew_cases-new_admissions.tex} \\
\addlinespace
\addlinespace
\addlinespace

Panel B:  \\
\input{fragment-absencelognew_cases-hospital_cases.tex} \\
\addlinespace
\addlinespace

Panel C: \\
\input{fragment-absencelognew_cases-occupied_mv_beds.tex} \\
\addlinespace
\addlinespace

Provider  FE &X & X & X &X & X  & X  \\
Time FE &X & X & X &X & X  & X\\ 
Community transmission &X & X & X &X & X  & X\\ 
 \% Population vaccinated      & & X & X &X & X  & X \\
Provider specific linear trends  &  &   & X &X & X  & X\\
   \hline
   \end{tabular}

    \begin{tablenotes} {\footnotesize
    \item Notes:  Regressions capture the changing effect of NHS trust hospital admissions on provider specific excess mortality documenting how the vaccination roll out across the NHS is moderating this relationship. The excess death measure captures month-on-month changes in excess death constructed from the twelve month cumulative windows. Community Transmission indicates that the regressions control for the log of the number of COVID-19 cases within the catchment areas of NHS providers. Standard errors are clustered at the provider level with stars indicating *** p$<$ 0.01, ** p$<$ 0.05, * p$<$ 0.1.} \end{tablenotes}
      \end{threeparttable}
      
 }
  }    
 
\end{table}
 %%%%%%%%%%%%%%%%%%%%%%%%%%%%%%

%%%%%%%%%%%%%%%%%%%%%%%%%%%%%%
 \begin{table}[h!]
\centering{
\scalebox{0.7}{
  \begin{threeparttable}
  \caption{Impact of COVID-19 pressures on staff absence rates: the effect of NHS vaccination uptake  \label{table:staff absences-vaccination}}
\begin{tabular}{lccccccccc}
  \hline

%   & \multicolumn{6}{c}{\emph{DV:  }}  \\
% \cmidrule(lr){2-7}     
\addlinespace
 &\multicolumn{1}{c}{(1)}   &\multicolumn{1}{c}{(2)}   &\multicolumn{1}{c}{(3)}   &\multicolumn{1}{c}{(4)}   &\multicolumn{1}{c}{(5)}  &\multicolumn{1}{c}{(6)} \\ %&\multicolumn{1}{c}{(7)}   &\multicolumn{1}{c}{(8)}  &\multicolumn{1}{c}{(9)}  \\

%  & \multicolumn{6}{c}{\emph{NHS staff vaccination uptake measure}} \\
%     \cmidrule(lr){2-7} 
  \emph{DV: staff absence rates}    & \multicolumn{6}{c}{\emph{NHS staff vaccination uptake measure}} \\
     \cmidrule(lr){2-7} 
  & \multicolumn{3}{c}{\emph{\% of NHS staff with with 2 doses}} & \multicolumn{3}{c}{\emph{\% of NHS staff with at least 1 dose}}  \\% & \multicolumn{3}{c}{\emph{Above 75\% percentile}}      \\
    \cmidrule(lr){2-4}        \cmidrule(lr){5-7}  % \cmidrule(lr){8-10}  
  \addlinespace
Panel A: \\
\input{fragment-absencelognew_cases-vaccination-new_admissions.tex} \\
\addlinespace
\addlinespace
\addlinespace

Panel B:  \\
\input{fragment-absencelognew_cases-vaccination-hospital_cases.tex} \\
\addlinespace
\addlinespace

Panel C: \\
\input{fragment-absencelognew_cases-vaccination-occupied_mv_beds.tex} \\
\addlinespace
\addlinespace

Provider  FE &X & X & X &X & X  & X  \\
Time FE &X & X & X &X & X  & X\\ 
Community transmission &X & X & X &X & X  & X\\ 
 \% Population vaccinated      &   & X & X &  & X  & X \\
Provider specific linear trends  &  &   & X & &   & X\\
   \hline
   \end{tabular}

    \begin{tablenotes} {\footnotesize
    \item Notes:  Regressions capture the changing effect of different measures of COVID-19 pressures on staff absence rates depending on the vaccination uptake of NHS staff.  Community Transmission indicates that the regressions control for the log of the number of COVID-19 cases within the catchment areas of NHS providers. Standard errors are clustered at the provider level with stars indicating *** p$<$ 0.01, ** p$<$ 0.05, * p$<$ 0.1.} \end{tablenotes}
      \end{threeparttable}
      
 }
  }    
 
\end{table}
 %%%%%%%%%%%%%%%%%%%%%%%%%%%%%%

%%%%%%%%%%%%%%%%%%%%%%%%%%%%%%
 \begin{table}[h!]
\centering{
\scalebox{0.725}{
  \begin{threeparttable}
  \caption{Impact of COVID-19 pressures on non-COVID-19 excess mortality: the moderating effect of NHS vaccination uptake  \label{table:excess-deaths-nhsvaxx}}
\begin{tabular}{lccccccccc}
  \hline

 & \multicolumn{6}{c}{}  \\
% \cmidrule(lr){2-7}     
\addlinespace
 &\multicolumn{1}{c}{(1)}   &\multicolumn{1}{c}{(2)}   &\multicolumn{1}{c}{(3)}   &\multicolumn{1}{c}{(4)}   &\multicolumn{1}{c}{(5)}  &\multicolumn{1}{c}{(6)} \\ %&\multicolumn{1}{c}{(7)}   &\multicolumn{1}{c}{(8)}  &\multicolumn{1}{c}{(9)}  \\

  \emph{DV: Non-COVID-19 Excess Death}    & \multicolumn{6}{c}{\emph{NHS staff vaccination uptake measure}} \\
     \cmidrule(lr){2-7} 
  & \multicolumn{3}{c}{\emph{\% of NHS staff with with 2 doses}} & \multicolumn{3}{c}{\emph{\% of NHS staff with at least 1 dose}}  \\% & \multicolumn{3}{c}{\emph{Above 75\% percentile}}      \\
    \cmidrule(lr){2-4}        \cmidrule(lr){5-7}  % \cmidrule(lr){8-10} 

%   &\multicolumn{1}{c}{ }   &\multicolumn{2}{c}{treatment}     &\multicolumn{1}{c}{referral}   & \multicolumn{2}{c}{treatment}   \\ 
   
 %  &\multicolumn{1}{c}{referrals}   &\multicolumn{1}{c}{decision}   &\multicolumn{1}{c}{start}   &\multicolumn{1}{c}{seen $>$ 14 days}   &\multicolumn{1}{c}{decision $>$ 31 days} &\multicolumn{1}{c}{start $>$ 62 days}  \\
 
  \addlinespace
\input{interaction-nhsvaxlognew_cases-shmi-dchexcess.tex} \\
\addlinespace
\addlinespace
\addlinespace

Provider  FE &X & X & X &X & X  & X  \\
Time FE &X & X & X &X & X  & X\\ 
Community transmission &X & X & X &X & X  & X\\ 
$ \text{Excess deaths}_{p,t-13} $  &X & X & X &X & X  & X\\
$\Delta \text{Spells}_{p,t}$     &X & X & X &X & X  & X \\
\% Population vaccination      &X & X & X &X & X  & X \\

   \hline
   \end{tabular}

    \begin{tablenotes} {\footnotesize
    \item Notes:  Regressions capture the changing effect of NHS trust hospital admissions on provider specific excess mortality documenting how the vaccination roll out across the NHS is moderating this relationship. The excess death measure captures month-on-month changes in excess death constructed from the twelve month cumulative windows.  Standard errors are clustered at the provider level with stars indicating *** p$<$ 0.01, ** p$<$ 0.05, * p$<$ 0.1.} \end{tablenotes}
      \end{threeparttable}
      
 }
  }    
 
\end{table}
 %%%%%%%%%%%%%%%%%%%%%%%%%%%%%%

\newpage
\startappendixtables{}
\startappendixfigures{}
 
\appendix
\clearpage

\setcounter{footnote}{0} \setcounter{page}{1} 

\begin{center}
{\LARGE Appendix to ``Pandemic Pressures and Public Health Care: Evidence from England''}

\bigskip

{\large For Online Publication}

\bigskip

\begin{tabular}[t]{c}
%{\large    Thiemo Fetzer  \hspace{.7em} Oliver Vanden Eynde \hspace{.7em}  Austin L. Wright \hspace{.7em} Pedro CL Souza}%
\end{tabular}
\\[0pt]
%\vspace{1cm} {\large \today}
\end{center}

%%%%%%%%%%%%%%%%%%%%%%%%%%%%%%
\begin{landscape}

\begin{figure}[h!]
\caption{Allocation of spatial areas to NHS Trusts \label{fig:spatialallocation}}
\begin{center}
$
\begin{array}{ll}
\text{\emph{Panel A}: MSOA's across England }& \text{\emph{Panel B}: Visits to Barking, Havering and Redbrige NHS Trust }  \\
  \includegraphics[width=0.4\columnwidth]{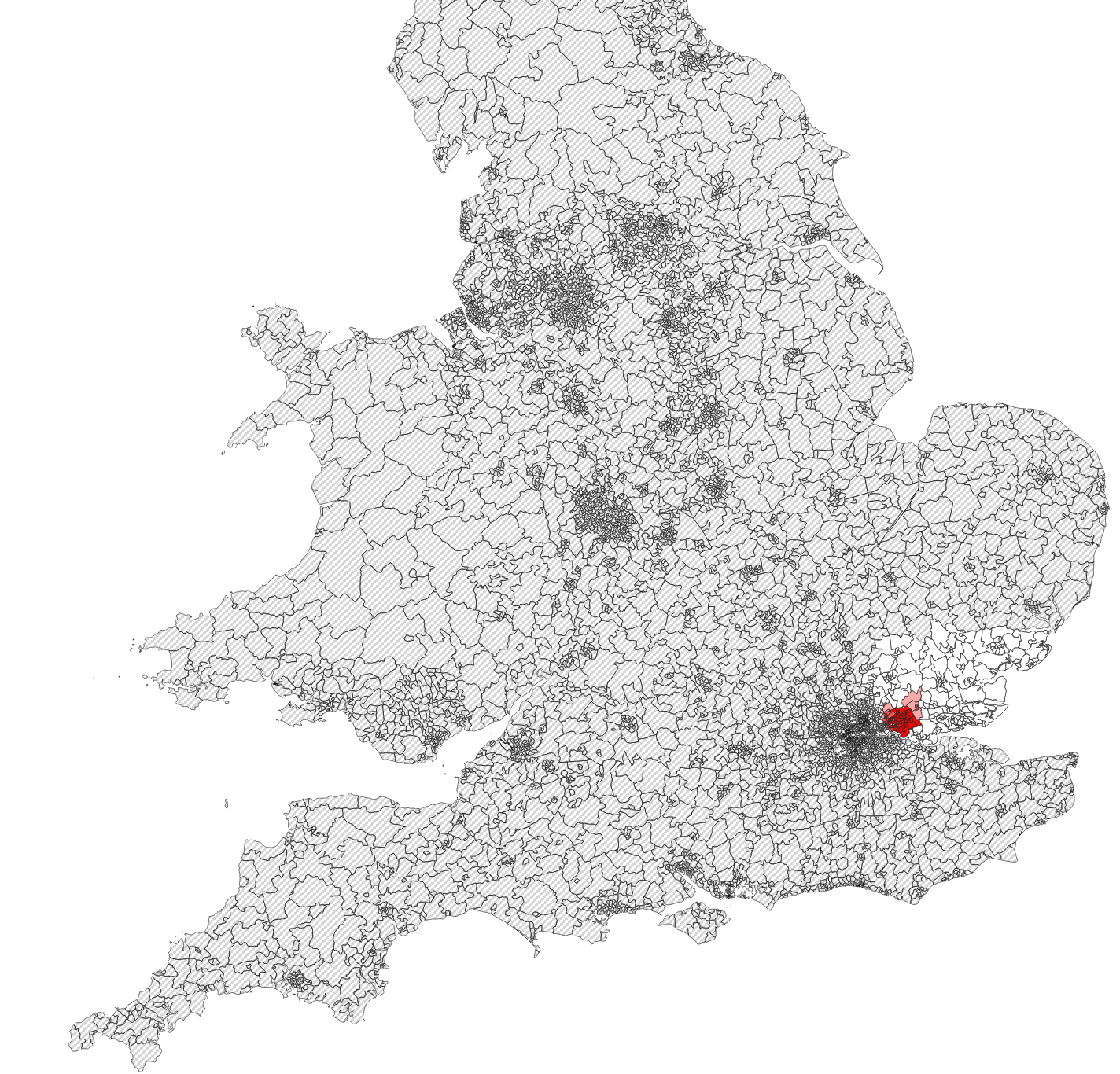} &   \includegraphics[width=0.55\columnwidth]{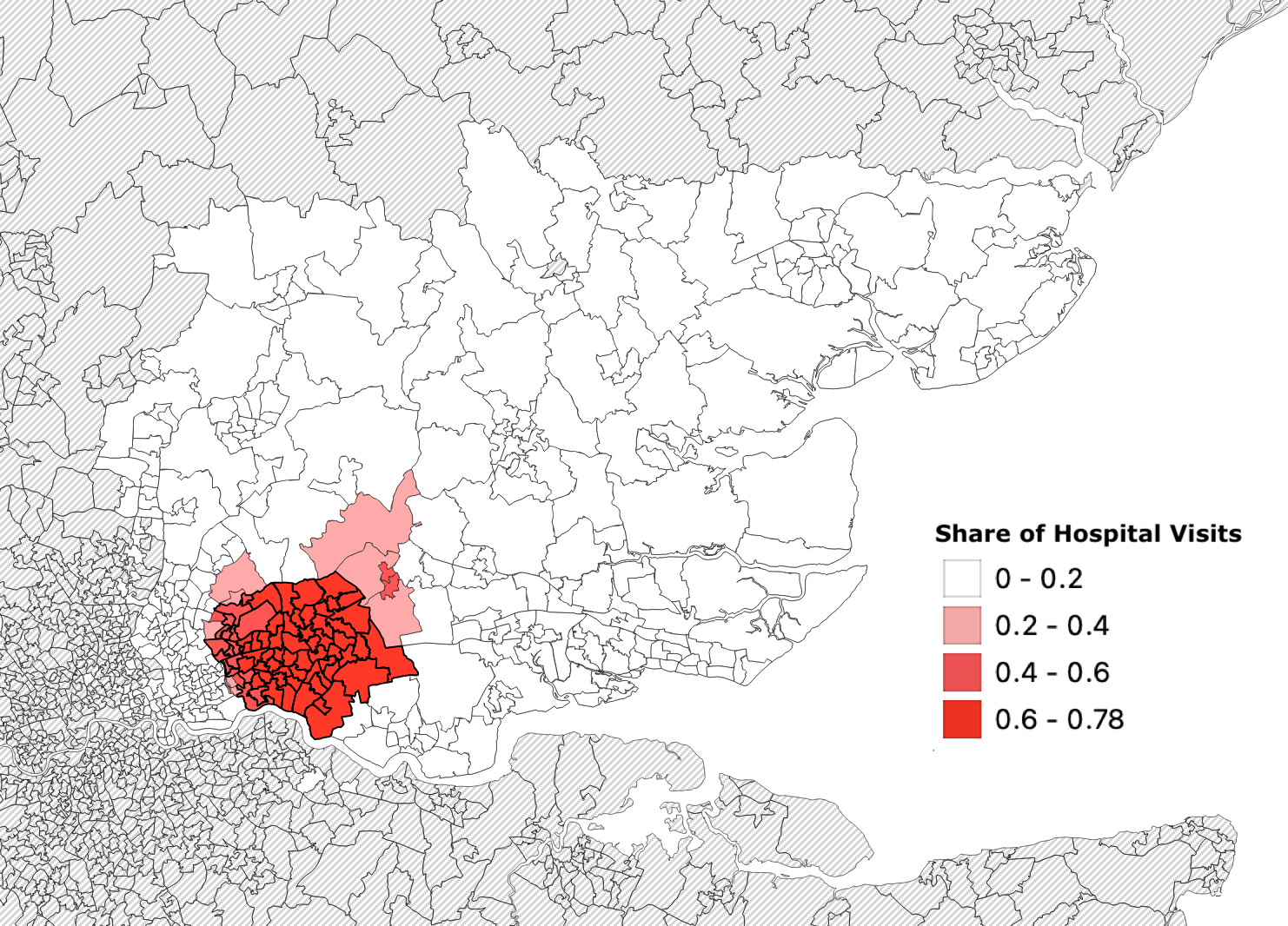} \\  
    \end{array}$
\end{center}
\scriptsize{\textbf{Notes:} Map displays the residential address of visitors to the Barking, Havering and Redbrige University Hospital NHS Trust in 2019. The left figure plots the distribution across England's 6791 MSOAs. The hospital trust saw hospital visits from patients coming from 412 MSOAs. The right figure provides a zoom in on the spatial distribution of patients visiting the Barking, Havering and Redbrige University Hospital NHS Trust in 2019 and what share of visits are made up by residents from different MSOAs. The vast majority 83\% come from 70 MSOAs that are immediately in the neighborhood of the trusts's main hospitals: the King George Hospital and the Queen's Hospital. The solid dark lines in the right panel indicate the MSOAs that are attributed to the NHS Trust by virtue of the trust's hospitals have been serving most of the patients that had a hospital spell in 2019 that reside in each MSOA.}

\end{figure}
\end{landscape}
%%%%%%%%%%%%%%%%%%%%%%%%%%%%%%

%%%%%%%%%%%%%%%%%%%%%%%%%%%%%%
\begin{figure}[h!]
\caption{Measuring  poor quality  with thresholds exceeding NHS set goals  over time  \label{fig:quality_long}}
\begin{center}
$
\begin{array}{ll}
%\multicolumn{2}{c}{\text{\emph{Panel A:} Referrals of Suspected Cancers }} \\  
\multicolumn{1}{c}{\text{ A\&E > 12h }} & \multicolumn{1}{c}{\text{Completed RT referral > 36 weeks}} \\  
    \includegraphics[width=.5\columnwidth]{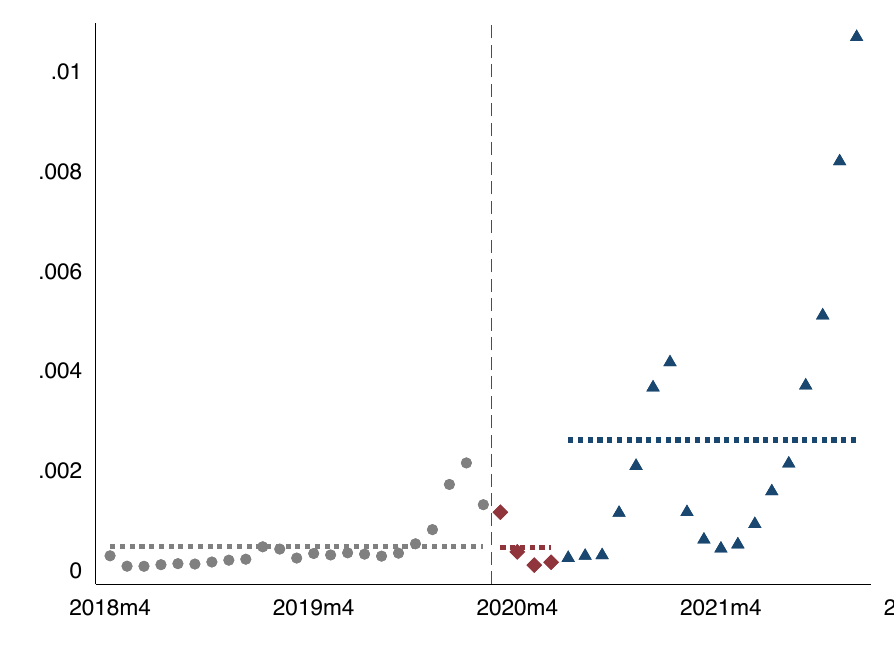}  &     \includegraphics[width=0.5\columnwidth]{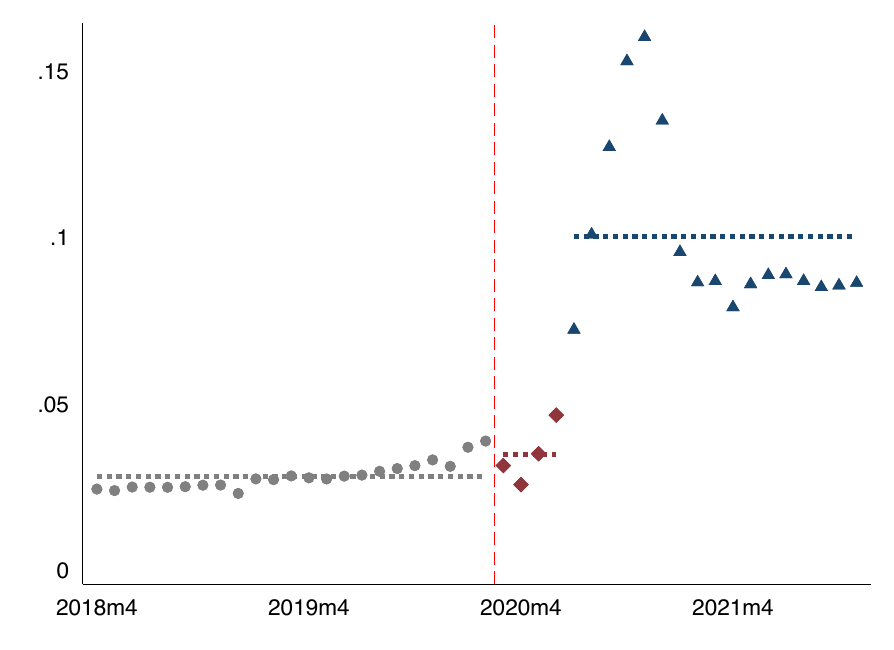} \\

%\multicolumn{2}{c}{\text{\emph{Panel B:} Decision to treat among referrals with cancers identified}} \\  
\multicolumn{1}{c}{\text{Diagnostic > 12 weeks }} & \multicolumn{1}{c}{\text{Cancer treatment > 104 days}} \\  
 \includegraphics[width=0.5\columnwidth]{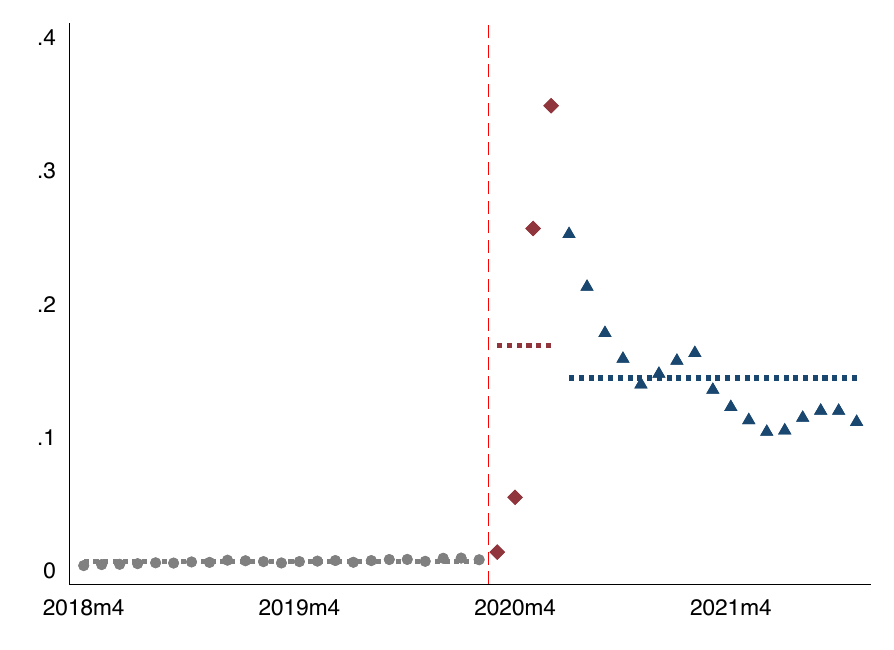} & \includegraphics[width=0.5\columnwidth]{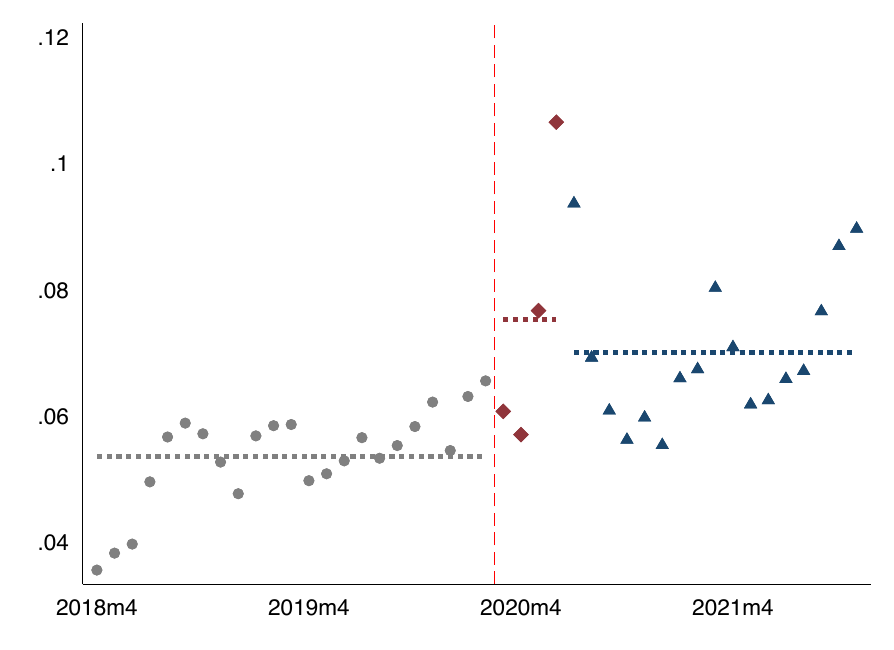}  \\
 
% \multicolumn{2}{c}{\text{\emph{Panel C:} Cancer Start of Treatment}} \\  
% \multicolumn{1}{c}{\text{\# of cases treated}} & \multicolumn{1}{c}{\text{ \% treated within 62 days of referral}} \\  
% \includegraphics[width=0.5\columnwidth]{n_cases_absolute_ae_waiting.pdf} & \includegraphics[width=0.5\columnwidth]{n_cases_absolute_ae_waiting.pdf}    \\

  %&   \includegraphics[width=0.55\columnwidth]{zoom-in-allocation-msoa-hes.pdf} \\  
    \end{array}$
\end{center}
\scriptsize{\textbf{Notes:} Figures present aggregate measures across a broad range of metrics indicative of extent as well as timeliness or accessibility of health care across NHS. The pre-pandemic mean is represented by the gray dotted line, during the first wave by red dotted line, and after the first wave by the blue dotted line. The panels  capture measures indicative of very poor quality or accessibility clockwise measuring the share of A\&E attendances seeing a doctor after more than 12 hours; the share of completed specialist referrals that had their referral completed after 36 weeks; the share of patients waiting more than 12 weeks for a diagnostic test; the share of cancer patients that  received first treatment after more than 104 days.}

\end{figure}
%%%%%%%%%%%%%%%%%%%%%%%%%%%%%%

%%%%%%%%%%%%%%%%%%%%%%%%%%%%%%
\begin{figure}[h!]
\caption{Measuring distribution of poor quality  with thresholds exceeding NHS set goals across NHS trusts over time  \label{fig:quality_long_time_fe}}
\begin{center}
$
\begin{array}{ll}
%\multicolumn{2}{c}{\text{\emph{Panel A:} Referrals of Suspected Cancers }} \\  
\multicolumn{1}{c}{\text{ A\&E > 12h }} & \multicolumn{1}{c}{\text{Completed RT referral > 36 weeks}} \\  
    \includegraphics[width=.5\columnwidth]{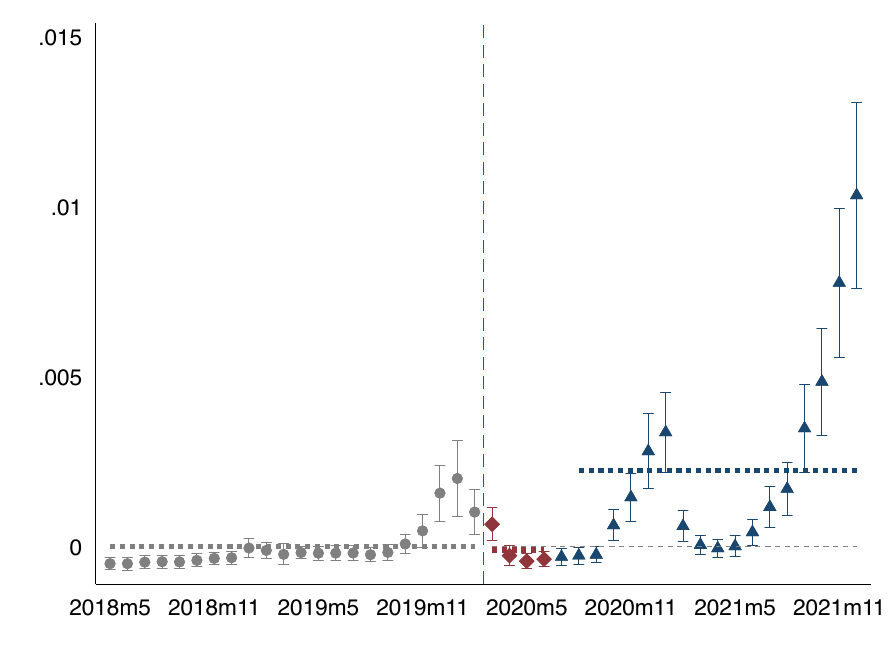}  &     \includegraphics[width=0.5\columnwidth]{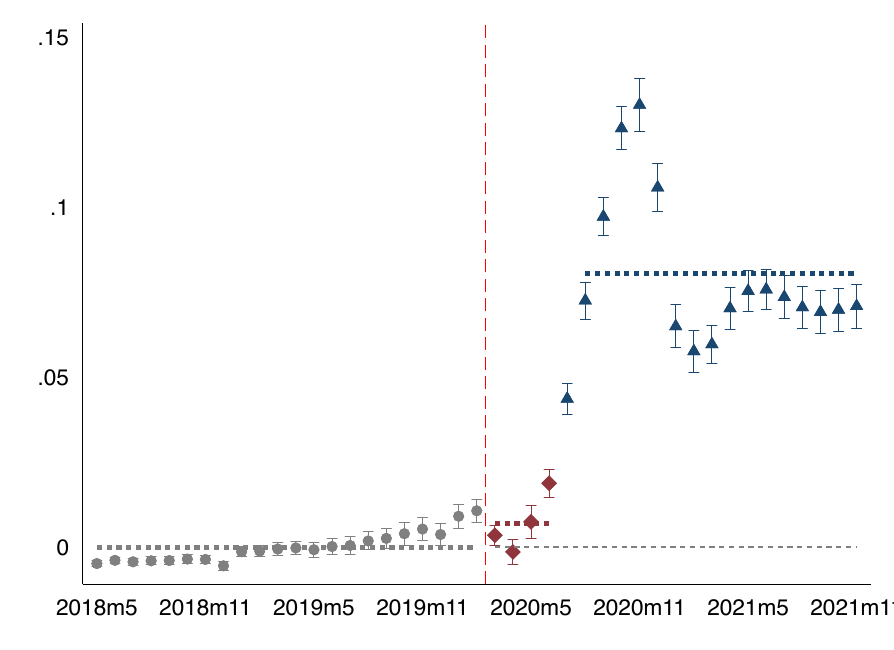} \\

%\multicolumn{2}{c}{\text{\emph{Panel B:} Decision to treat among referrals with cancers identified}} \\  
\multicolumn{1}{c}{\text{Diagnostic > 12 weeks }} & \multicolumn{1}{c}{\text{Cancer treatment > 104 days}} \\  
 \includegraphics[width=0.5\columnwidth]{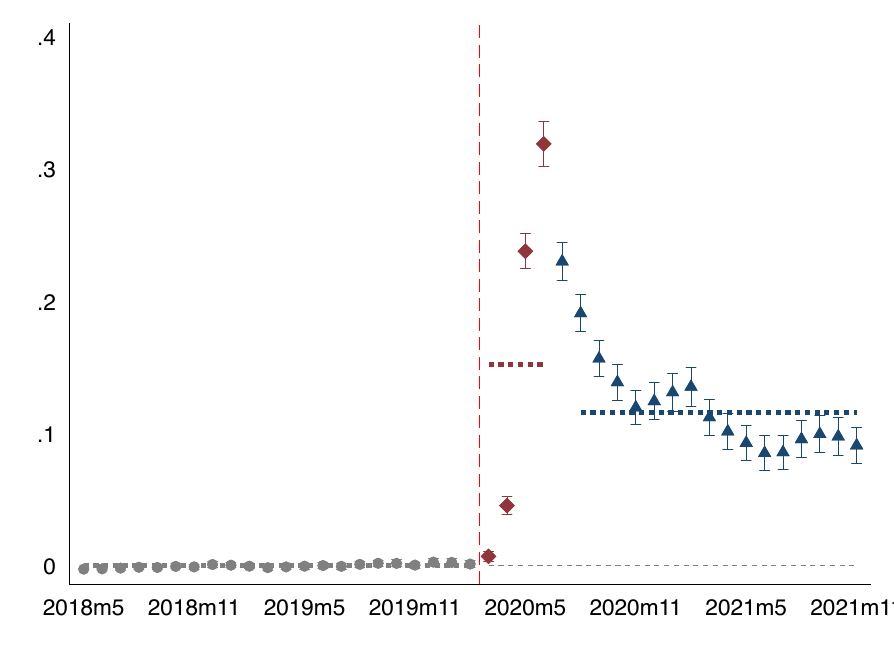} & \includegraphics[width=0.5\columnwidth]{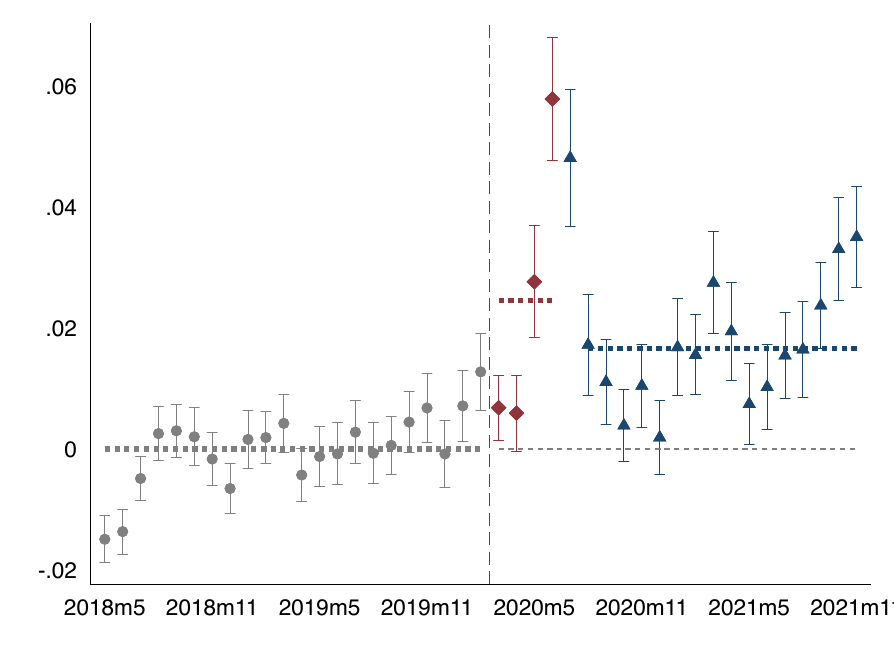}  \\
 
% \multicolumn{2}{c}{\text{\emph{Panel C:} Cancer Start of Treatment}} \\  
% \multicolumn{1}{c}{\text{\# of cases treated}} & \multicolumn{1}{c}{\text{ \% treated within 62 days of referral}} \\  
% \includegraphics[width=0.5\columnwidth]{n_cases_absolute_ae_waiting.pdf} & \includegraphics[width=0.5\columnwidth]{n_cases_absolute_ae_waiting.pdf}    \\

  %&   \includegraphics[width=0.55\columnwidth]{zoom-in-allocation-msoa-hes.pdf} \\  
    \end{array}$
\end{center}
\scriptsize{\textbf{Notes:} Figures plot out estimated time effects and 90\% confidence intervals capturing both the time average as well as the distribution of that time average across a broad range of metrics indicative of extent as well as timeliness or accessibility of health care across different NHS providers. The pre-pandemic mean is represented by the gray dotted line, during the first wave by red dotted line, and after the first wave by the blue dotted line. All regressions include provider fixed effects centering the pre COVID-19 arrival data around zero. The panels  capture measures indicative of very poor quality or accessibility clockwise measuring the share of A\&E attendances seeing a doctor after more than 12 hours; the share of completed specialist referrals that had their referral completed after 36 weeks; the share of patients waiting more than 12 weeks for a diagnostic test; the share of cancer patients that  received first treatment after more than 104 days.}

\end{figure}
%%%%%%%%%%%%%%%%%%%%%%%%%%%%%%

%%%%%%%%%%%%%%%%%%%%%%%%%%%%%%
\begin{figure}[h!]
\caption{Distributions of quality  over time  \label{fig:distrib_quality}}
\begin{center}
$
\begin{array}{ll}
%\multicolumn{2}{c}{\text{\emph{Panel A:} Referrals of Suspected Cancers }} \\  
 \multicolumn{1}{c}{\text{Completed RT referral }} & \multicolumn{1}{c}{\text{Diagnostic waiting list}}  \\  
    \includegraphics[width=.5\columnwidth]{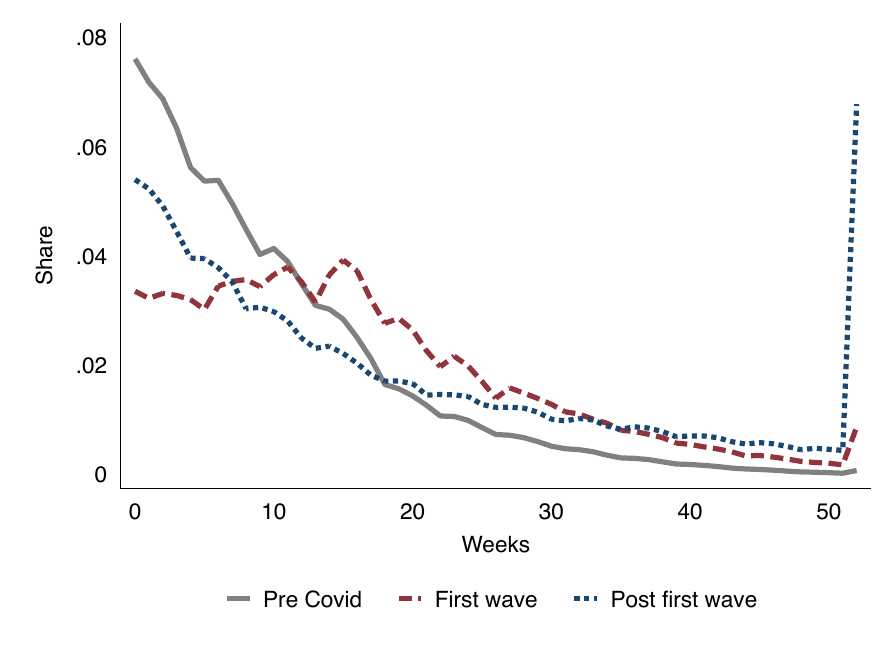}  &     \includegraphics[width=.5\columnwidth]{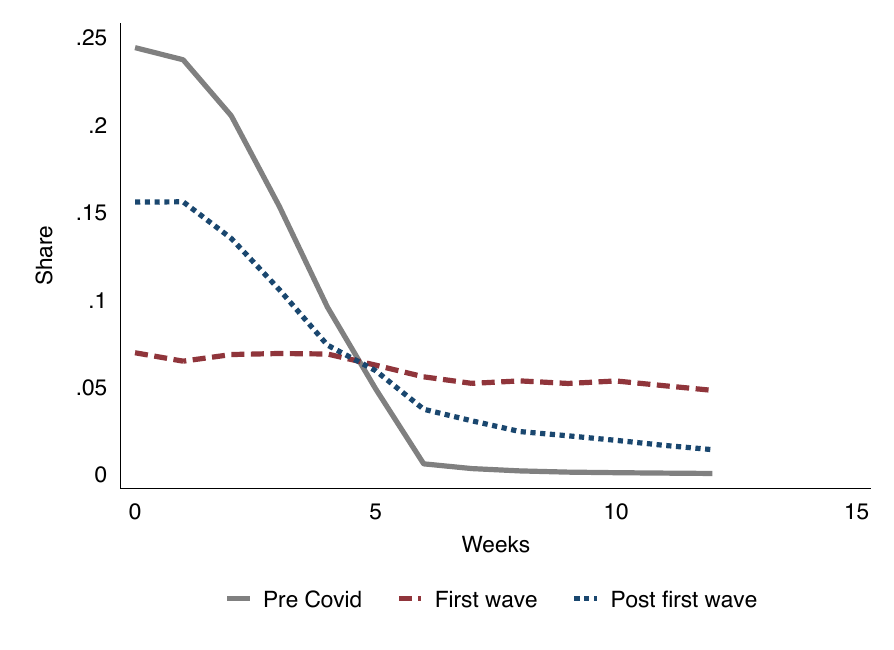}  \\

%\multicolumn{2}{c}{\text{\emph{Panel B:} Decision to treat among referrals with cancers identified}} \\  
\multicolumn{1}{c}{\text{Cancer treatment after referral }} & \multicolumn{1}{c}{\text{Cancer consultation after referral}} \\  
\includegraphics[width=.5\columnwidth]{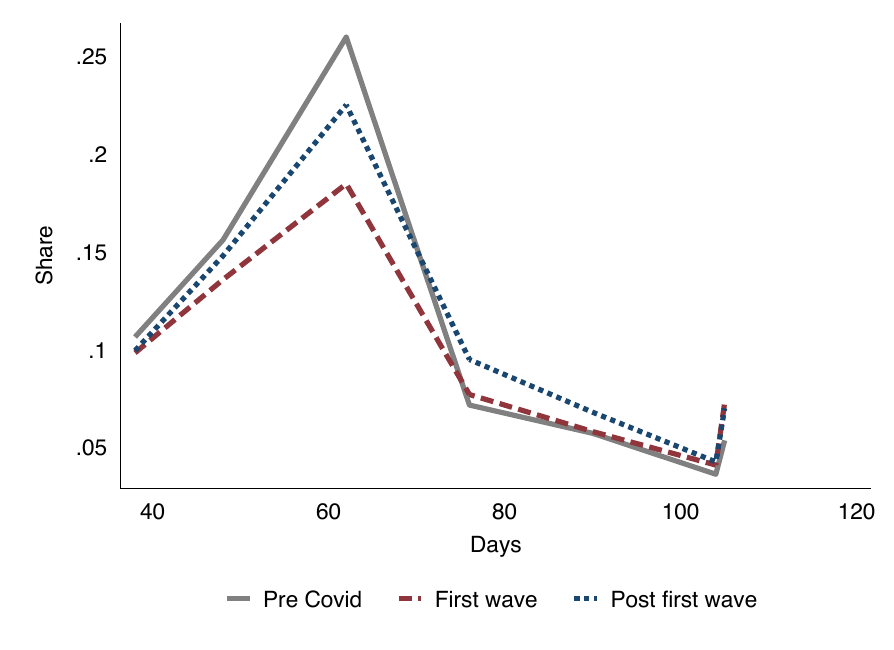}   & \includegraphics[width=.5\columnwidth]{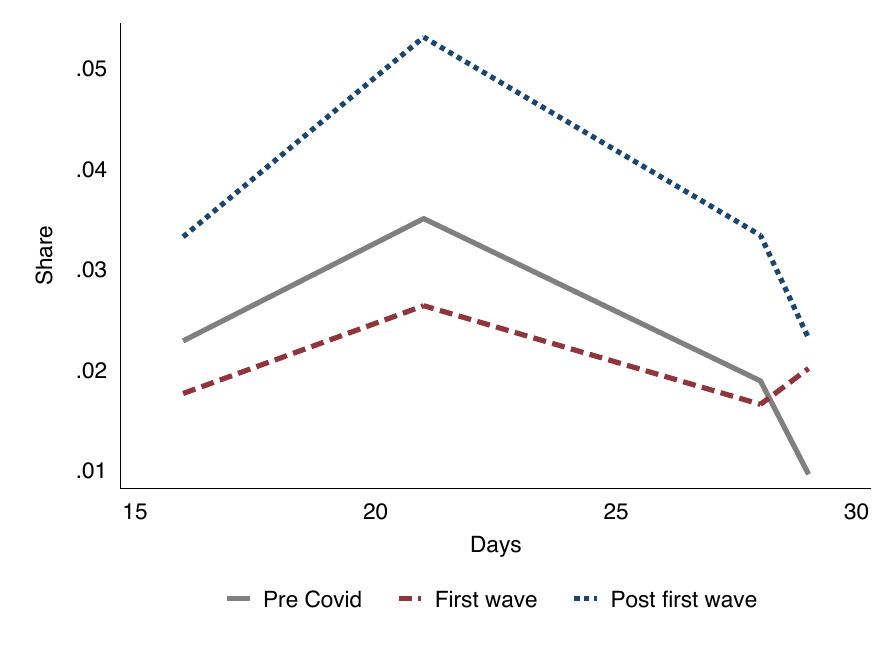}   \\
 
% \multicolumn{2}{c}{\text{\emph{Panel C:} Cancer Start of Treatment}} \\  
% \multicolumn{1}{c}{\text{\# of cases treated}} & \multicolumn{1}{c}{\text{ \% treated within 62 days of referral}} \\  
% \includegraphics[width=0.5\columnwidth]{n_cases_absolute_ae_waiting.pdf} & \includegraphics[width=0.5\columnwidth]{n_cases_absolute_ae_waiting.pdf}    \\

  %&   \includegraphics[width=0.55\columnwidth]{zoom-in-allocation-msoa-hes.pdf} \\  
    \end{array}$
\end{center}
\scriptsize{\textbf{Notes:} Figures plot the distributions of measures of completion. The pre-pandemic mean is represented by the gray  line, during the first wave by red dashed line, and after the first wave by the blue dotted line.  The panels  capture measures indicative of the distribution of quality or accessibility clockwise measuring  the share of completed specialist referrals that had their referral completed within a certain number of weeks; the share of patients waiting more than a certain number of weeks for a diagnostic test; the share of cancer patients that  received urgent first treatment after referral within a given number of days; the share of urgent suspected cancer referrals that had a consultation with a specialist  within a given number of days. For cancer treatments and consultations the x-axis begins at NHS set goals as the majority of cases are completed within this time frame and therefore variations in the tails are not easily distinguishable due to the differences in the levels. For A\&E visits we do not have enough time thresholds to plot meaningful distributions. }

\end{figure}
%%%%%%%%%%%%%%%%%%%%%%%%%%%%%%

%%%%%%%%%%%%%%%%%%%%%%%%%%%%%%

\begin{figure}[h!]
\caption{Quantity and quality of cancer detection and treatment across time  \label{fig:quantity-qual-cancer}}
\begin{center}
$
\begin{array}{ll}
\multicolumn{2}{c}{\text{\emph{Panel A:} Measures of quantity }}  \\  \\
\multicolumn{1}{c}{\text{ Referral to consultation }} & \multicolumn{1}{c}{\text{Decision to treatment}}  \\  

    \includegraphics[width=.5\columnwidth]{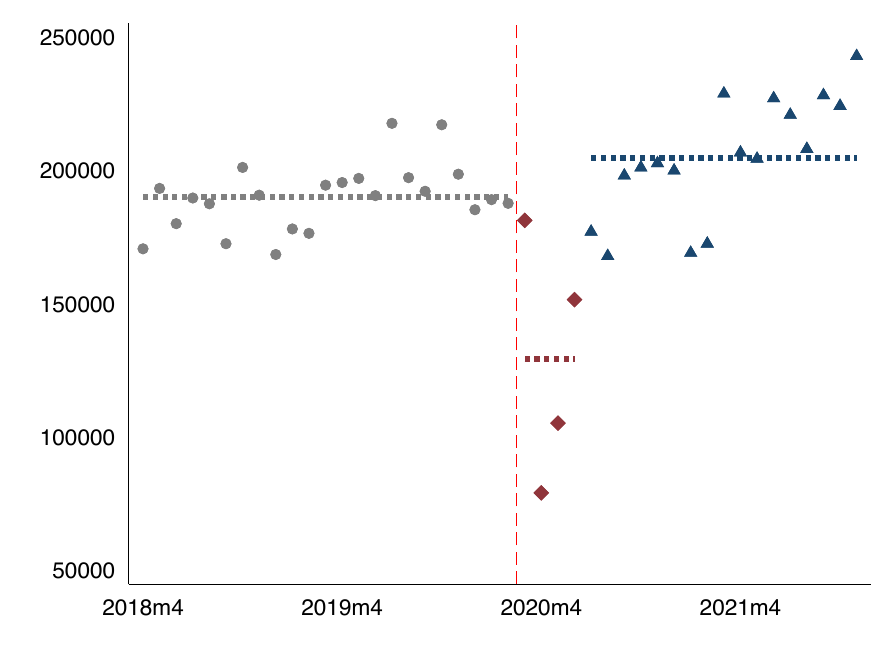}  &       \includegraphics[width=.5\columnwidth]{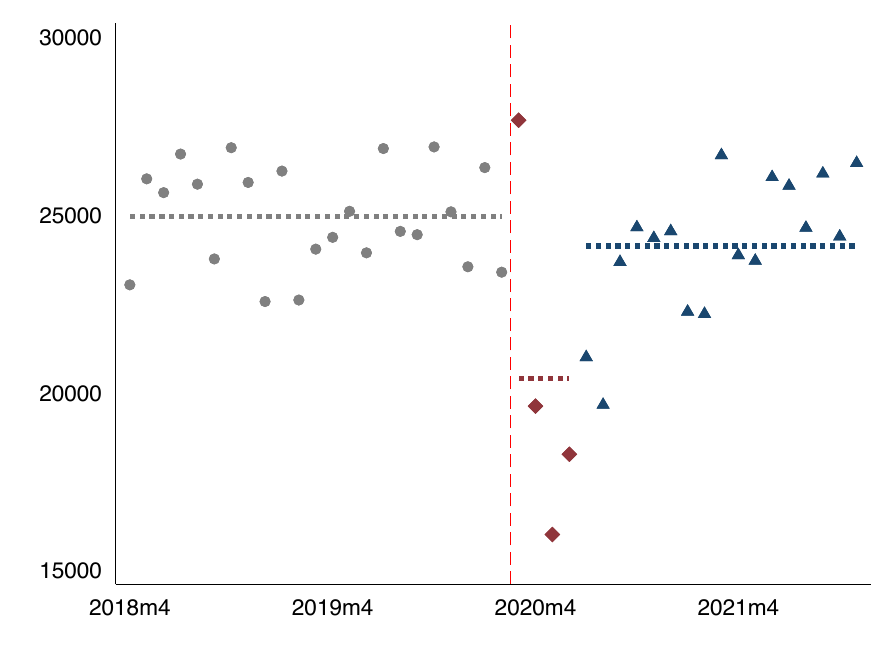}   \\
    \\
\multicolumn{2}{c}{\text{\emph{Panel B:} Measures of quality }}  \\  \\
\multicolumn{1}{c}{\text{ Referral to consultation  < 14 days}} & \multicolumn{1}{c}{\text{Decision to treatment < 31 days}}  \\  
    \includegraphics[width=.5\columnwidth]{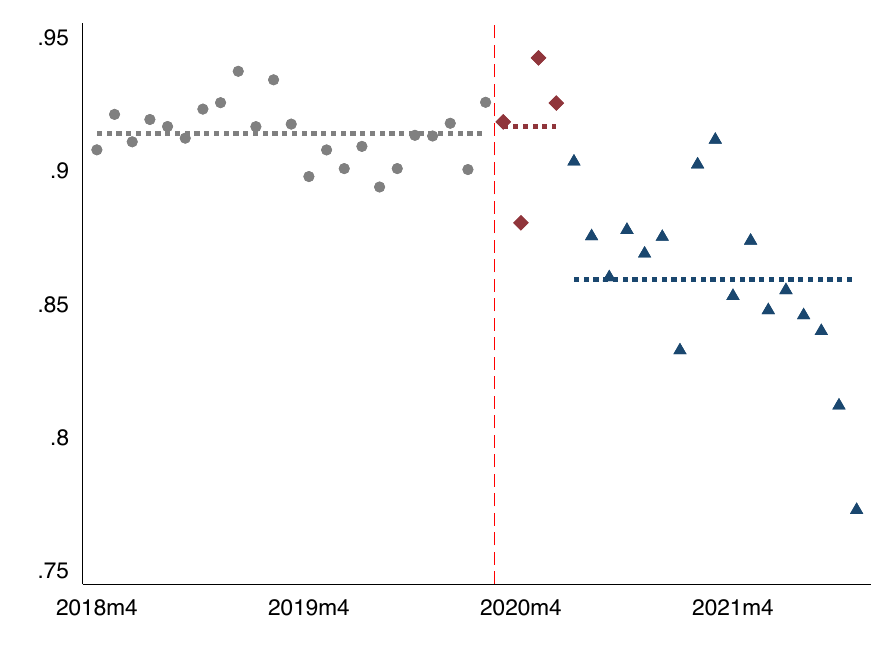}  &       \includegraphics[width=.5\columnwidth]{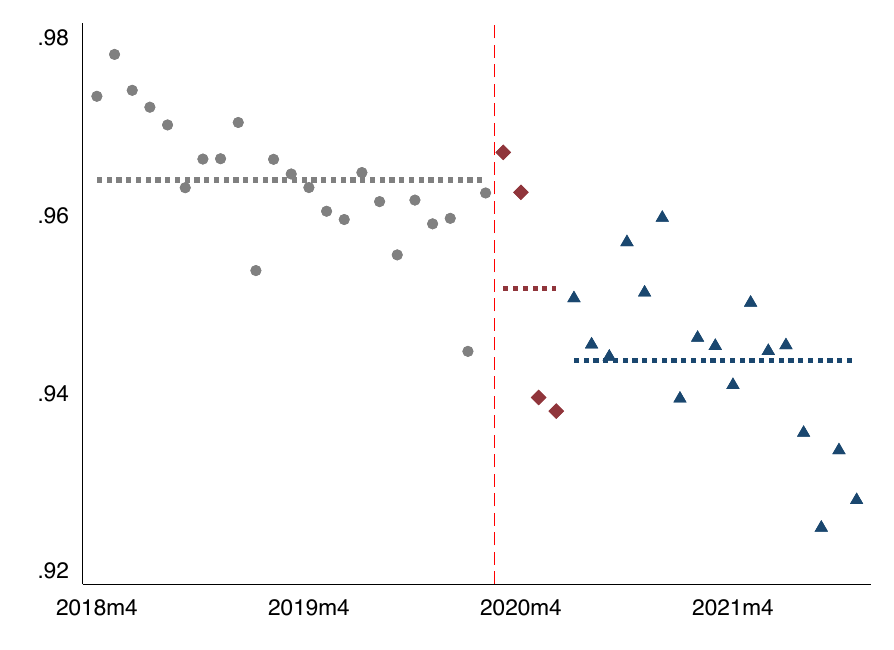}   \\

    \end{array}$
\end{center}
\scriptsize{\textbf{Notes:} Figures present aggregate measures  concerning  timeliness or accessibility of  cancer detection and treatment across the NHS. The pre-pandemic mean is represented by the gray dotted line, during the first wave by red dotted line, and after the first wave by the blue dotted line. From left to right panel A studies quantity metrics  from capturing the number of urgent cancer referrals seen by a specialist and decisions to treat leading to first treatment. Panel B captures measures indicative of quality or accessibility from left to right measuring the share of urgent cancer referrals seen by a specialist within 14 days and decisions to treat leading to first treatment within 31 days.}

\end{figure}

%%%%%%%%%%%%%%%%%%%%%%%%%%%%%%

%%%%%%%%%%%%%%%%%%%%%%%%%%%%%%
%\begin{landscape}
\begin{figure}[h!]
\caption{Referrals and treatment of suspected cancers by type of cancer  \label{fig:suspected-type-cancer-quantity}}
\begin{center}
$
\begin{array}{l}
\multicolumn{1}{l}{\text{\emph{Panel A}: \# of referrals  to treatment}} \\
  \includegraphics[width=1\columnwidth]{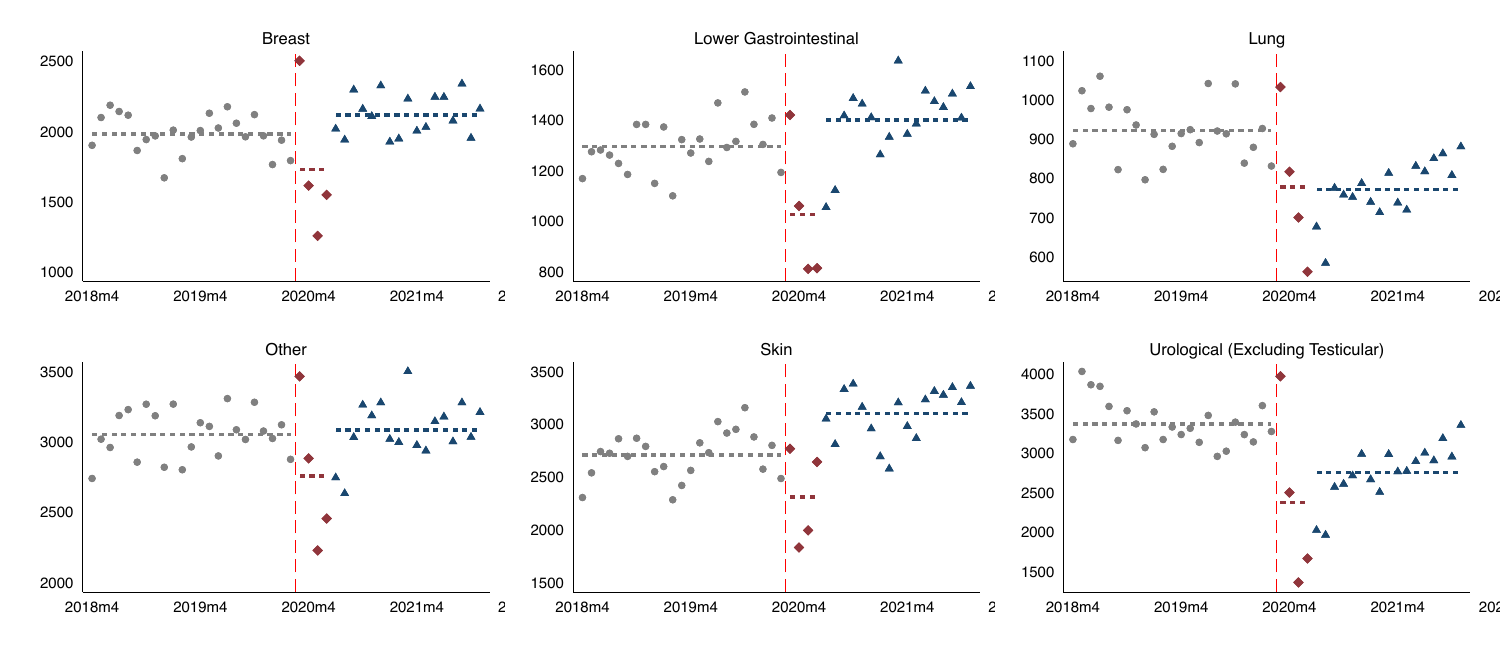}  \\
  \\
    \multicolumn{1}{l}{\text{ \emph{Panel B}:  \# of referrals  to consultation}} \\
  \includegraphics[width=1\columnwidth]{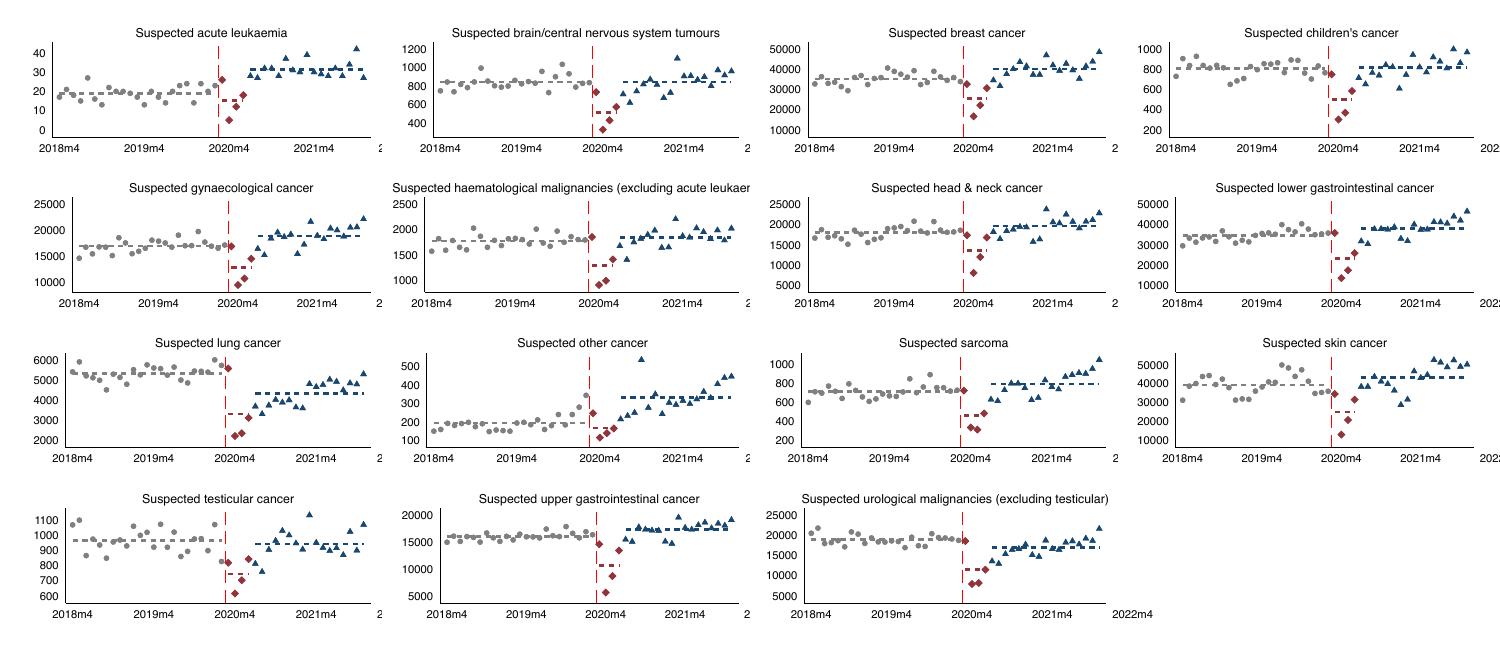}  \\

  %&   \includegraphics[width=0.55\columnwidth]{zoom-in-allocation-msoa-hes.pdf} \\  
    \end{array}$
\end{center}
\scriptsize{\textbf{Notes:}  Figures present  measures  concerning  quantity of   detection and treatment of different types of cancer across the NHS. The pre-pandemic mean is represented by the gray dotted line, during the first wave by red dotted line, and after the first wave by the blue dotted line.  Panel A studies quantity metrics  from capturing the number of urgent referrals leading to first treatment and panel B urgent cancer referrals seen by a specialist.}

\end{figure}
%%%%%%%%%%%%%%%%%%%%%%%%%%%%%%

%%%%%%%%%%%%%%%%%%%%%%%%%%%%%%
\begin{figure}[h!]
\caption{Performance targets of referrals and treatment of suspected cancers by type of cancer  \label{fig:suspected-type-cancer-quality}}
\begin{center}
$
\begin{array}{l}
\multicolumn{1}{l}{\text{\emph{Panel A}: Referrals  to treatment < 62 days}} \\
  \includegraphics[width=1\columnwidth]{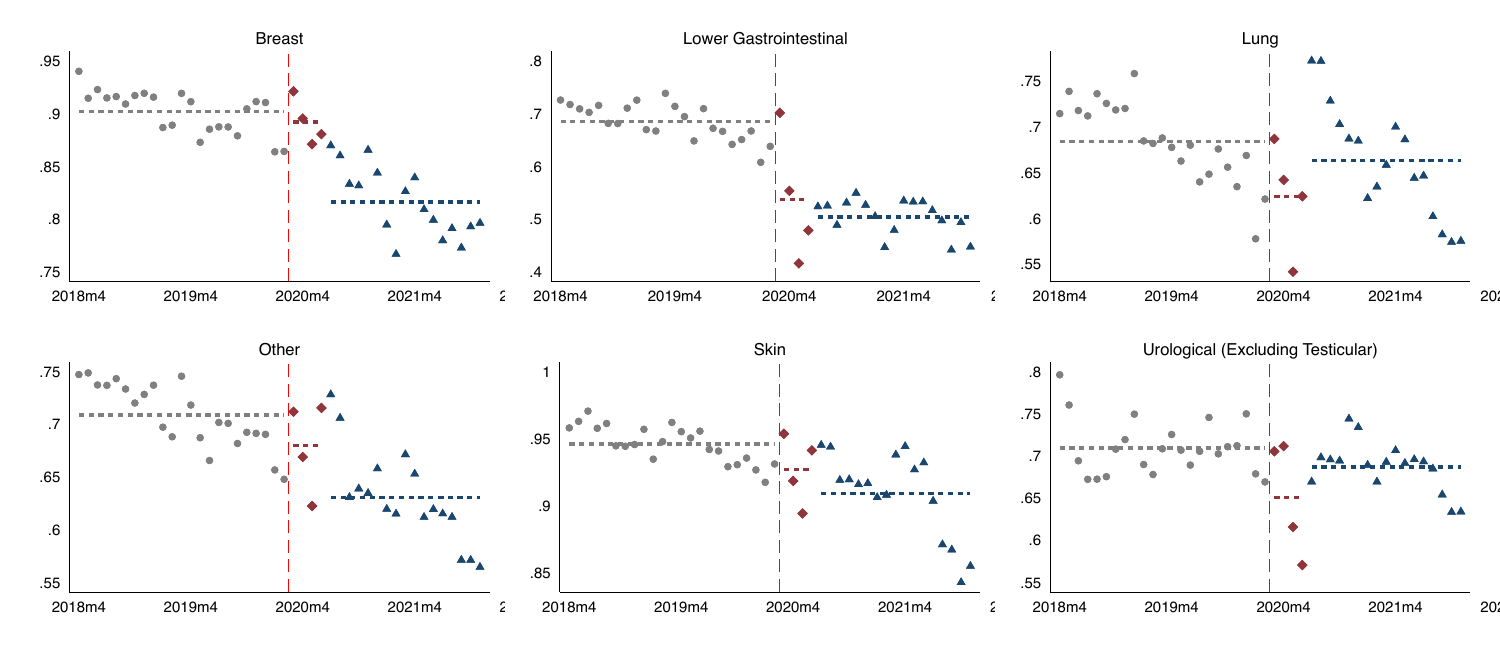}  \\
  \\
    \multicolumn{1}{l}{\text{ \emph{Panel B}:  Referrals  to consultation < 14 days}} \\
  \includegraphics[width=1\columnwidth]{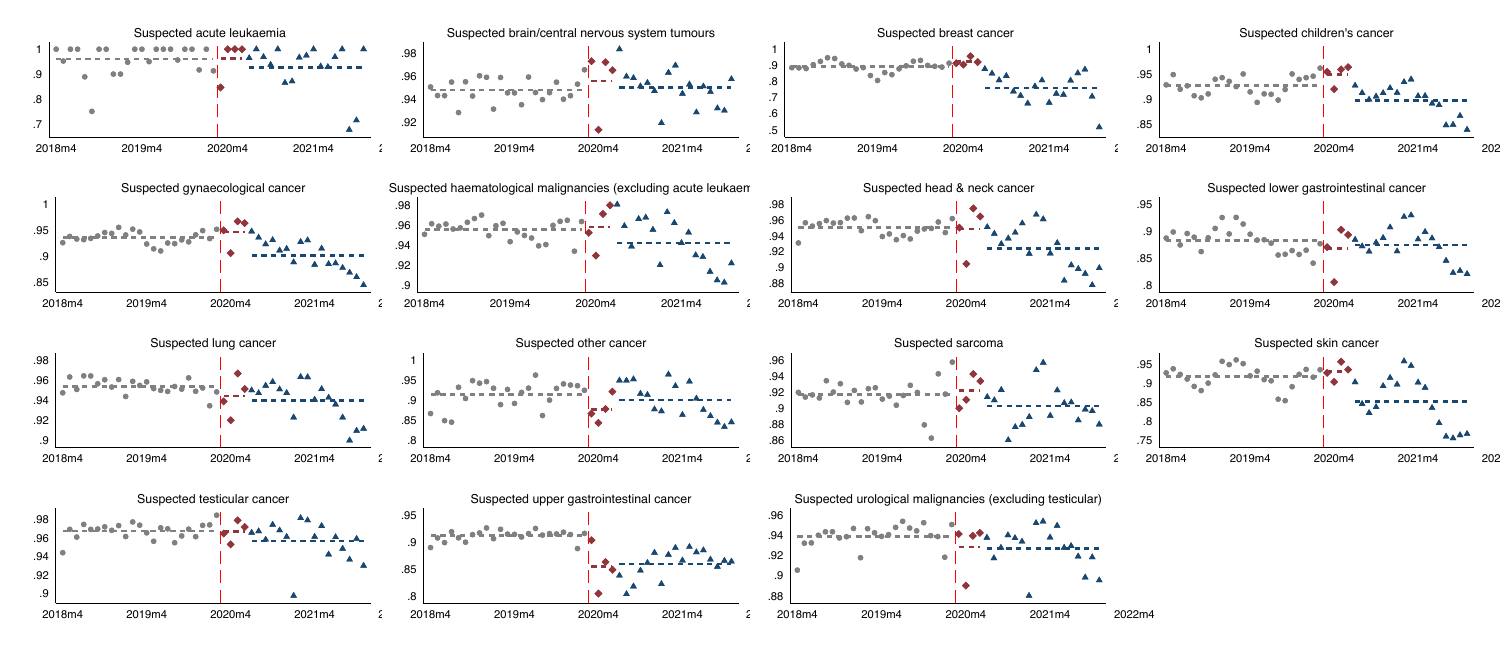}  \\

  %&   \includegraphics[width=0.55\columnwidth]{zoom-in-allocation-msoa-hes.pdf} \\  
    \end{array}$
\end{center}
\scriptsize{\textbf{Notes:}   Figures present  measures  concerning  quantity of   detection and treatment of different types of cancer across the NHS. The pre-pandemic mean is represented by the gray dotted line, during the first wave by red dotted line, and after the first wave by the blue dotted line.  Panel A captures measures indicative of quality or accessibility from left to right measuring the share of urgent referrals leading to first treatment within 62 days and panel B shows urgent cancer referrals seen by a specialist within 14 days .}
\end{figure}
%%%%%%%%%%%%%%%%%%%%%%%%%%%%%%

%%%%%%%%%%%%%%%%%%%%%%%%%%%%%%
\begin{figure}[h!]
\caption{Referrals and treatment of suspected cancers by type of cancer  \label{fig:suspected-type-cancer-quantity-timefe}}
\begin{center}
$
\begin{array}{l}
\multicolumn{1}{l}{\text{\emph{Panel A}: Log \# of referrals  to treatment}} \\
  \includegraphics[width=1\columnwidth]{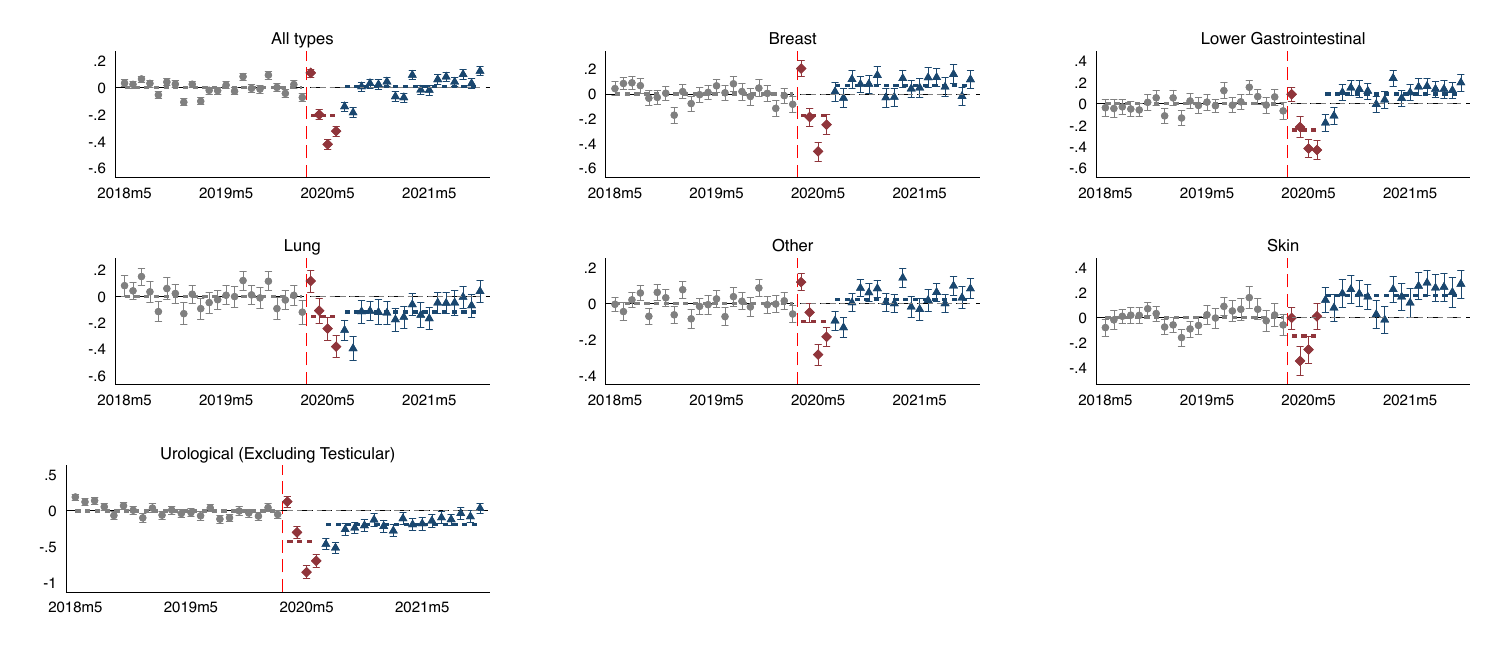}  \\
  \\
    \multicolumn{1}{l}{\text{ \emph{Panel B}: Log \# of referrals  to consultation}} \\
  \includegraphics[width=1\columnwidth]{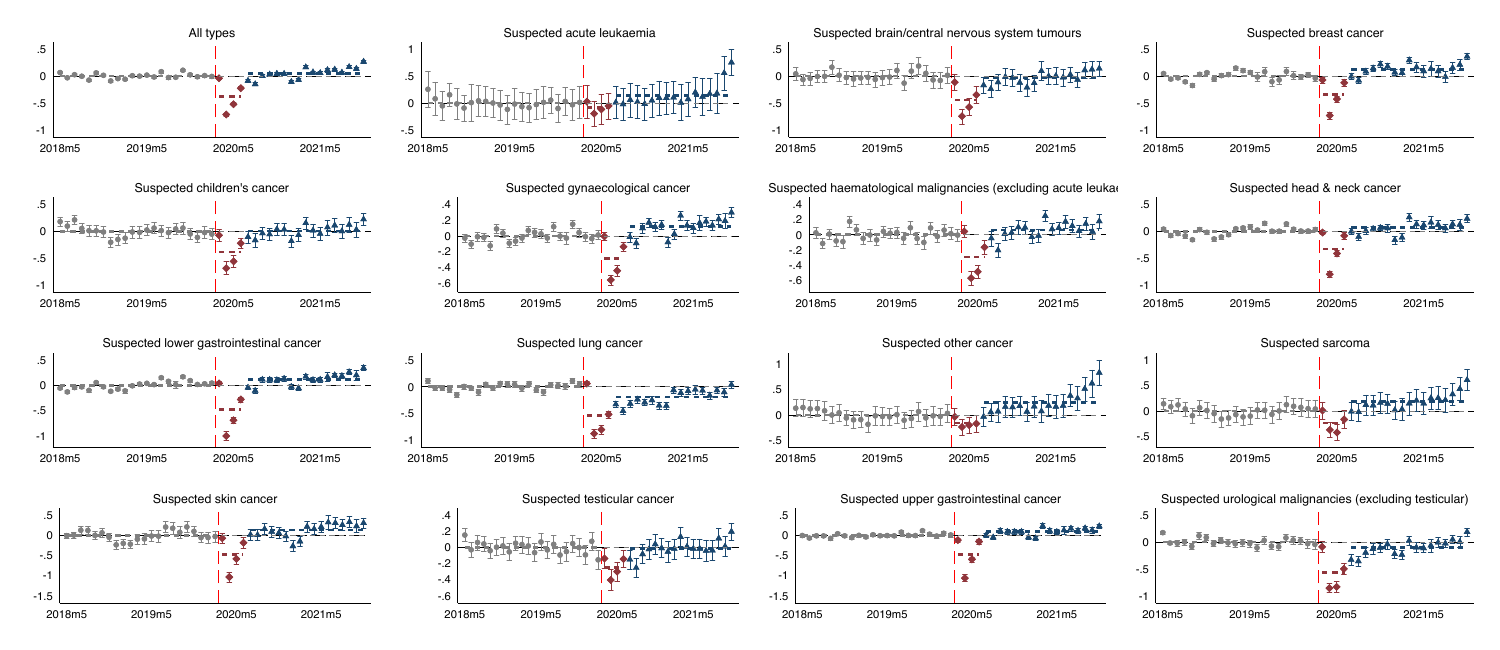}  \\

  %&   \includegraphics[width=0.55\columnwidth]{zoom-in-allocation-msoa-hes.pdf} \\  
    \end{array}$
\end{center}
\scriptsize{\textbf{Notes:} Figures plot out estimated time effects and 90\% confidence intervals capturing both the time average as well as the distribution of that time average across   treatment and detection metrics indicative of extent  or accessibility of health care across different NHS providers. The pre-pandemic mean is represented by the gray dotted line, during the first wave by red dotted line, and after the first wave by the blue dotted line.  Panel A studies quantity metrics  from capturing the number of urgent referrals leading to first treatment and panel B urgent cancer referrals seen by a specialist. The dependent variable is the log of the  number of cases + 1 as for some sorts of cancers the number of cases drops to zero. }

\end{figure}
%%%%%%%%%%%%%%%%%%%%%%%%%%%%%%

%%%%%%%%%%%%%%%%%%%%%%%%%%%%%%
%\end{landscape}
%%%%%%%%%%%%%%%%%%%%%%%%%%%%%%

%%%%%%%%%%%%%%%%%%%%%%%%%%%%%%
\begin{figure}[h!]
\caption{Distribution of month on month changes in log(new COVID-19 admissions) at the trust level \label{fig:month_on_month_distrib_new_admissions}}
\centering
$
\begin{array}{l}
     \includegraphics[width=1\columnwidth]{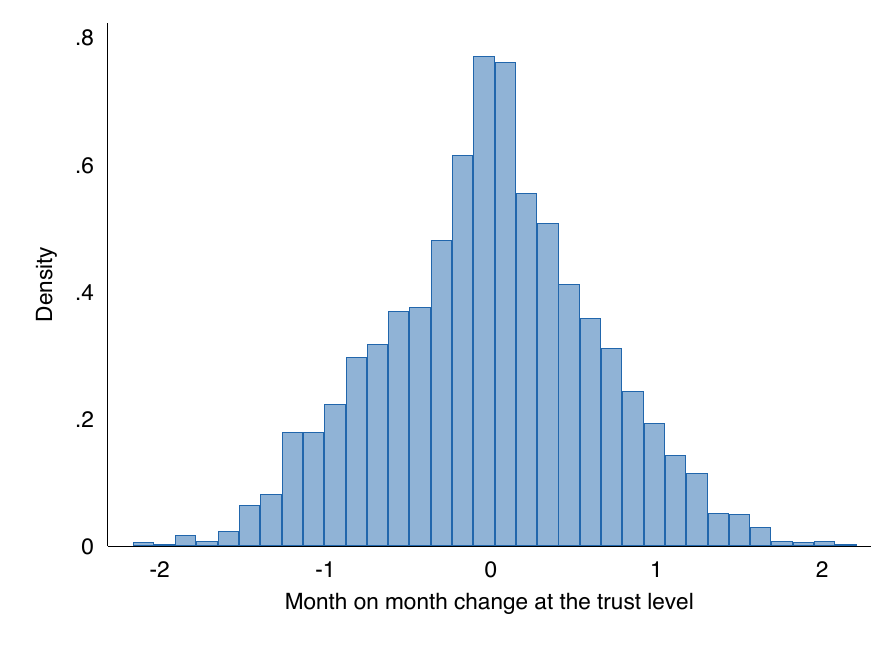}   

    \end{array}$

\scriptsize{\textbf{Notes:} Figure plots the binned distribution of month on month changes in log(new COVID-19 admissions) at the trust level. }

\end{figure}
%%%%%%%%%%%%%%%%%%%%%%%%%%%%%%

%%%%%%%%%%%%%%%%%%%%%%%%%%%%%%
%\begin{landscape}
\begin{figure}[h!]
\caption{Impact of COVID-19 pressures on specialist referrals by specialisation \label{fig:coefplot-rtt-treatment-function}}
\begin{center}
$
\begin{array}{lll}
\text{\emph{Panel A}:  log(Referrals) } \\
 \multicolumn{1}{c}{}  &\multicolumn{2}{c}{\text{Completed} }  \\
  \multicolumn{1}{c}{ \text{New}}  &\multicolumn{1}{c}{\text{Admitted}} &\multicolumn{1}{c}{\text{Non-admitted}} \\
  \includegraphics[width=0.33\columnwidth]{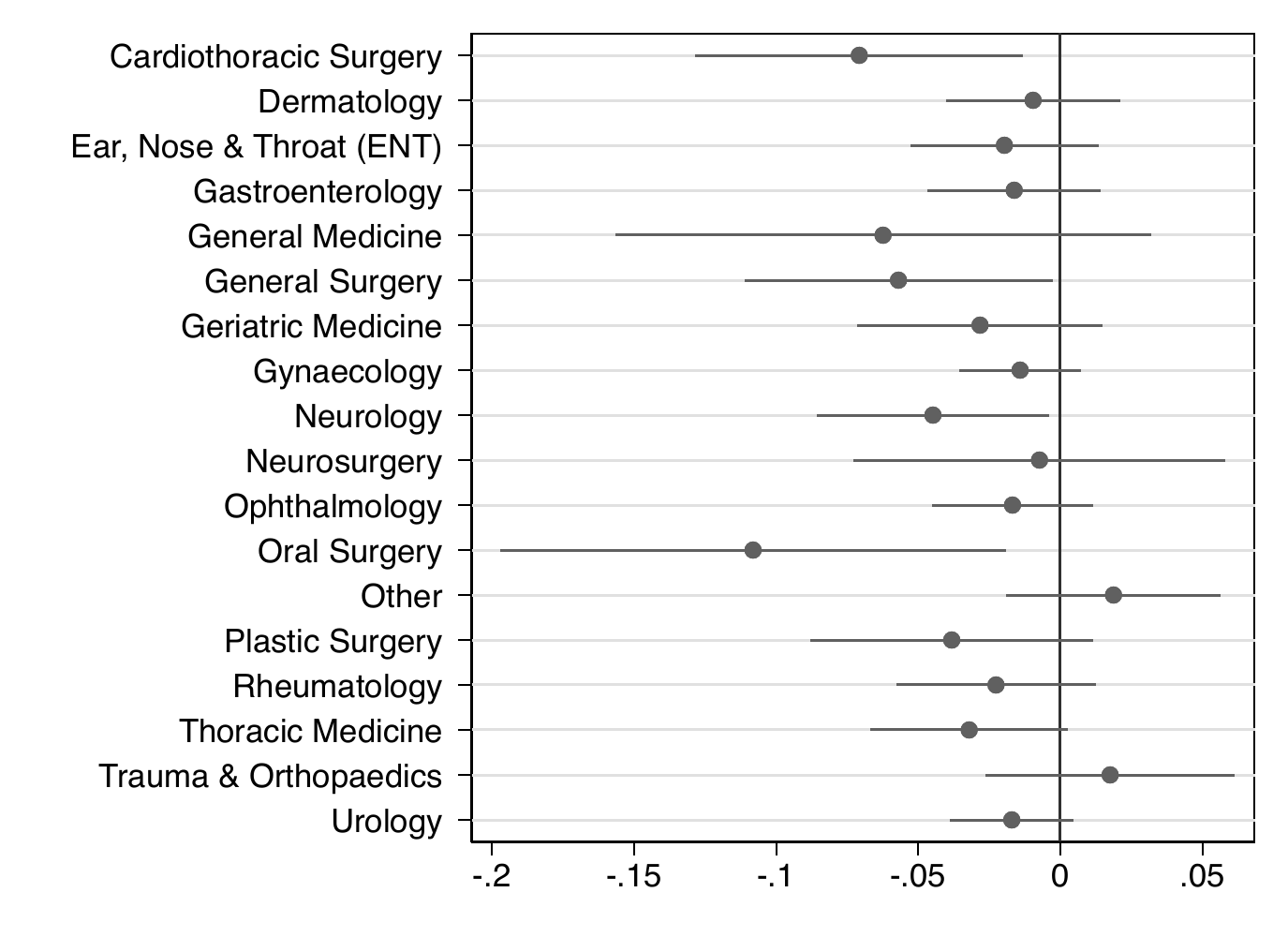} &   \includegraphics[width=0.33\columnwidth]{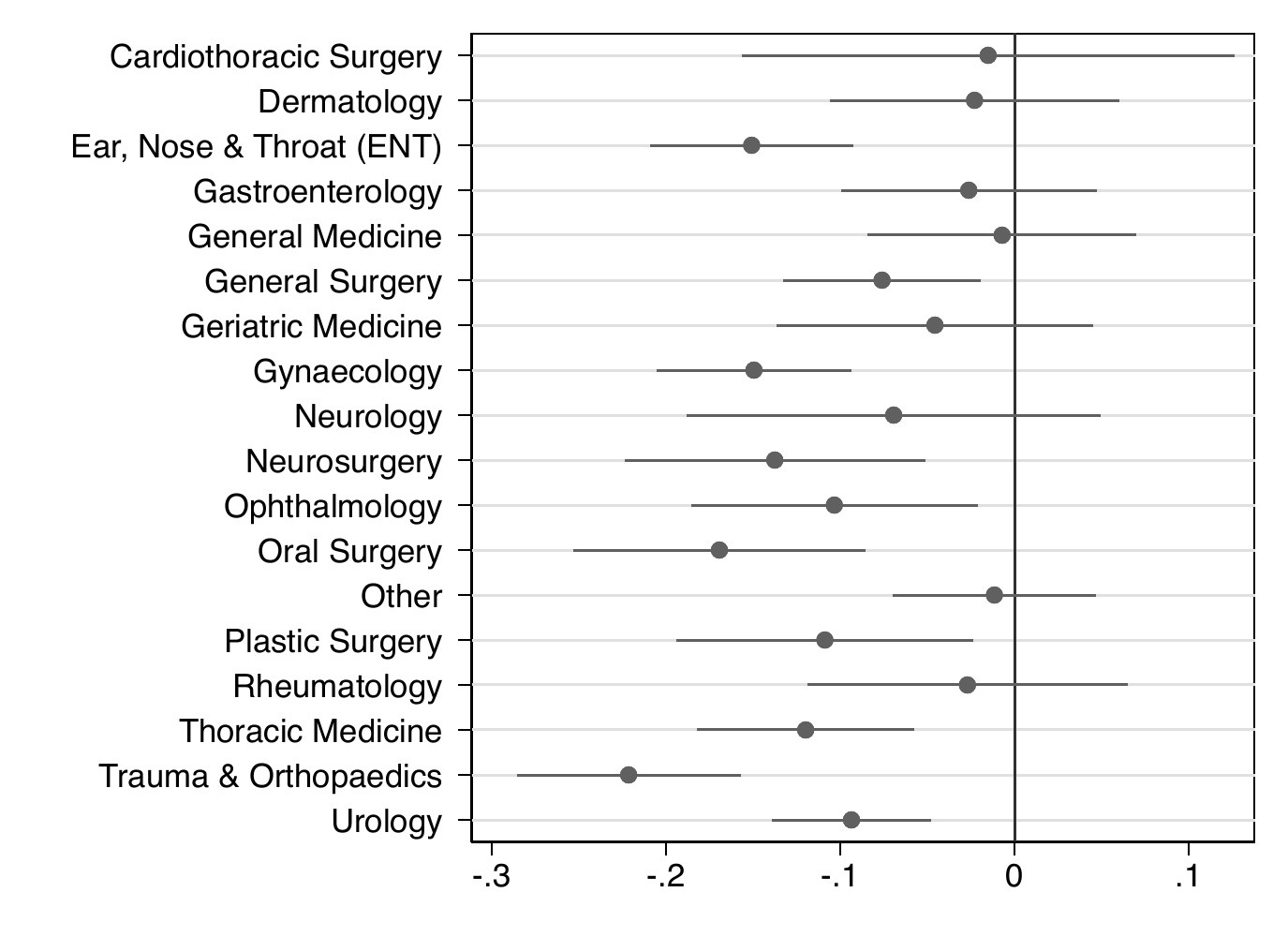} &  \includegraphics[width=0.33\columnwidth]{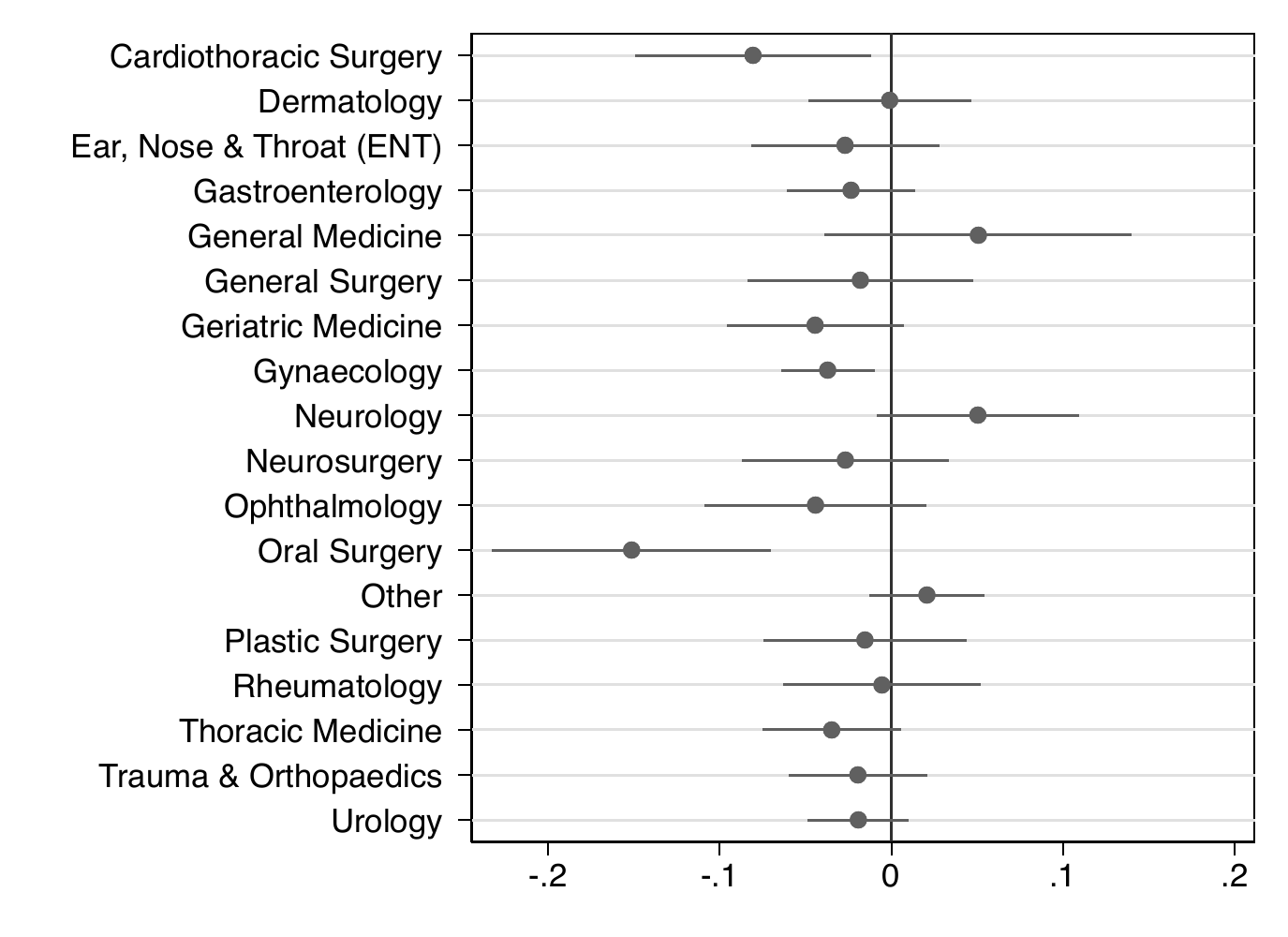}   \\
  \\
\text{\emph{Panel B}: log(Waiting list) } \\
\multicolumn{1}{c}{\text{Length}} &\multicolumn{1}{c}{\text{Aggregate wait}}  &\multicolumn{1}{c}{\text{Avg.\ wait}} \\
  \includegraphics[width=0.33\columnwidth]{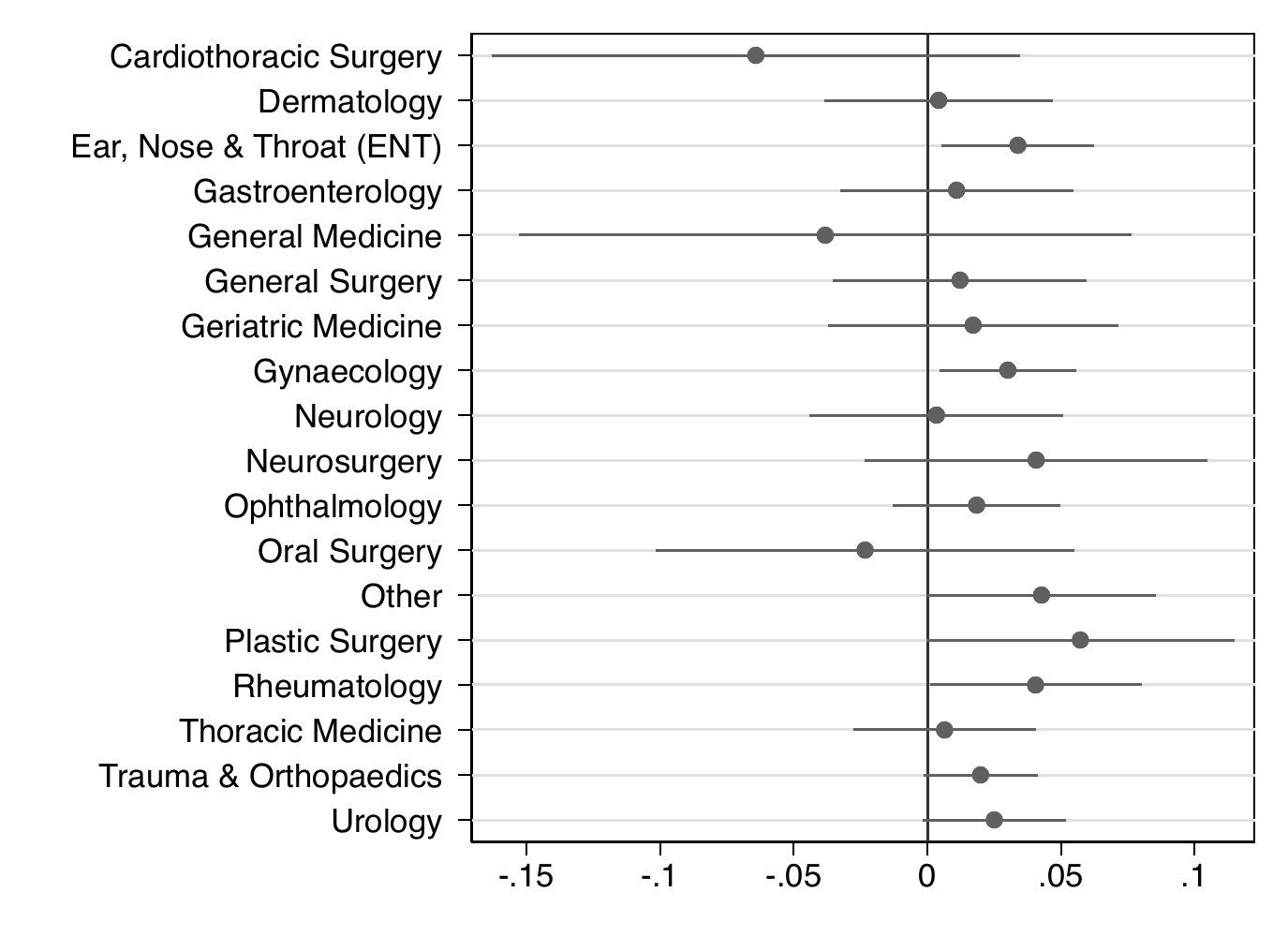} &   \includegraphics[width=0.33\columnwidth]{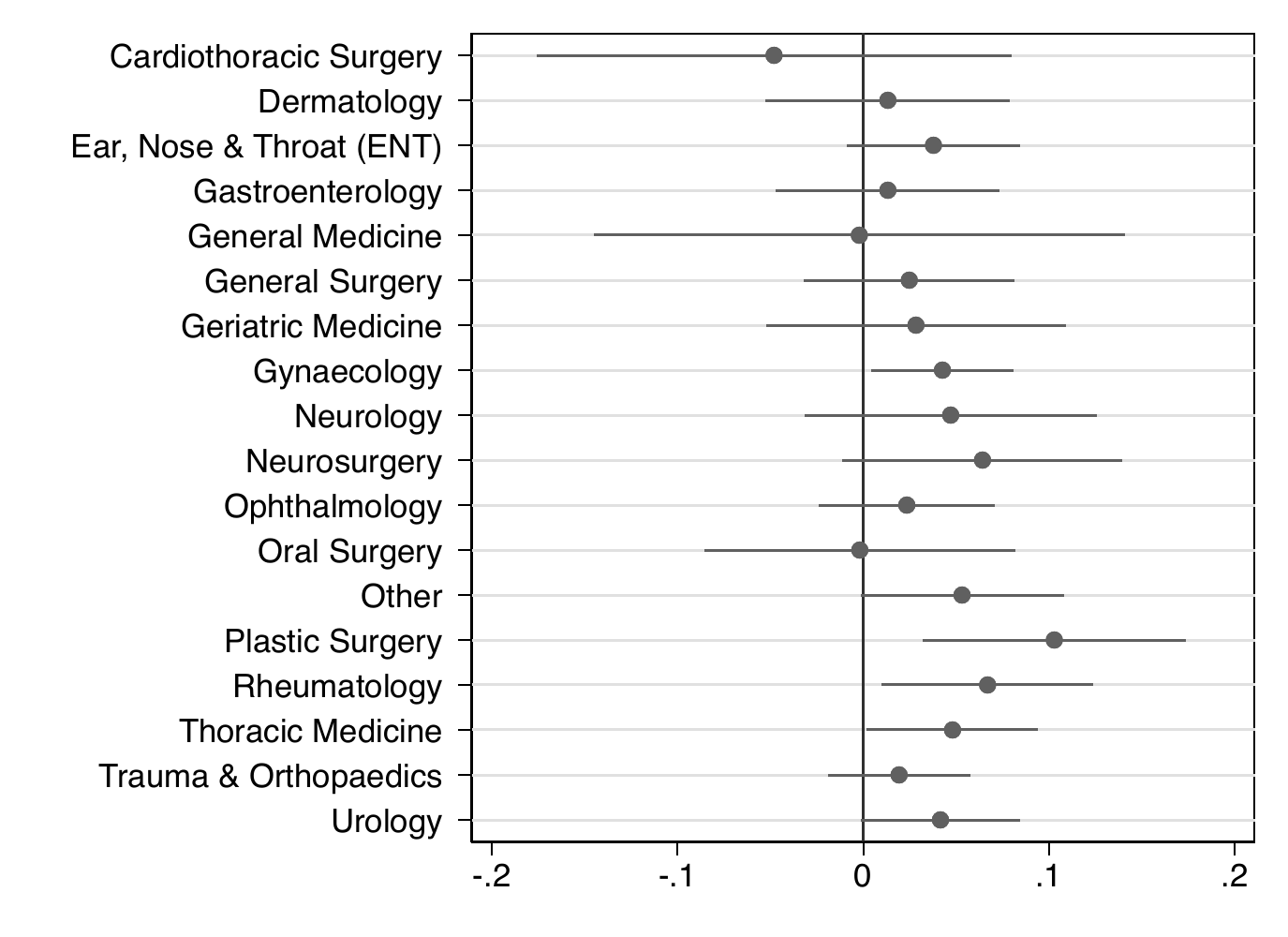} &  \includegraphics[width=0.33\columnwidth]{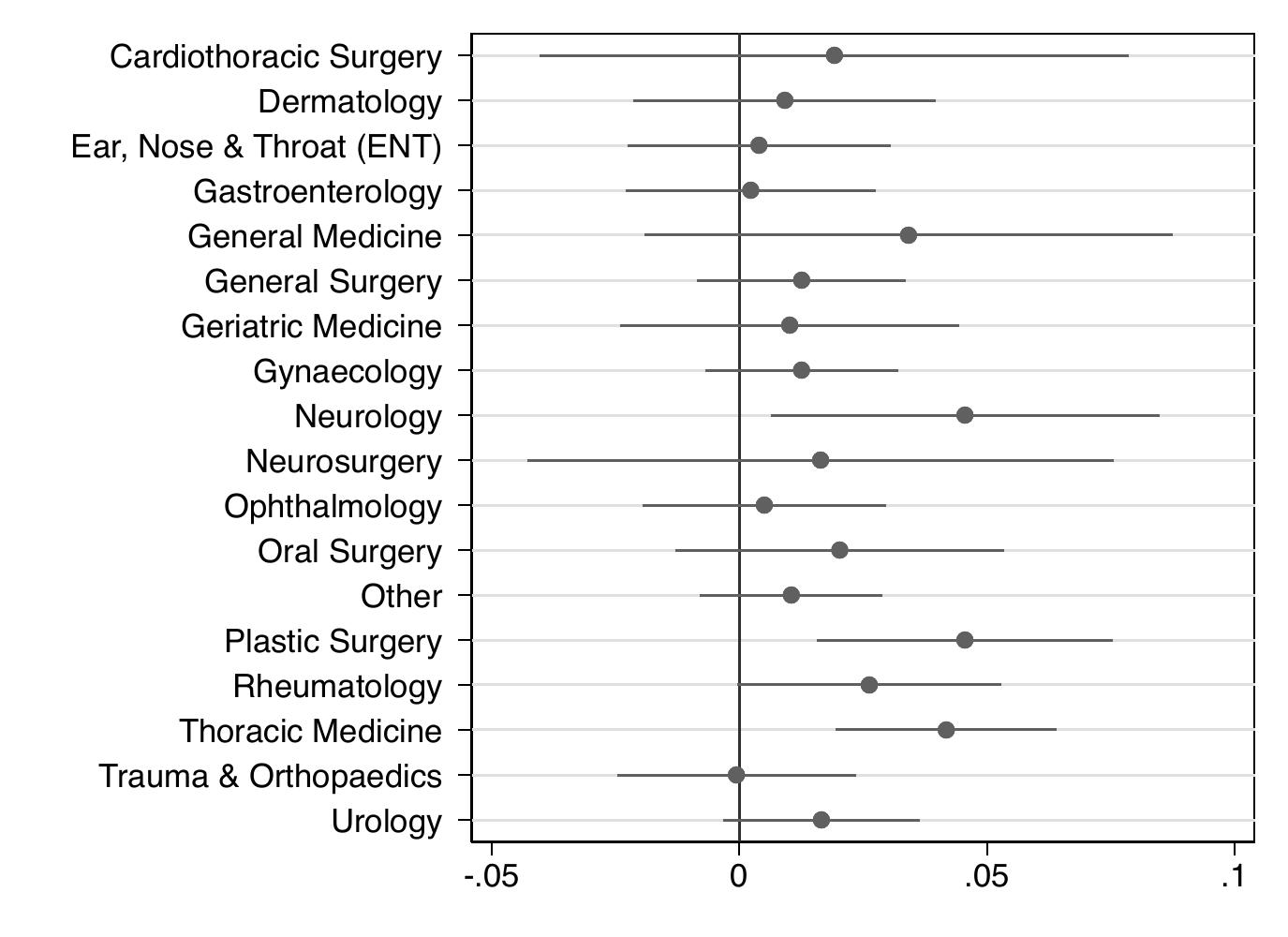}   \\
\\
\text{\emph{Panel C}: \% waiting } \\
\multicolumn{1}{c}{\text{ $>$ 4 weeks}} &\multicolumn{1}{c}{\text{ $> 8$ weeks}}  &\multicolumn{1}{c}{\text{$> 12$ weeks}} \\
  \includegraphics[width=0.33\columnwidth]{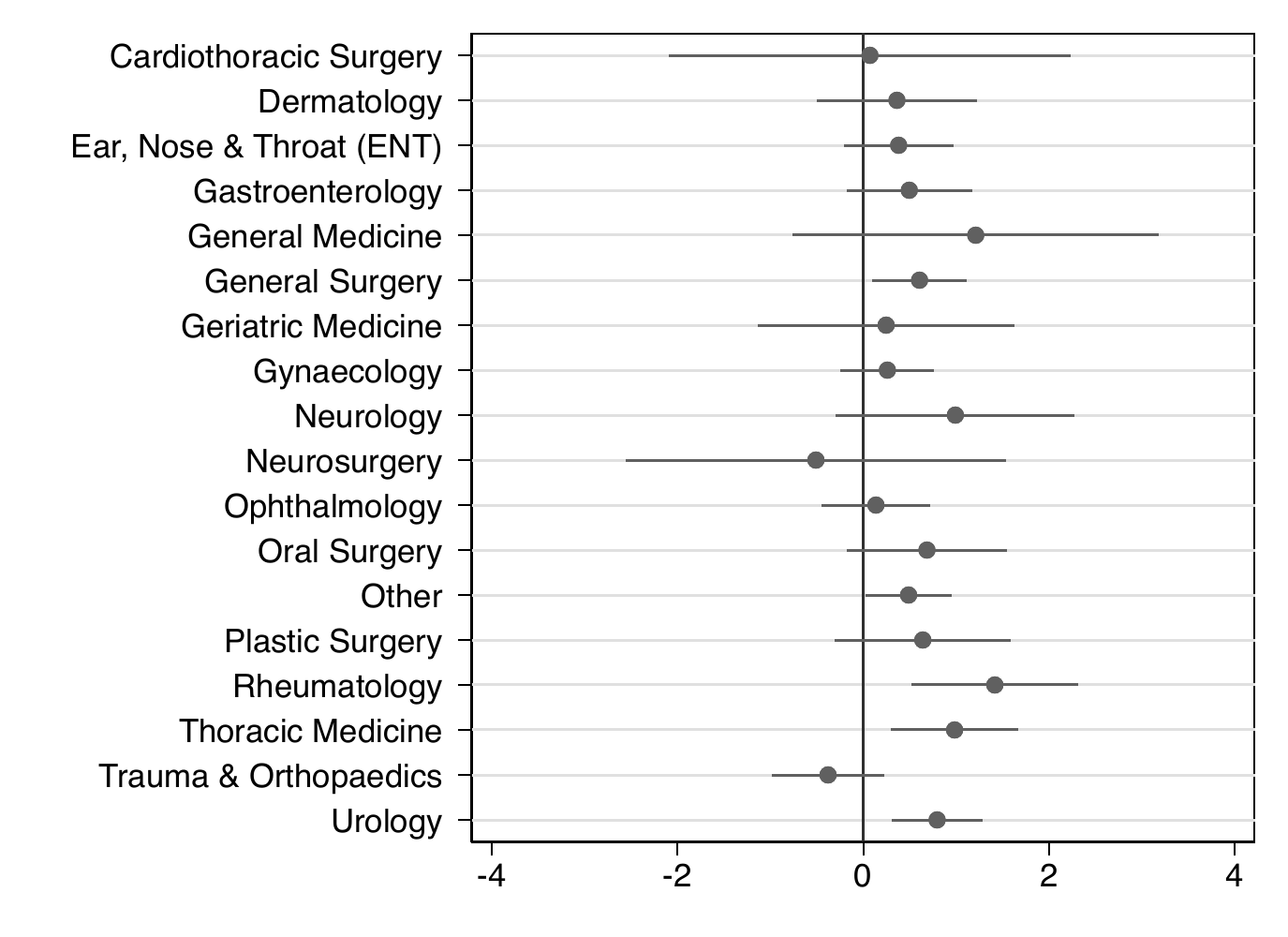} &   \includegraphics[width=0.33\columnwidth]{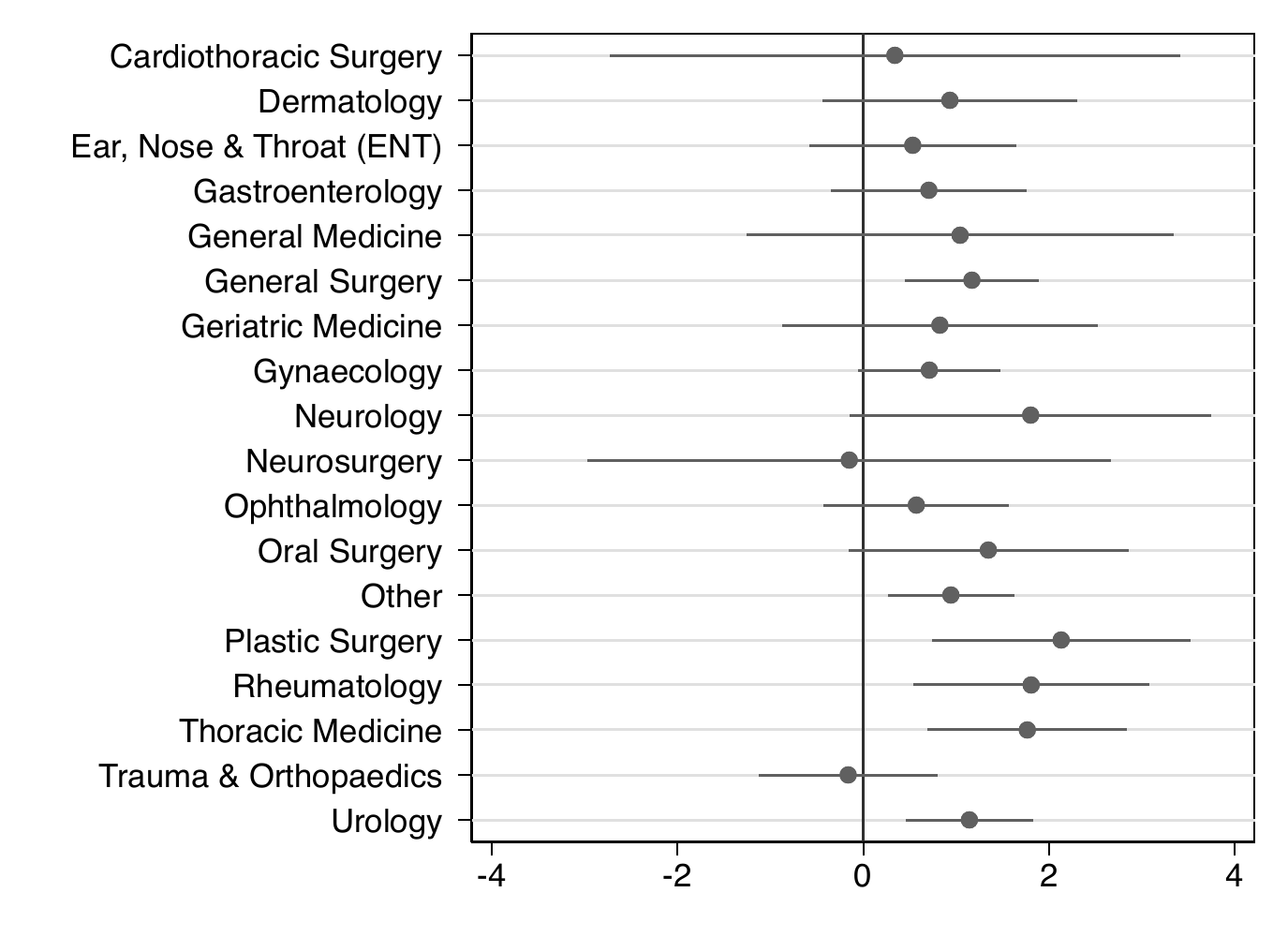} &  \includegraphics[width=0.33\columnwidth]{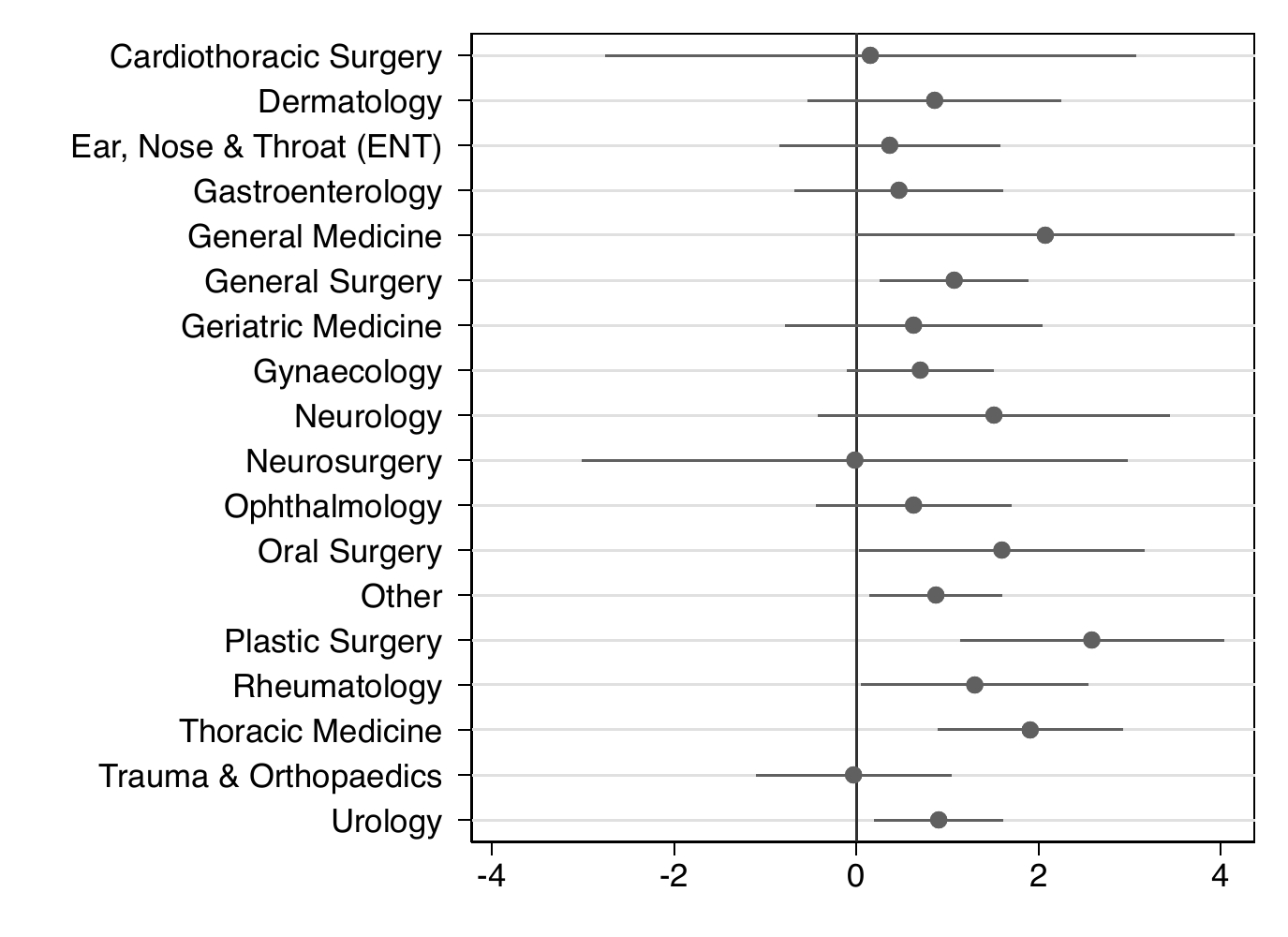}   \\

    \end{array}$
\end{center}
\scriptsize{\textbf{Notes:} Figure presents heterogenous treatment effects capturing the impact of COVID-19 pressures on quantity and waiting times for specialist referrals across different specialist treatment functions. The figures capture heterogenous effects pertaining to Panel A of Table \ref{table:rtt-pressures}. All regressions control for provider by specialist treatment function fixed effects and specialist treatment function by time fixed effects. 90\% confidence intervals obtained from clustering standard errors at the provider level are indicated.}

\end{figure}
%%%%%%%%%%%%%%%%%%%%%%%%%%%%%%

%%%%%%%%%%%%%%%%%%%%%%%%%%%%%%
%\begin{landscape}
\begin{figure}[h!]
\caption{Impact of COVID-19 pressures on specialist referrals by specialisation: effect across different deciles of the COVID-pressure intensity \label{fig:coefplot-nonlinear-rtt}}
\begin{center}
$
\begin{array}{lll}
\text{\emph{Panel A}:  log(Referrals) } \\
 \multicolumn{1}{c}{}  &\multicolumn{2}{c}{\text{Completed} }  \\
  \multicolumn{1}{c}{ \text{New}}  &\multicolumn{1}{c}{\text{Admitted}} &\multicolumn{1}{c}{\text{Non-admitted}} \\
  \includegraphics[width=0.33\columnwidth]{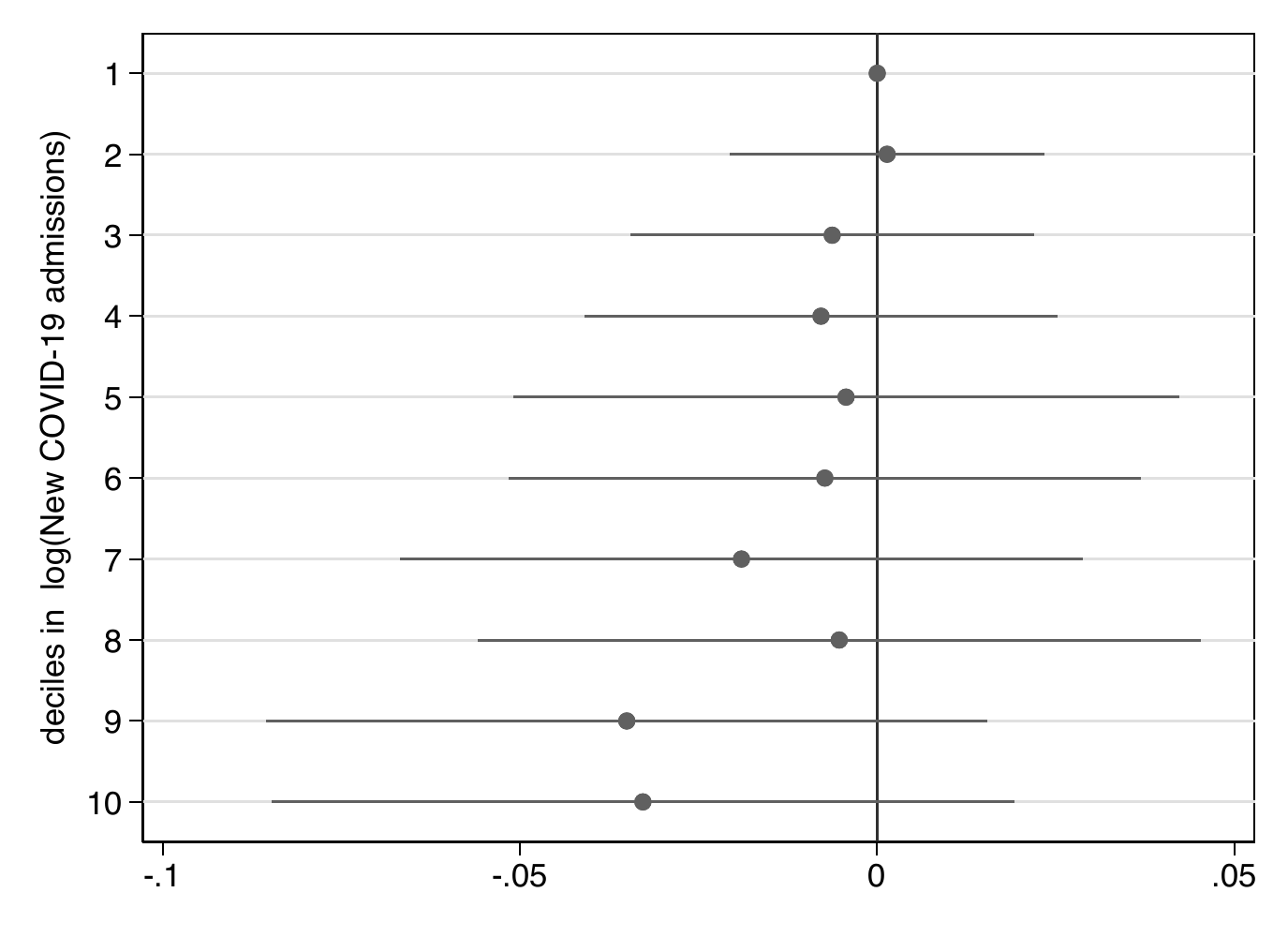} &   \includegraphics[width=0.33\columnwidth]{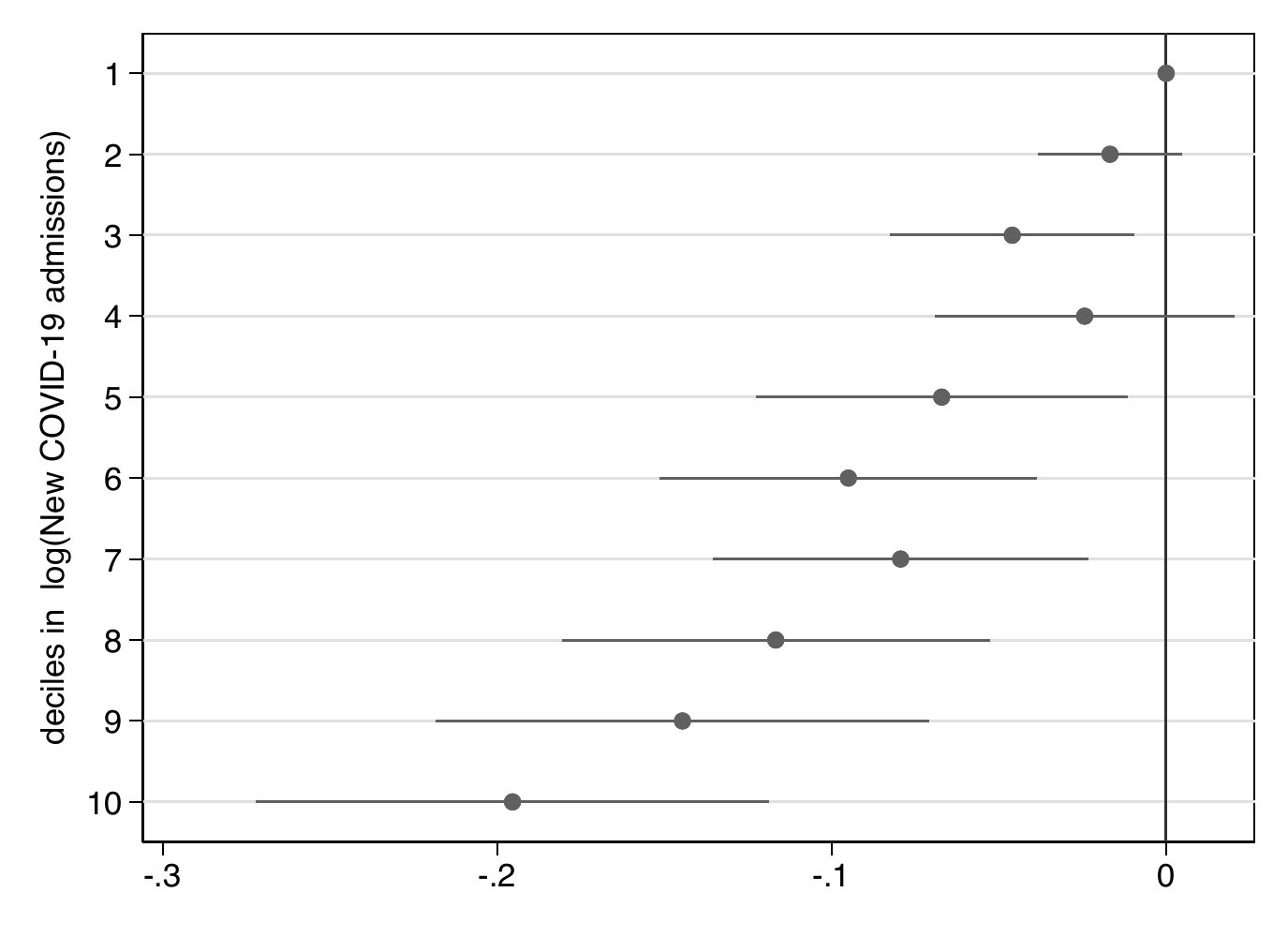} &  \includegraphics[width=0.33\columnwidth]{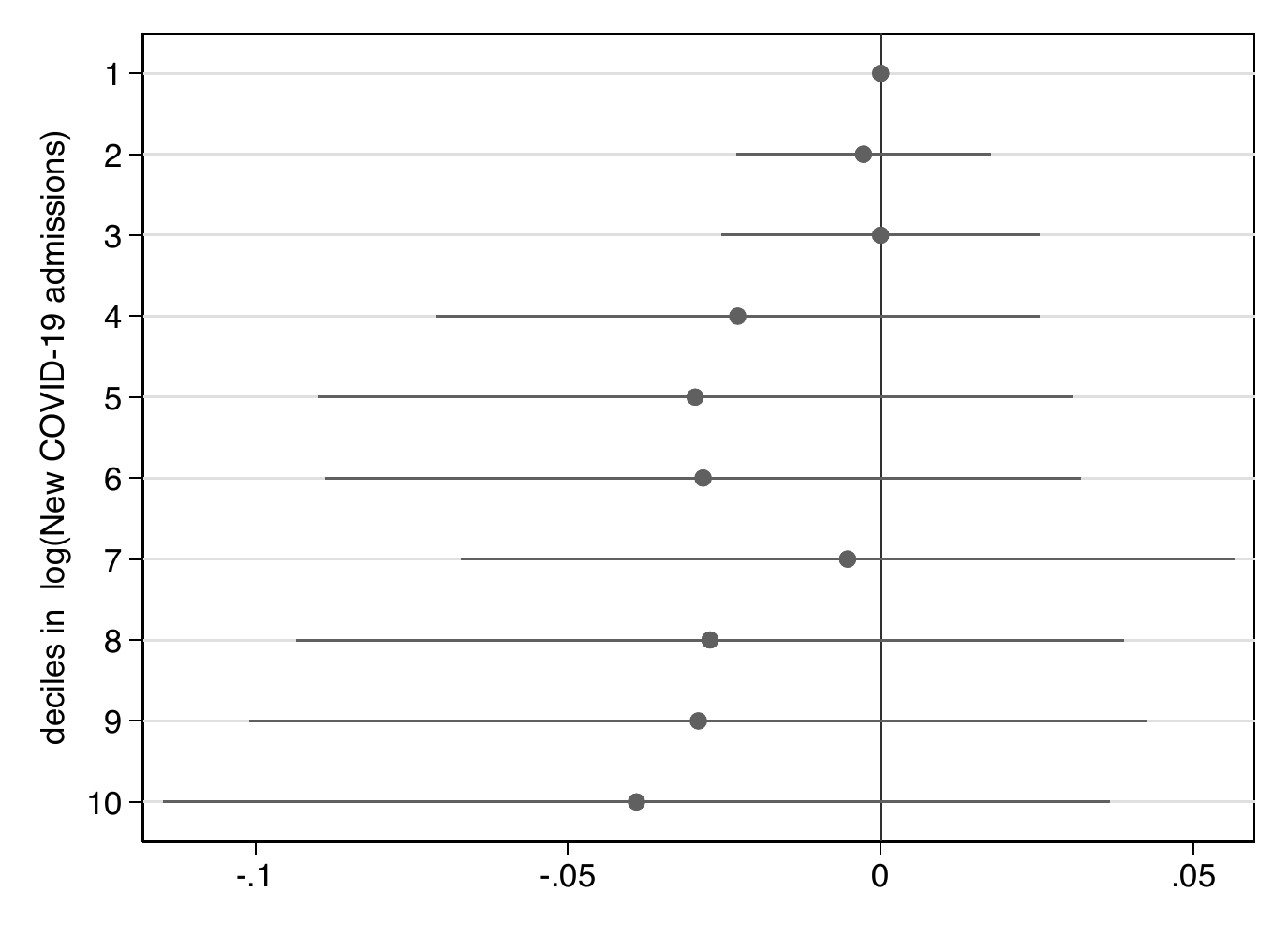}   \\
  \\
\text{\emph{Panel B}: log(Waiting list) } \\
\multicolumn{1}{c}{\text{Length}} &\multicolumn{1}{c}{\text{Aggregate wait}}  &\multicolumn{1}{c}{\text{Avg.\ wait}} \\
  \includegraphics[width=0.33\columnwidth]{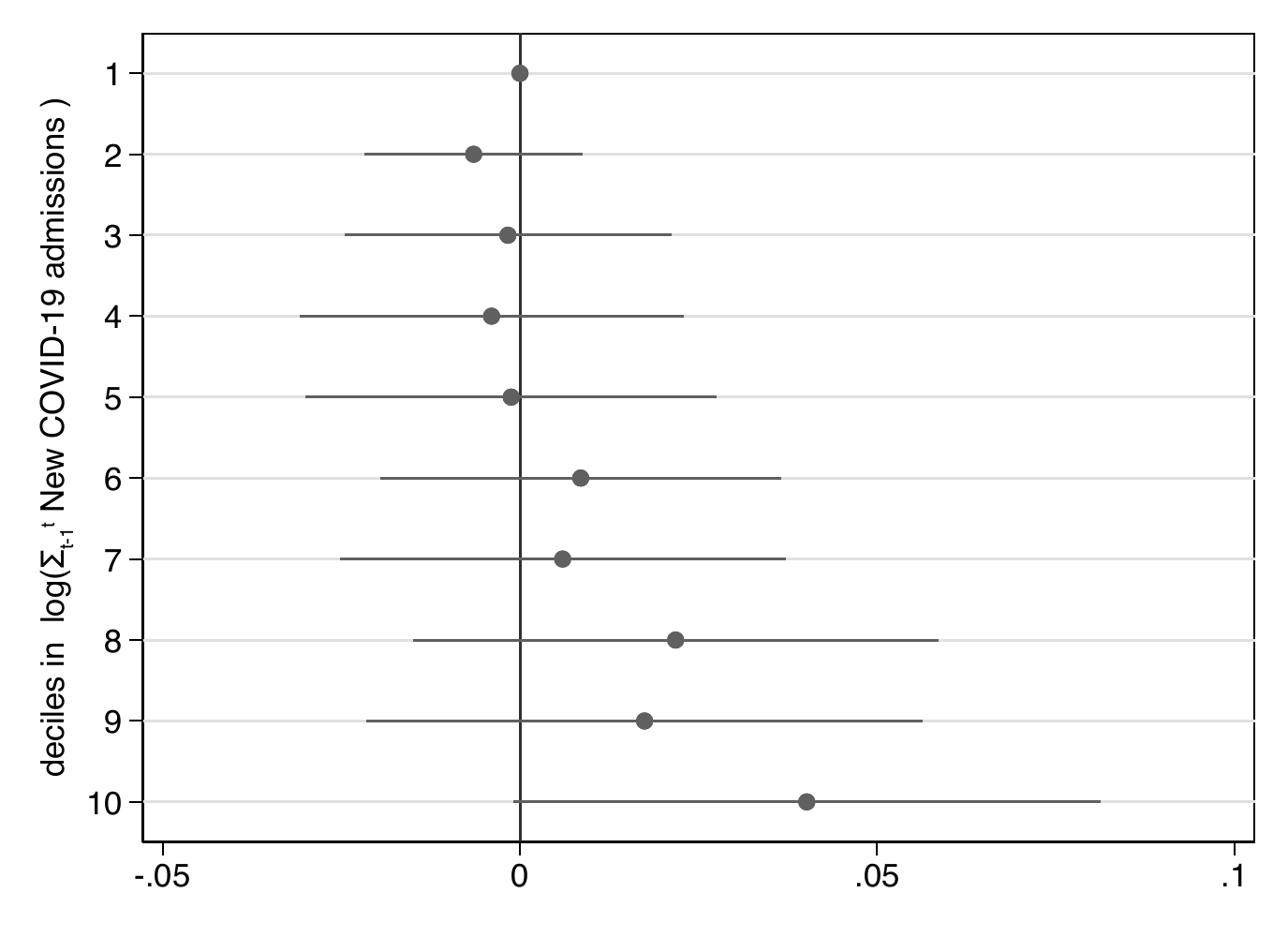} &   \includegraphics[width=0.33\columnwidth]{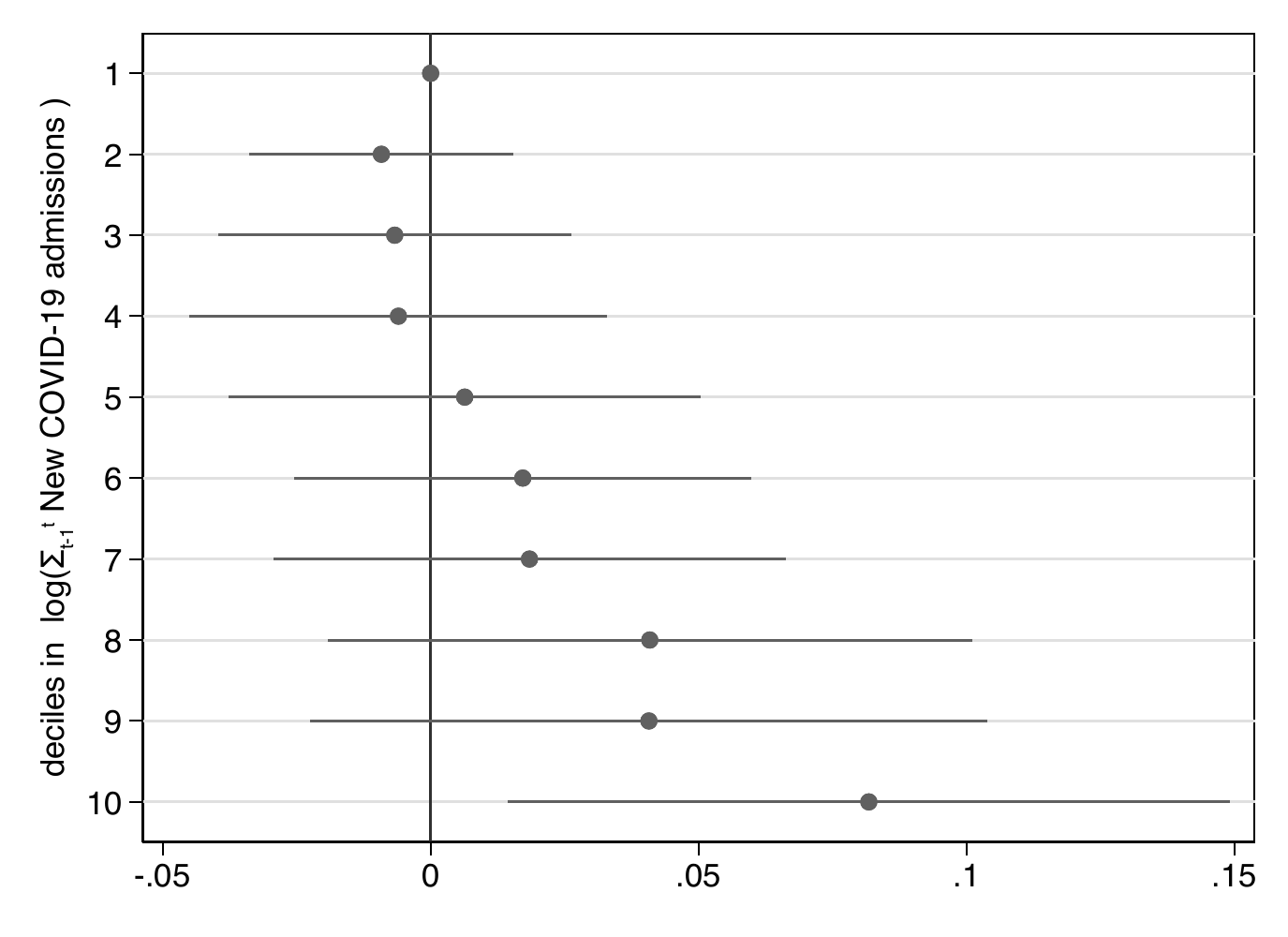} &  \includegraphics[width=0.33\columnwidth]{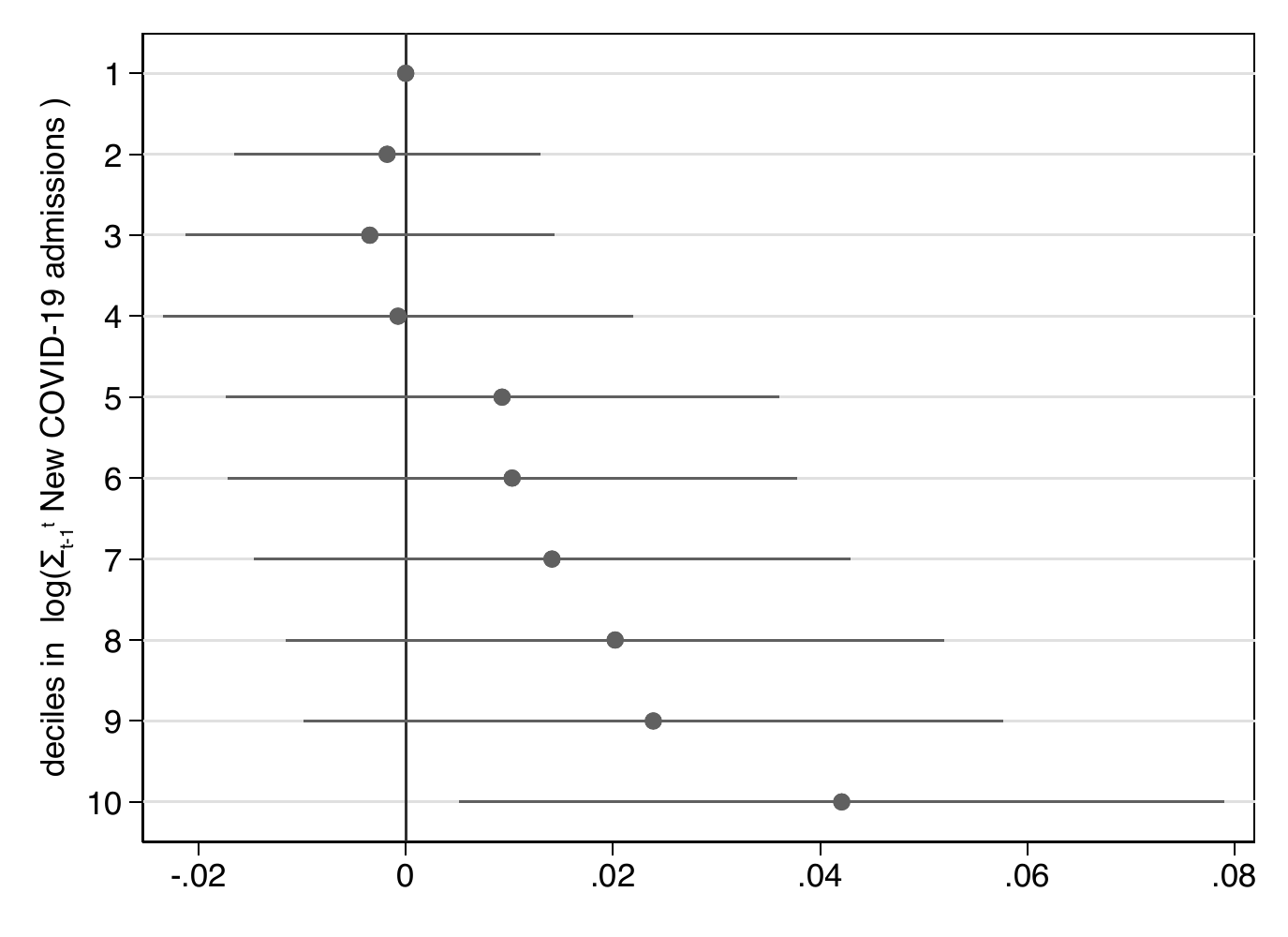}   \\
\\
\text{\emph{Panel C}: \% waiting } \\
\multicolumn{1}{c}{\text{ $>$ 4 weeks}} &\multicolumn{1}{c}{\text{ $> 8$ weeks}}  &\multicolumn{1}{c}{\text{$> 12$ weeks}} \\
  \includegraphics[width=0.33\columnwidth]{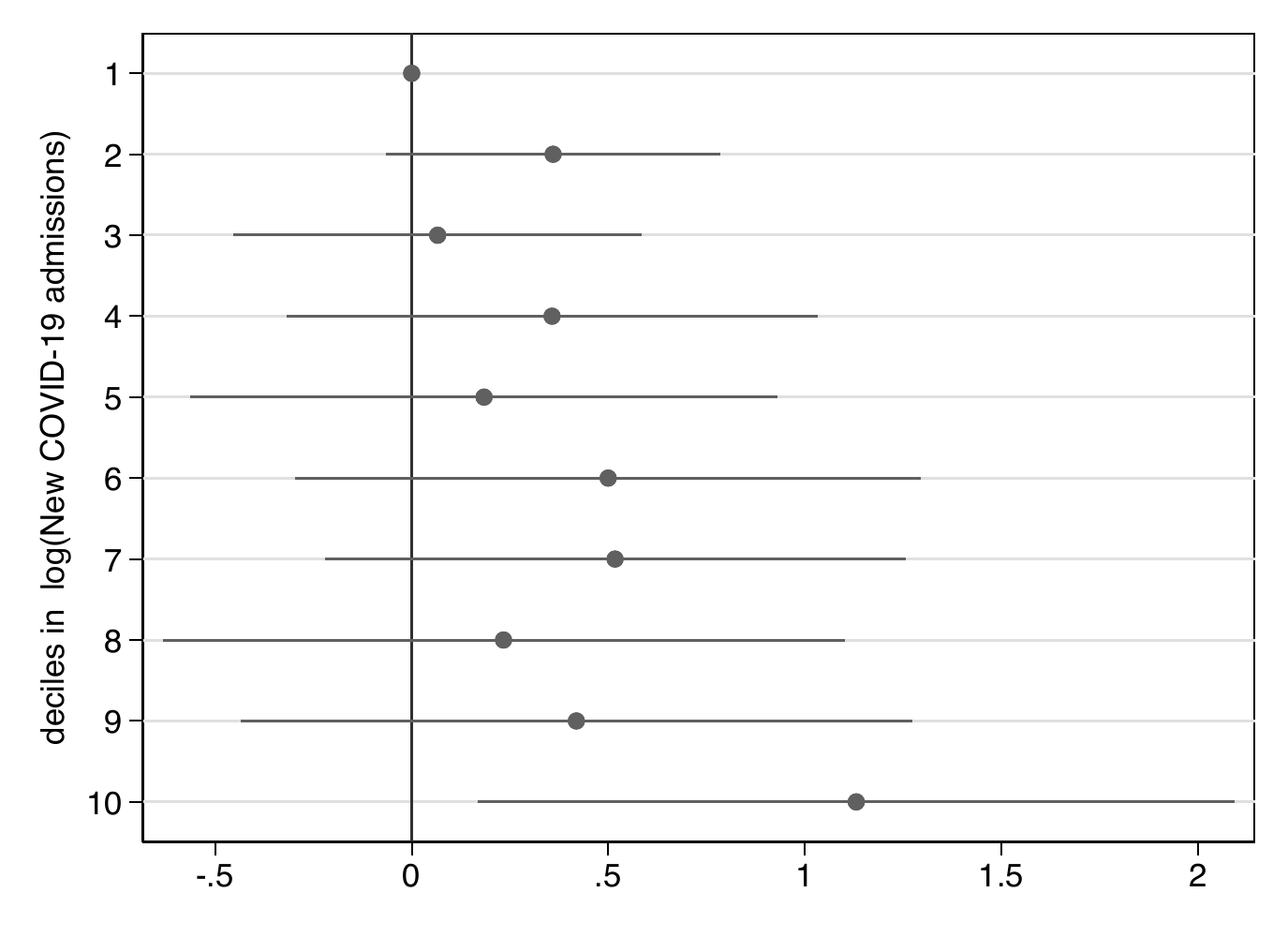} &   \includegraphics[width=0.33\columnwidth]{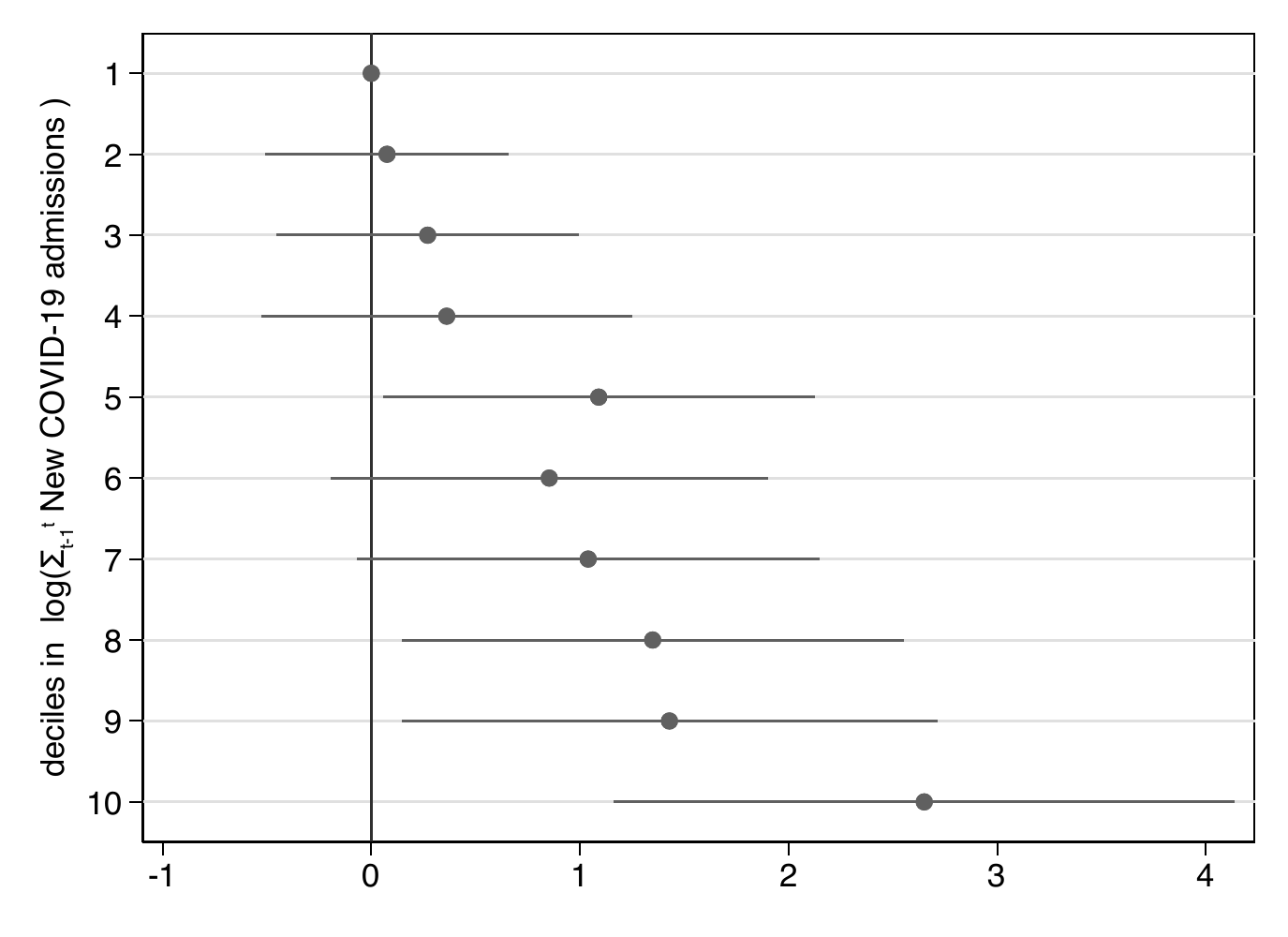} &  \includegraphics[width=0.33\columnwidth]{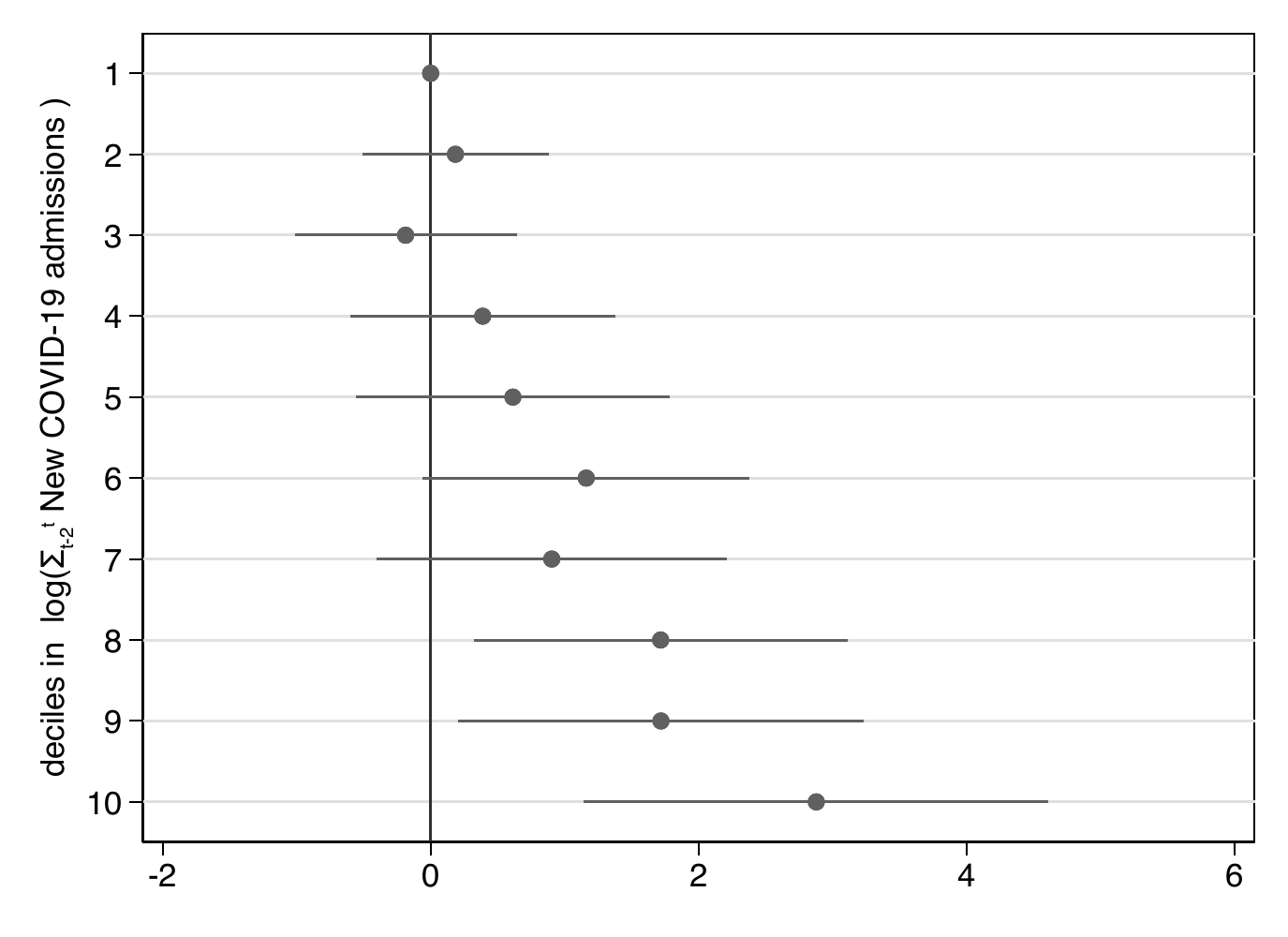}   \\

    \end{array}$
\end{center}
\scriptsize{\textbf{Notes:} Figure presents heterogenous treatment effects capturing the impact of COVID-19 pressures measured as provider-specific deciles of the new admissions affect both the quantity and waiting times for specialist referrals across different specialist treatment functions. The figures capture heterogenous effects pertaining to Panel A of Table \ref{table:rtt-pressures} where new admissions are converted to deciles by provider. All regressions control for provider by specialist treatment function fixed effects and specialist treatment function by time fixed effects. 90\% confidence intervals obtained from clustering standard errors at the provider level are indicated.}

\end{figure}
%%%%%%%%%%%%%%%%%%%%%%%%%%%%%%

%%%%%%%%%%%%%%%%%%%%%%%%%%%%%%
\begin{landscape}
\begin{figure}[h!]
\caption{Impact of COVID-19 pressures on diagnostic activity and waiting time \label{fig:coefplot-diagnostic-tests}}
\begin{center}
$
\begin{array}{lll}
\text{\emph{Panel A}:  log(Total activity) }  & \text{\emph{Panel B}:  log(Average wait) }  & \text{\emph{Panel C}:  \% waiting $>$  6 weeks  } \\
  \includegraphics[width=0.33\columnwidth]{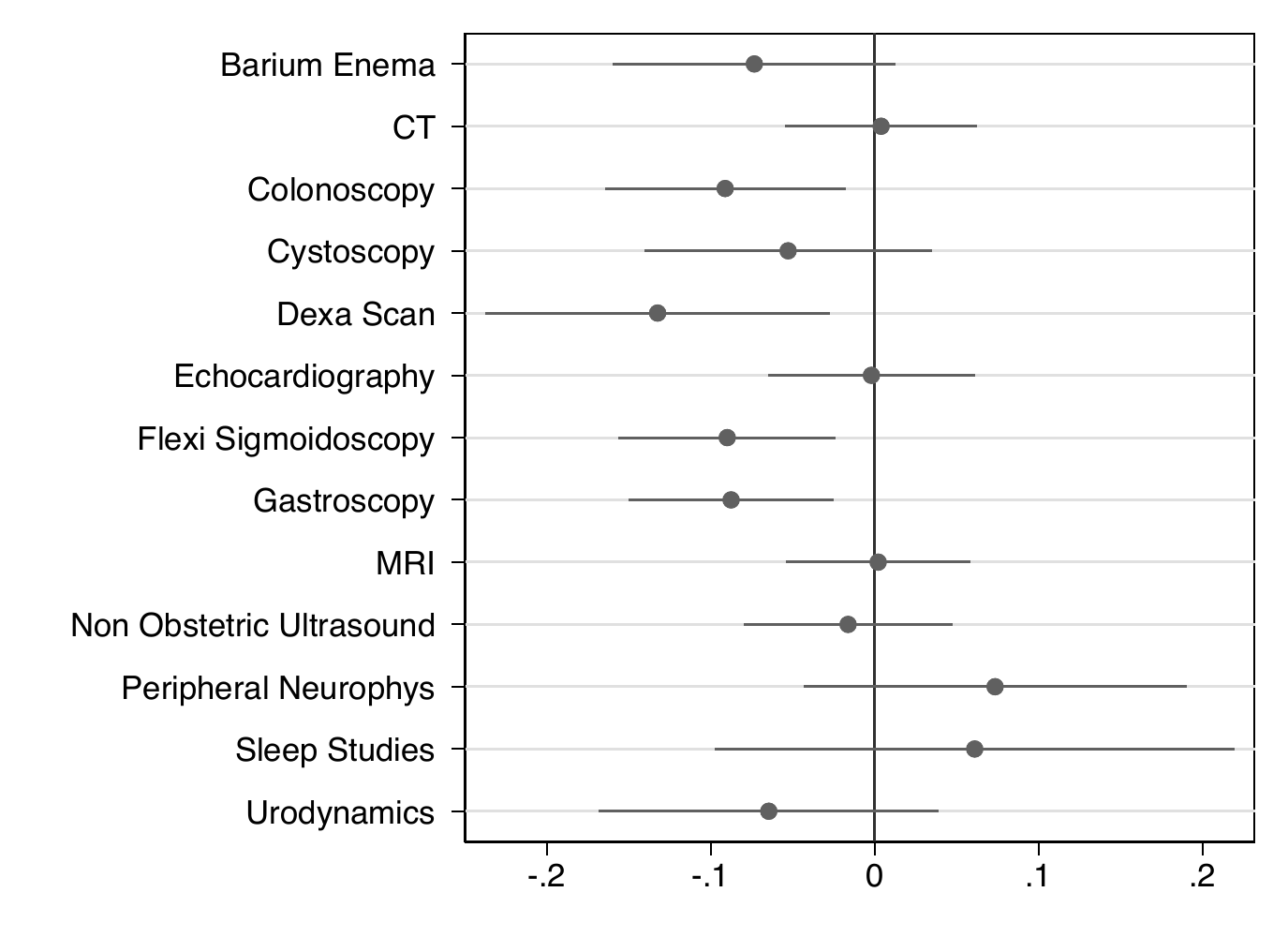} &   \includegraphics[width=0.33\columnwidth]{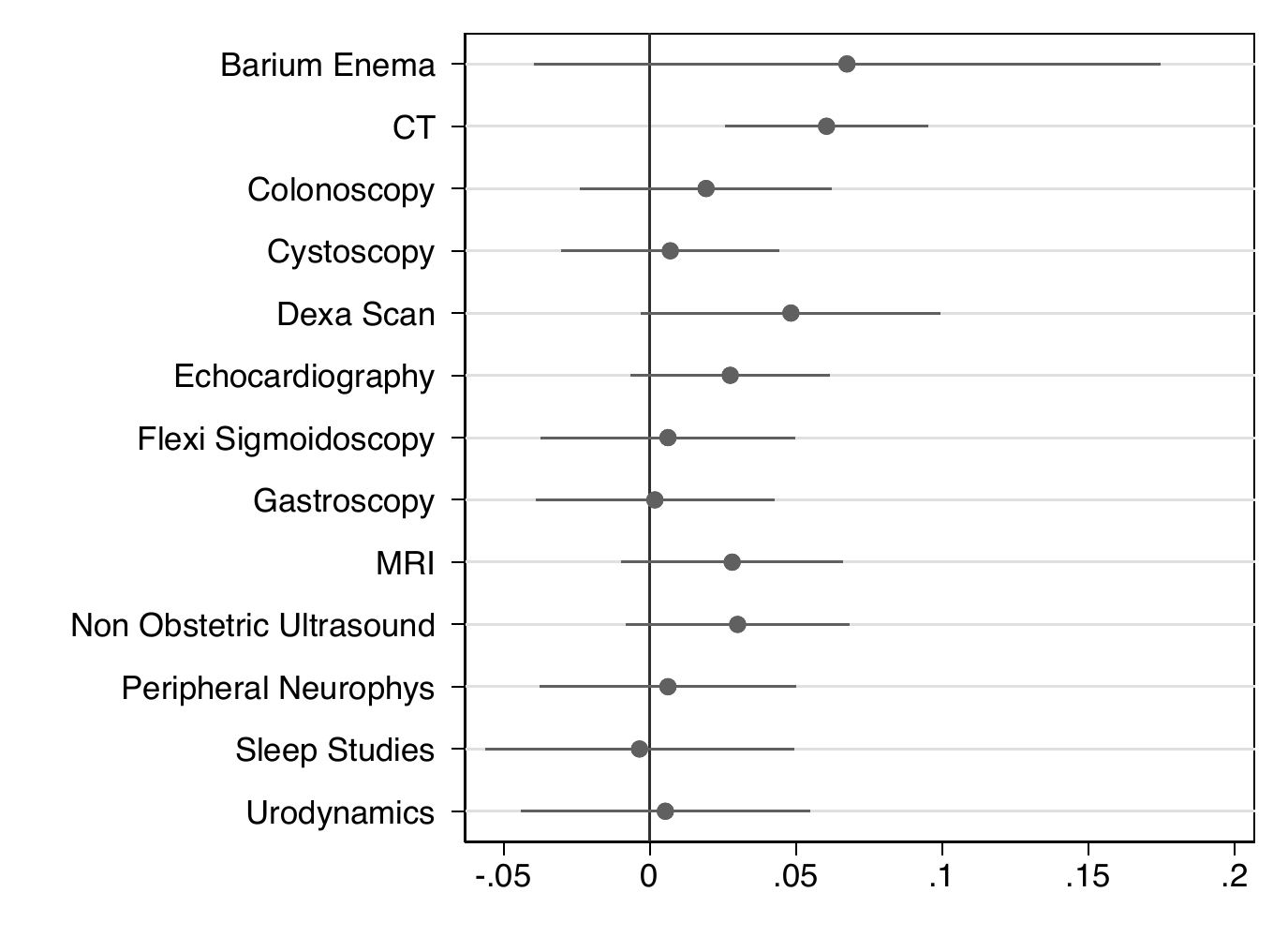} &  \includegraphics[width=0.33\columnwidth]{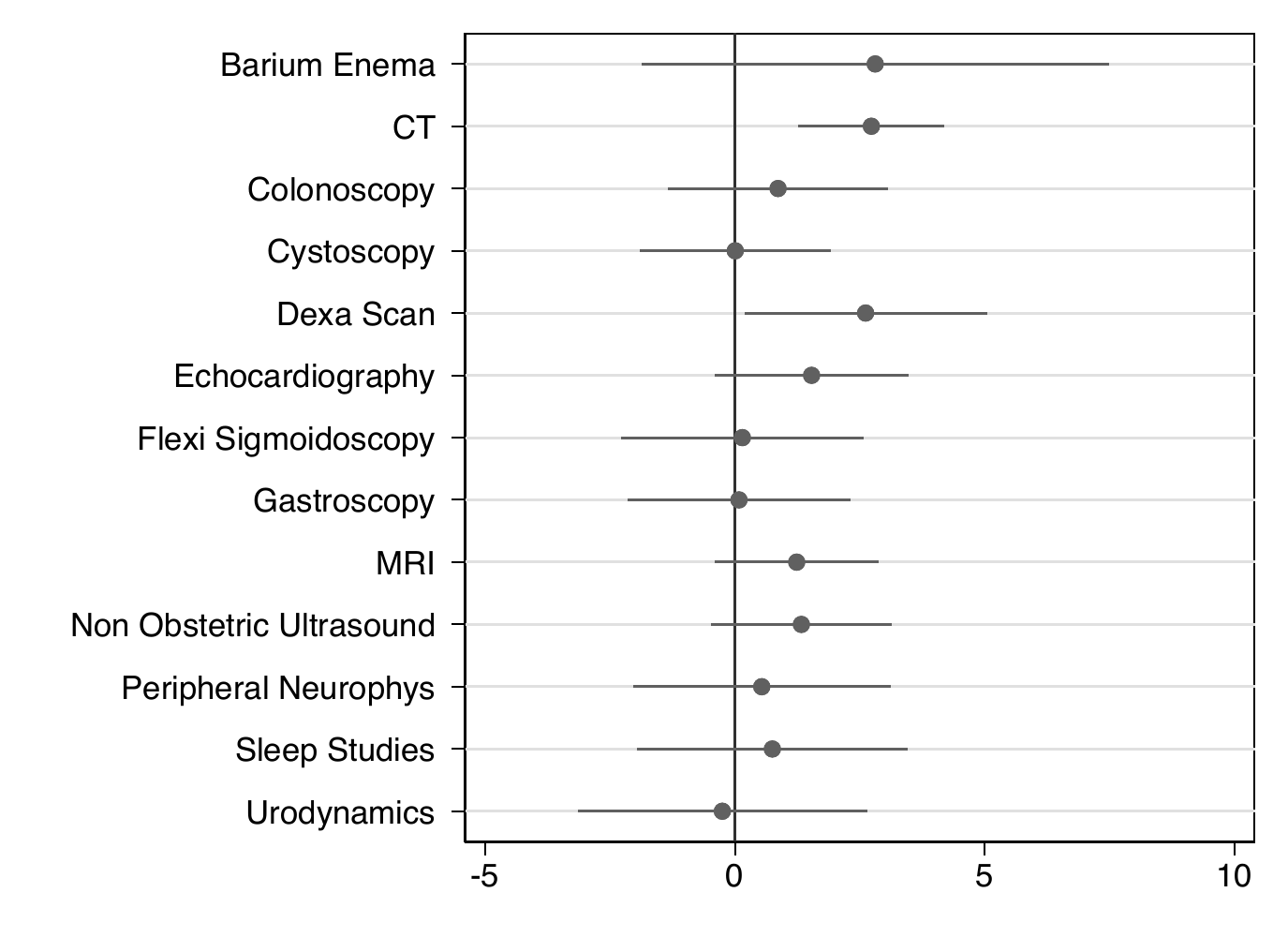}   \\
     \end{array}$
\end{center}
\scriptsize{\textbf{Notes:} Figure presents heterogenous treatment effects capturing the impact of COVID-19 pressures on quantity and waiting times for different diagnostic activity. The figures capture heterogenous effects pertaining to Panel A of Table \ref{table:diagnostic-pressures}. All regressions control for provider by diagnostic function fixed effects and diagnostic test by time fixed effects. 90\% confidence intervals obtained from clustering standard errors at the provider level are indicated.}

\end{figure}
\end{landscape}
%%%%%%%%%%%%%%%%%%%%%%%%%%%%%%

%%%%%%%%%%%%%%%%%%%%%%%%%%%%%%
\begin{landscape}
\begin{figure}[h!]
\caption{Impact of COVID-19 pressures on diagnostic activity and waiting time: effect across different deciles of the COVID-pressure intensity \label{fig:coefplot-nonlinear-diagnostic-tests}}
\begin{center}
$
\begin{array}{lll}
\text{\emph{Panel A}:  log(Total activity) }  & \text{\emph{Panel B}:  log(Average wait) }  & \text{\emph{Panel C}:  \% waiting $>$  6 weeks  } \\
  \includegraphics[width=0.33\columnwidth]{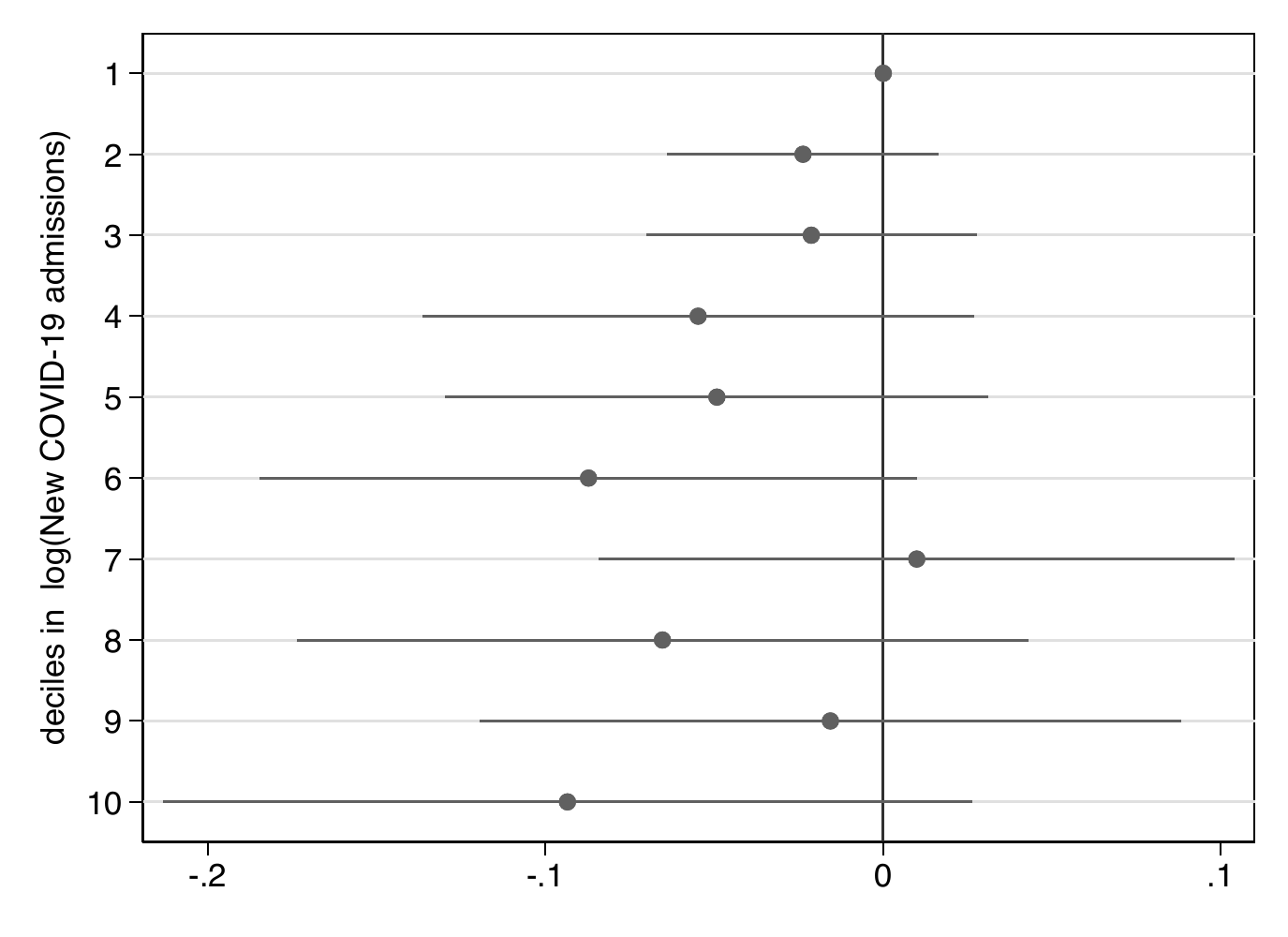} &   \includegraphics[width=0.33\columnwidth]{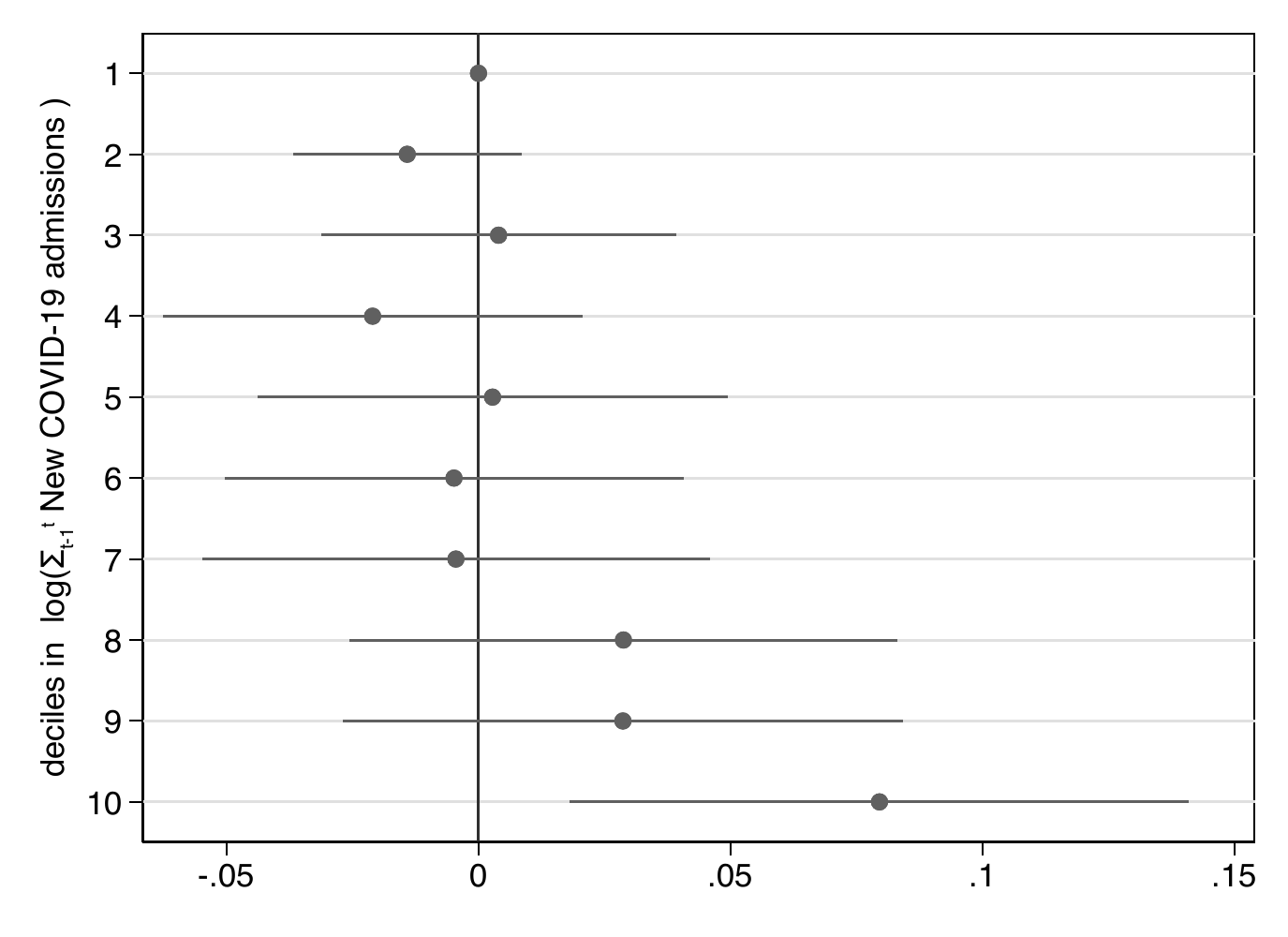} &  \includegraphics[width=0.33\columnwidth]{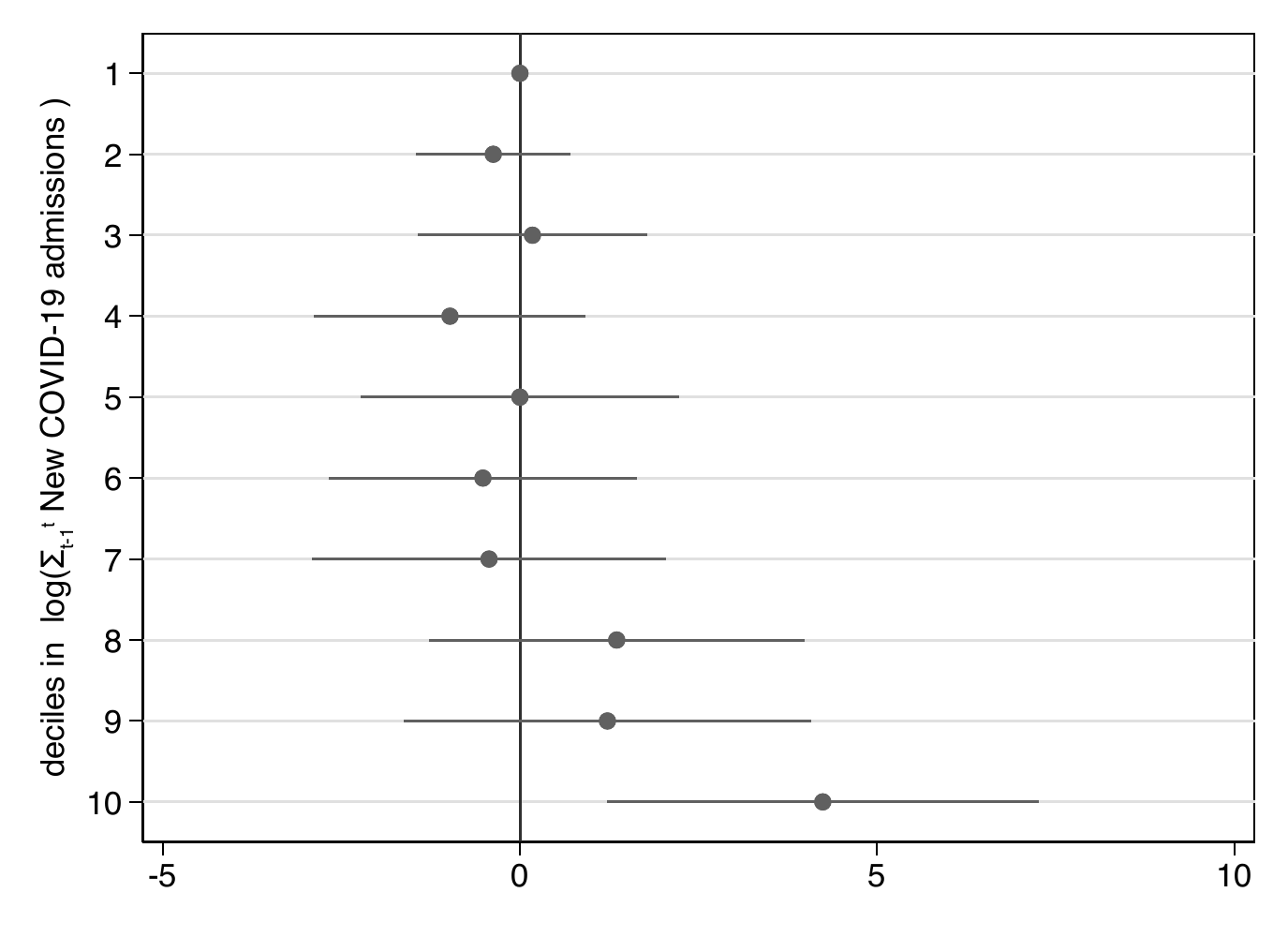}   \\
     \end{array}$
\end{center}
\scriptsize{\textbf{Notes:} Figure presents heterogenous treatment effects capturing the impact of COVID-19 pressures measured by deciles of the provider-specific empirical distribution on quantity and waiting times for different diagnostic activity. The figures capture heterogenous effects pertaining to Panel A of Table \ref{table:diagnostic-pressures} where new admissions are converted to deciles by provider. All regressions control for provider by diagnostic function fixed effects and diagnostic test by time fixed effects. 90\% confidence intervals obtained from clustering standard errors at the provider level are indicated.}

\end{figure}
\end{landscape}
%%%%%%%%%%%%%%%%%%%%%%%%%%%%%%

%%%%%%%%%%%%%%%%%%%%%%%%%%%%%%
\begin{landscape}
\begin{figure}[h!]
\caption{Impact of COVID-19 pressures on cancer treatment by cancer type \label{fig:coefplot-cancer}}
\begin{center}
$
\begin{array}{ll}
\text{\emph{Panel A}: log(Total \# of cancer patients treated)   }  & \text{\emph{Panel B}:  \% patients treated after 62 days  }  \\
  \includegraphics[width=0.5\columnwidth]{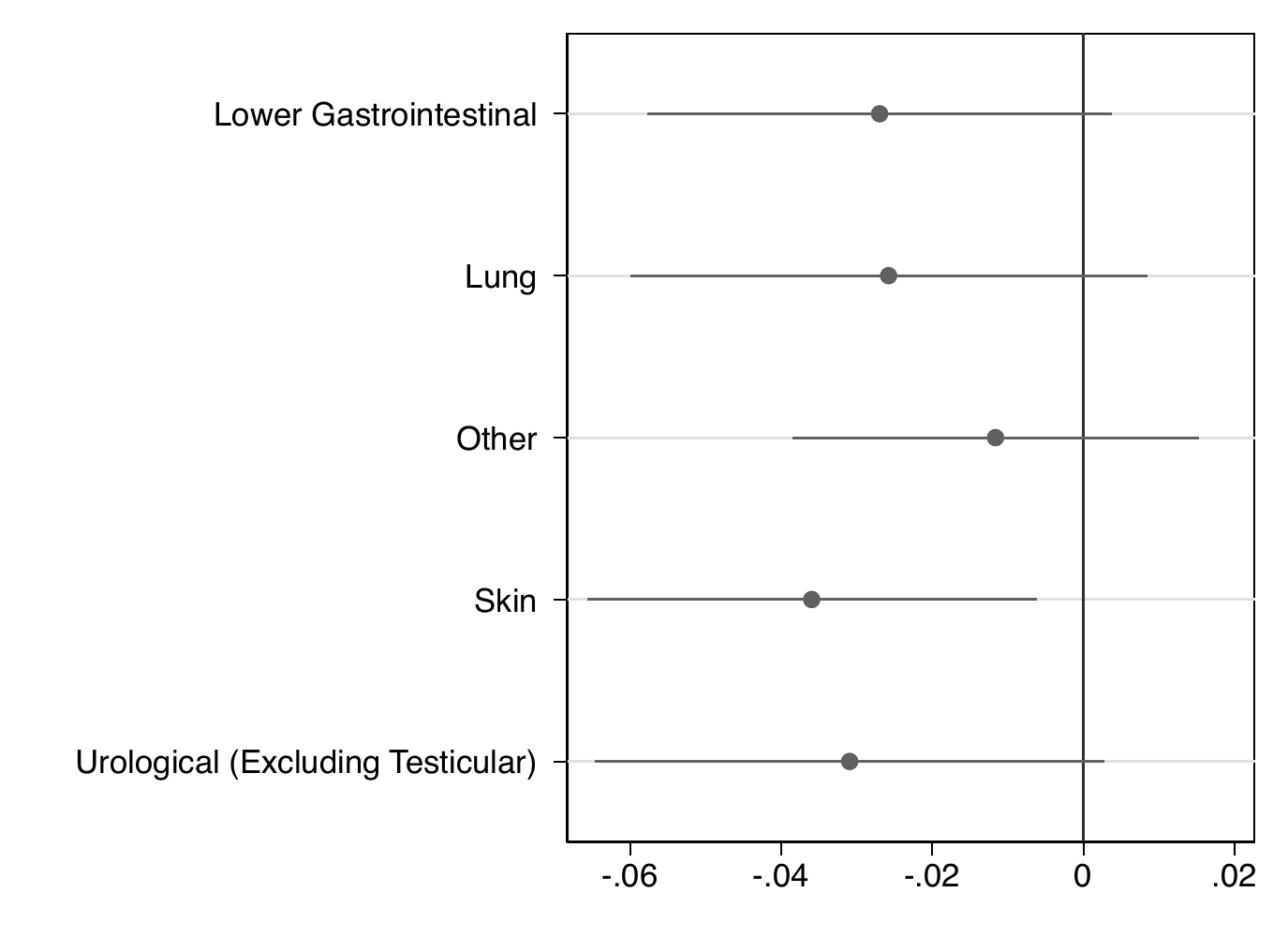} &   \includegraphics[width=0.5\columnwidth]{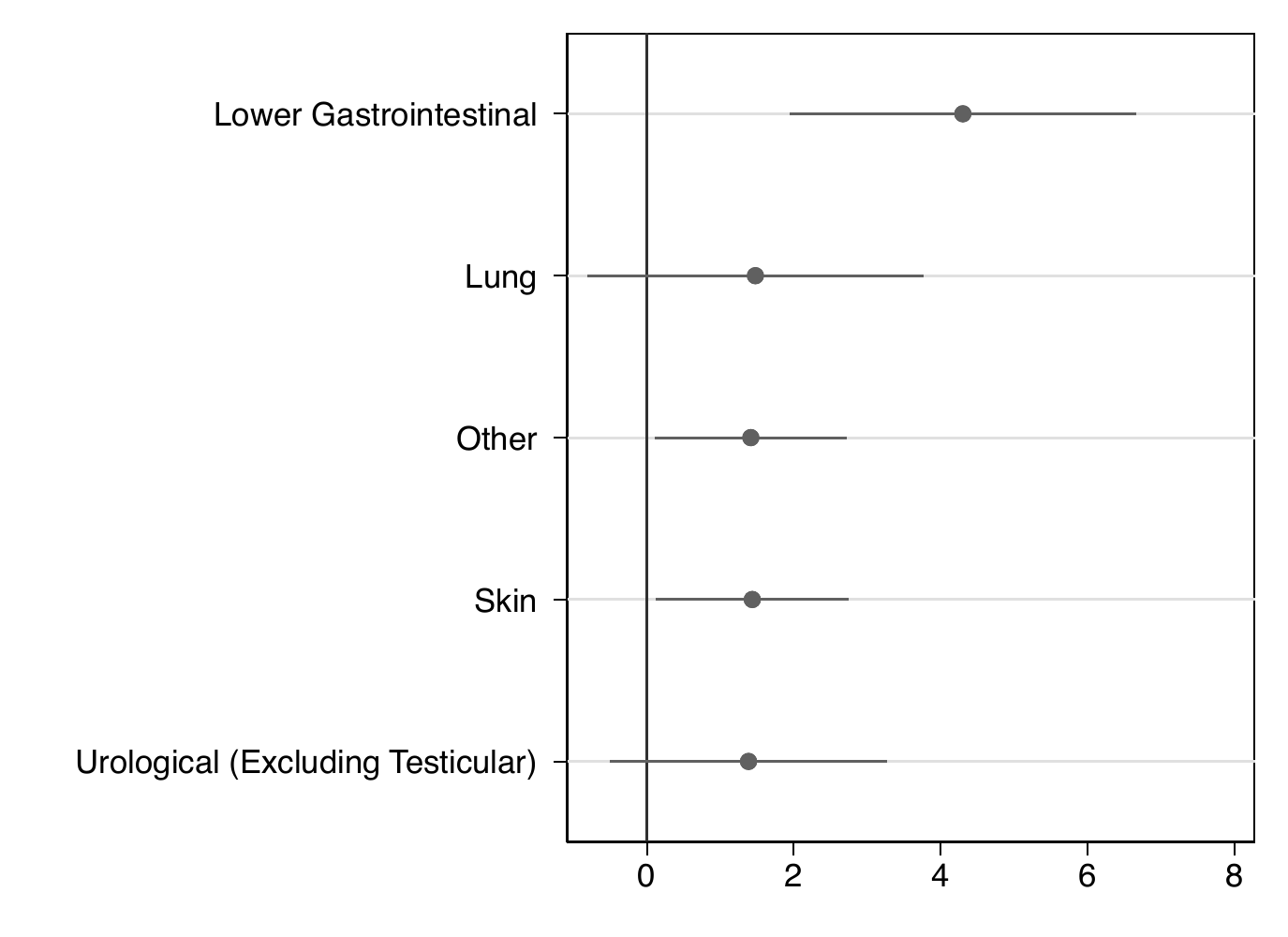}  \\
     \end{array}$
\end{center}
\scriptsize{\textbf{Notes:} Figure presents heterogenous treatment effects capturing the impact of COVID-19 pressures on cancer treatment measured as number of patients referred for treatment (Panel A) and the share of said cancer treatment receiving cancer treatment after 62 days of urgent referral (Panel B). All regressions control for provider by cancer by care setting (admitted or non-admitted) fixed effects and cancer by care setting by time fixed effects. 90\% confidence intervals obtained from clustering standard errors at the provider level are indicated.}

\end{figure}
\end{landscape}
%%%%%%%%%%%%%%%%%%%%%%%%%%%%%%

%%%%%%%%%%%%%%%%%%%%%%%%%%%%%%
\begin{landscape}
\begin{figure}[h!]
\caption{Impact of COVID-19 pressures on cancer treatment by cancer type: effect across different deciles of the COVID-pressure intensity \label{fig:coefplot-nonlinear-cancer}}
\begin{center}
$
\begin{array}{ll}
\text{\emph{Panel A}: log(Total \# of cancer patients treated)   }  & \text{\emph{Panel B}:  \% patients treated after 62 days  }  \\
  \includegraphics[width=0.5\columnwidth]{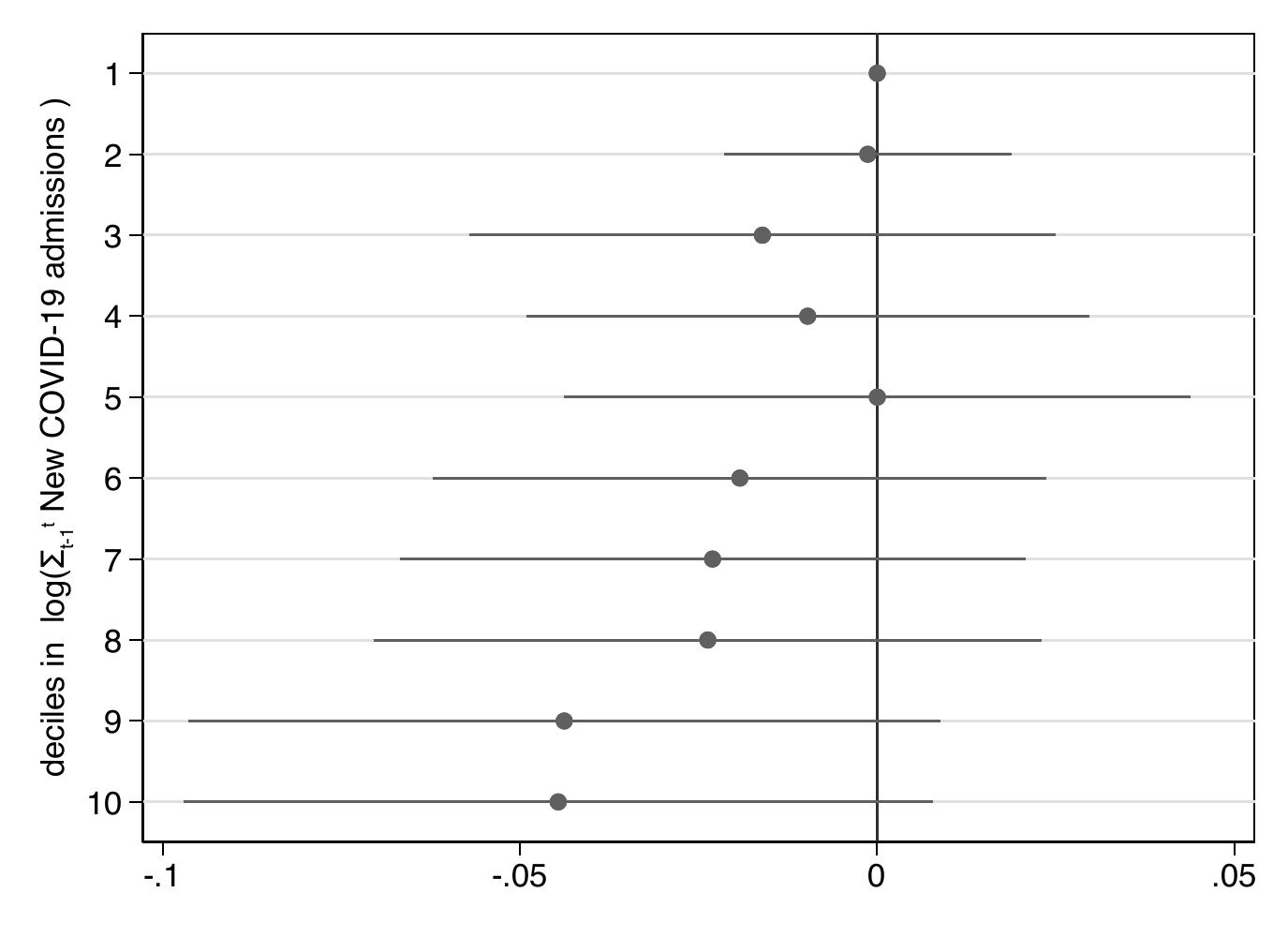} &   \includegraphics[width=0.5\columnwidth]{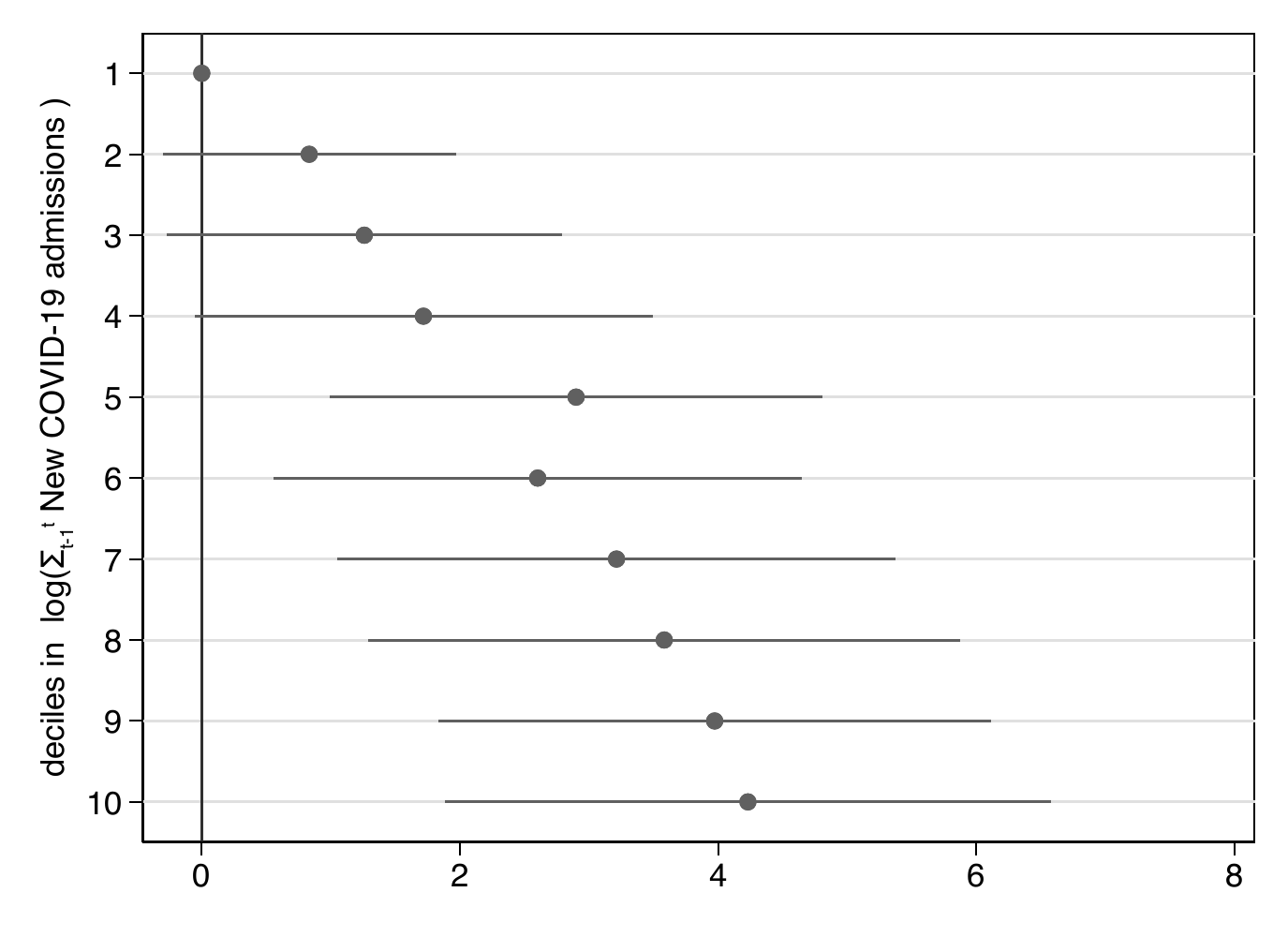}  \\
     \end{array}$
\end{center}
\scriptsize{\textbf{Notes:} Figure presents heterogenous treatment effects capturing the impact of COVID-19 pressures  measured by deciles of the provider-specific empirical distribution on cancer treatment measured as number of patients referred for treatment (Panel A) and the share of said cancer treatment receiving cancer treatment after 62 days of urgent referral (Panel B). All regressions control for provider by cancer by care setting (admitted or non-admitted) fixed effects and cancer by care setting by time fixed effects. 90\% confidence intervals obtained from clustering standard errors at the provider level are indicated.}

\end{figure}
\end{landscape}
%%%%%%%%%%%%%%%%%%%%%%%%%%%%%%

%%%%%%%%%%%%%%%%%%%%%%%%%%%%%%
%%%%%%%%%%%%%%%%%%%%%%%%%%%%%%

%%%%%%%%%%%%%%%%%%%%%%%%%%%%%%
%%%%%%%%%%%%%%%%%%%%%%%%%%%%%%

%%%%%%%%%%%%%%%%%%%%%%%%%%%%%%
 \begin{table}[h!]
\centering{
\scalebox{0.725}{
  \begin{threeparttable}
  \caption{Impact of COVID-19 health care system pressures on A\&E activity and performance  \emph{without controlling for community COVID-19 transmission}  \label{table:ae_statistics-pressures-nocomm}}
\begin{tabular}{lccccc}
  \hline

%   & \multicolumn{6}{c}{\emph{DV:  }}  \\
% \cmidrule(lr){2-7}     
\addlinespace
 &\multicolumn{1}{c}{(1)}   &\multicolumn{1}{c}{(2)}   &\multicolumn{1}{c}{(3)}   &\multicolumn{1}{c}{(4)}   &\multicolumn{1}{c}{(5)}  \\

  & \multicolumn{2}{c}{\emph{log(A\&E visits$_t$) }} & \multicolumn{3}{c}{\emph{Patients waiting to be treated or admitted for}}     \\
   \cmidrule(lr){2-3}        \cmidrule(lr){4-6} 
      & \multicolumn{1}{c}{All }  &\multicolumn{1}{c}{ Resulting admissions} &\multicolumn{1}{c}{$>$ 4h } &\multicolumn{1}{c}{  4 - 12 h } &\multicolumn{1}{c}{$ >$ 12 h } \\ 
   \hline
  \addlinespace
\multicolumn{5}{l}{\emph{Panel A}:    } \\
\input{fragment-ae_statistics-new_admissions.tex} \\
\addlinespace
\addlinespace
\multicolumn{5}{l}{\emph{Panel B}:  }\\
\input{fragment-ae_statistics-hospital_cases.tex} \\
\addlinespace
\addlinespace
\multicolumn{5}{l}{\emph{Panel C}:   } \\
\input{fragment-ae_statistics-occupied_mv_beds.tex} \\
\addlinespace
\addlinespace

Provider   FE &X & X & X &X & X\\
Time FE &X & X & X &X & X\\ 

   \hline
   \end{tabular}

    \begin{tablenotes} {\footnotesize
    \item Notes:  Regressions present results at the NHS provider level documenting the relationship between different measures of COVID-19 pressures at the provider level across panels on the performance of the A\&E departments and waiting times in a given month. This does not distinguish between COVID-19 and non-COVID-19 related A\&E visits. Columns (1) and (2) measure the quantity of A\&E visits and resulting admissions respectively, while columns (3)-(5) capture the impact that pressures COVID pressures have on waiting times for A\&E visits. Standard errors are clustered at the provider level with stars indicating *** p$<$ 0.01, ** p$<$ 0.05, * p$<$ 0.1.} \end{tablenotes}
      \end{threeparttable}
      
 }
  }    
  
\end{table}
 %%%%%%%%%%%%%%%%%%%%%%%%%%%%%%

%%%%%%%%%%%%%%%%%%%%%%%%%%%%%%
\begin{landscape}
 \begin{table}[h!]
\centering{
\scalebox{0.7}{
  \begin{threeparttable}
  \caption{Impact of COVID-19 health care system pressures on specialist referral  \emph{without controlling for community COVID-19 transmission} \label{table:rtt-pressures-nocomm}}
\begin{tabular}{lcccccccccc}
  \hline

%   & \multicolumn{6}{c}{\emph{DV:  }}  \\
% \cmidrule(lr){2-7}     
\addlinespace
 &\multicolumn{1}{c}{(1)}   &\multicolumn{1}{c}{(2)}   &\multicolumn{1}{c}{(3)}   &\multicolumn{1}{c}{(4)}   &\multicolumn{1}{c}{(5)} &\multicolumn{1}{c}{(6)} &\multicolumn{1}{c}{(7)} &\multicolumn{1}{c}{(8)} &\multicolumn{1}{c}{(9)} \\

  & \multicolumn{3}{c}{\emph{log(Referrals$_t$) }} & \multicolumn{3}{c}{\emph{log(Waiting list$_t$) }}   & \multicolumn{3}{c}{\emph{Share waiting$_t$ }}     \\
   \cmidrule(lr){2-4}        \cmidrule(lr){5-7}  \cmidrule(lr){8-10}

   & \multicolumn{1}{c}{New}  &\multicolumn{2}{c}{ Completed } &\multicolumn{1}{c}{ Length } &\multicolumn{1}{c}{ Aggregate wait}  &\multicolumn{1}{c}{ Avg.\ wait} &\multicolumn{1}{c}{ $>$ 4 weeks } &\multicolumn{1}{c}{ $> 8$ weeks}  &\multicolumn{1}{c}{ $> 12$ weeks} \\
      & \multicolumn{1}{c}{ }  &\multicolumn{1}{c}{ Admitted } &\multicolumn{1}{c}{Non-admitted} &\multicolumn{1}{c}{   } &\multicolumn{1}{c}{  }  &\multicolumn{1}{c}{  } &\multicolumn{1}{c}{  } &\multicolumn{1}{c}{}  &\multicolumn{1}{c}{ } \\ 
   \hline
  \addlinespace
\multicolumn{5}{l}{\emph{Panel A}:    } \\
\input{fragment-rtt-new_admissions.tex}\\ 
\addlinespace
\addlinespace
\multicolumn{5}{l}{\emph{Panel B}:  }\\
\input{fragment-rtt-hospital_cases.tex} \\
\addlinespace
\addlinespace
\multicolumn{5}{l}{\emph{Panel C}:   } \\
\input{fragment-rtt-occupied_mv_beds.tex} \\
\addlinespace
\addlinespace

Provider x Treatment function FE &X & X & X &X & X & X&X & X & X\\
Treatment function x Time FE &  X& X &  X &  X & X & X &X & X & X\\ 

   \hline
   \end{tabular}

    \begin{tablenotes} {\footnotesize
    \item Notes:  Regressions present results at the NHS provider level documenting the relationship between different measures of COVID-19 pressures at the provider level across panels on the accessibility and quality of referrals to specialist treatment. This does not distinguish between COVID-19 and non-COVID-19 related referrals. Columns (1) - (3) focus on measures of output capturing new referrals to specialists and completion of referral pathways. Columns (4) - (6) study broad characteristics of the  waiting list for non-completed  specialist referrals. Columns (7)-(9) provide breakdown of stock of average wait on waiting list for non-completed referrals. Standard errors are clustered at the provider level with stars indicating *** p$<$ 0.01, ** p$<$ 0.05, * p$<$ 0.1.} \end{tablenotes}
      \end{threeparttable}
      
 }
  }    
  
\end{table}
\end{landscape}
 %%%%%%%%%%%%%%%%%%%%%%%%%%%%%%

%%%%%%%%%%%%%%%%%%%%%%%%%%%%%%
 \begin{table}[h!]
\centering{
\scalebox{0.725}{
  \begin{threeparttable}
  \caption{Impact of COVID-19 health care system pressures on diagnostic  activity and performance  \emph{without controlling for community COVID-19 transmission} \label{table:diagnostic-pressures-nocomm}}
\begin{tabular}{lccccc}
  \hline

%   & \multicolumn{6}{c}{\emph{DV:  }}  \\
% \cmidrule(lr){2-7}     
\addlinespace
 &\multicolumn{1}{c}{(1)}   &\multicolumn{1}{c}{(2)}   &\multicolumn{1}{c}{(3)}   &\multicolumn{1}{c}{(4)}   &\multicolumn{1}{c}{(5)}  \\

  & \multicolumn{1}{c}{\emph{log(Diagnostic activity$_t$) }} & \multicolumn{2}{c}{\emph{log(Average wait$_t$) }}   & \multicolumn{2}{c}{\emph{Share waiting$_t$  > 8 weeks}}     \\
   \cmidrule(lr){2-2}        \cmidrule(lr){3-4}   \cmidrule(lr){5-6}

      & \multicolumn{1}{c}{ }  &\multicolumn{1}{c}{ All } &\multicolumn{1}{c}{CT} &\multicolumn{1}{c}{  All } &\multicolumn{1}{c}{ CT } \\ 
   \hline
  \addlinespace
\multicolumn{5}{l}{\emph{Panel A}:    } \\
\input{fragment-diagnostic-new_admissions.tex} \\
\addlinespace
\addlinespace
\multicolumn{5}{l}{\emph{Panel B}:  }\\
\input{fragment-diagnostic-hospital_cases.tex} \\
\addlinespace
\addlinespace
\multicolumn{5}{l}{\emph{Panel C}:   } \\
\input{fragment-diagnostic-occupied_mv_beds.tex} \\
\addlinespace
\addlinespace

Provider x Diagnostic  FE &X & X & X &X & X\\
Diagnostic x Time FE &X & X & X &X & X\\ 

   \hline
   \end{tabular}

    \begin{tablenotes} {\footnotesize
    \item Notes:  Regressions present results at the NHS provider level documenting the relationship between different measures of COVID-19 pressures at the provider level across panels on the diagnostic performance and diagnostic waiting times. This does not distinguish between COVID-19 and non-COVID-19 related diagnostic activity. Columns (1) measures total diagnostic activity across 15 diagnostic functions performed. Columns (2) - (3) study average waiting times for all diagnostic activity (column 2) and CT diagnostic (column 3). Columns (4)-(5) study as dependent variable the share of individuals waiting more than 6 weeks across all diagnostic activity (column 4) and specifically for CT diagnostic (column 5). Standard errors are clustered at the provider level with stars indicating *** p$<$ 0.01, ** p$<$ 0.05, * p$<$ 0.1.} \end{tablenotes}
      \end{threeparttable}
      
 }
  }    
  
\end{table}
 %%%%%%%%%%%%%%%%%%%%%%%%%%%%%%

%%%%%%%%%%%%%%%%%%%%%%%%%%%%%%
\begin{landscape}
 \begin{table}[h!]
\centering{
\scalebox{0.75}{
  \begin{threeparttable}
  \caption{Impact of COVID-19 health care system pressures on cancer treatment pathways and performance  \emph{without controlling for community COVID-19 transmission} \label{table:cancer-pressures-nocomm}}
\begin{tabular}{lcccccc}
  \hline

%   & \multicolumn{6}{c}{\emph{DV:  }}  \\
% \cmidrule(lr){2-7}     
\addlinespace
 &\multicolumn{1}{c}{(1)}   &\multicolumn{1}{c}{(2)}   &\multicolumn{1}{c}{(3)}   &\multicolumn{1}{c}{(4)}   &\multicolumn{1}{c}{(5)} &\multicolumn{1}{c}{(6)}  \\

  & \multicolumn{3}{c}{\emph{log(cases$_t$) with }} & \multicolumn{3}{c}{\emph{\% with time taken to }}      \\
     \cmidrule(lr){2-4}        \cmidrule(lr){5-7} 

   &\multicolumn{1}{c}{ }   &\multicolumn{2}{c}{treatment}     &\multicolumn{1}{c}{referral}   & \multicolumn{2}{c}{treatment}   \\ 
   
   &\multicolumn{1}{c}{referrals}   &\multicolumn{1}{c}{decision}   &\multicolumn{1}{c}{start}   &\multicolumn{1}{c}{seen $>$ 14 days}   &\multicolumn{1}{c}{decision $>$ 31 days} &\multicolumn{1}{c}{start $>$ 62 days}  \\

  %    & \multicolumn{1}{c}{ }  &\multicolumn{1}{c}{ All } &\multicolumn{1}{c}{CT} &\multicolumn{1}{c}{  All } &\multicolumn{1}{c}{ CT } \\ 
   \hline
  \addlinespace
\multicolumn{5}{l}{\emph{Panel A}:    } \\
\input{fragment-cancer-new_admissions.tex} \\
\addlinespace
\addlinespace
\multicolumn{5}{l}{\emph{Panel B}:  }\\
\input{fragment-cancer-hospital_cases.tex} \\
\addlinespace
\addlinespace
\multicolumn{5}{l}{\emph{Panel C}:   } \\
\input{fragment-cancer-occupied_mv_beds.tex} \\
\addlinespace
\addlinespace

Provider x  Care setting x Cancer  FE &X & X & X &X & X  & X\\
Cancer x Care setting x Time FE &X & X & X &X & X  & X\\ 

   \hline
   \end{tabular}

    \begin{tablenotes} {\footnotesize
    \item Notes:  Regressions present results at the NHS provider level documenting the relationship between different measures of COVID-19 pressures at the provider level across panels on the diagnostic performance and diagnostic waiting times. This does not distinguish between COVID-19 and non-COVID-19 related diagnostic activity. Columns (1) measures total diagnostic activity across 15 diagnostic functions performed. Columns (2) - (3) study average waiting times for all diagnostic activity (column 2) and CT diagnostic (column 3). Columns (4)-(5) study as dependent variable the share of individuals waiting more than 6 weeks across all diagnostic activity (column 4) and specifically for CT diagnostic (column 5). Standard errors are clustered at the provider level with stars indicating *** p$<$ 0.01, ** p$<$ 0.05, * p$<$ 0.1.} \end{tablenotes}
      \end{threeparttable}
      
 }
  }    
  
\end{table}
\end{landscape}

 %%%%%%%%%%%%%%%%%%%%%%%%%%%%%%

%%%%%%%%%%%%%%%%%%%%%%%%%%%%%%
 \begin{table}[h!]
\centering{
\scalebox{0.75}{
  \begin{threeparttable}
  \caption{Impact of COVID-19 health care system pressures on non-COVID-19 excess deaths  \emph{without controlling for community COVID-19 transmission} \label{table:excess-deaths-main-nocomm}}
\begin{tabular}{lcccccc}
  \hline

%   & \multicolumn{6}{c}{\emph{DV:  }}  \\
% \cmidrule(lr){2-7}     
\addlinespace
 &\multicolumn{1}{c}{(1)}   &\multicolumn{1}{c}{(2)}   &\multicolumn{1}{c}{(3)}   &\multicolumn{1}{c}{(4)}   &\multicolumn{1}{c}{(5)}  \\

%  & \multicolumn{3}{c}{\emph{log(cases$_t$) with }} & \multicolumn{3}{c}{\emph{\% with time taken to }}      \\
%     \cmidrule(lr){2-4}        \cmidrule(lr){5-7} 

%   &\multicolumn{1}{c}{ }   &\multicolumn{2}{c}{treatment}     &\multicolumn{1}{c}{referral}   & \multicolumn{2}{c}{treatment}   \\ 
   
 %  &\multicolumn{1}{c}{referrals}   &\multicolumn{1}{c}{decision}   &\multicolumn{1}{c}{start}   &\multicolumn{1}{c}{seen $>$ 14 days}   &\multicolumn{1}{c}{decision $>$ 31 days} &\multicolumn{1}{c}{start $>$ 62 days}  \\
   \hline
  \addlinespace
\multicolumn{5}{l}{\emph{Panel A}:    } \\
\input{fragment-shmi-dchexcess-new_admissions.tex} \\
\addlinespace
\addlinespace
\multicolumn{5}{l}{\emph{Panel B}:  }\\
\input{fragment-shmi-dchexcess-hospital_cases.tex} \\
\addlinespace
\addlinespace
\multicolumn{5}{l}{\emph{Panel C}:   } \\
\input{fragment-shmi-dchexcess-occupied_mv_beds.tex} \\
\addlinespace
\addlinespace
\addlinespace

Provider  FE &X & X & X &X & X \\
Time FE &X & X & X &X & X\\ 
$\Delta \text{Spells}_{p,t}$   & & X & X &X & X\\ 
$ \text{Excess deaths}_{p,t-13} $   & &  & X &  &  \\ 
$ \text{Obs}_{p,t-13}$ and  $\text{Exp}_{p,t-13}$    & &  &  &X & X\\ 
Provider specific linear time trend  & &  &  & & X\\ 

   \hline
   \end{tabular}

    \begin{tablenotes} {\footnotesize
    \item Notes:  Regressions present results at the NHS provider level documenting the relationship between different measures of COVID-19 pressures at the provider level and overall excess deaths reported in a given month. The excess death measure captures month-on-month changes in excess death constructed from the twelve month cumulative windows. Across columns subsequently more control variables are added that aim to capture the potential confounding effect that base effects could have on the estimates.  Standard errors are clustered at the provider level with stars indicating *** p$<$ 0.01, ** p$<$ 0.05, * p$<$ 0.1.} \end{tablenotes}
      \end{threeparttable}
      
 }
  }    
  
\end{table}
 %%%%%%%%%%%%%%%%%%%%%%%%%%%%%%

%%%%%%%%%%%%%%%%%%%%%%%%%%%%%%
 \begin{table}[h!]
\centering{
\scalebox{0.725}{
  \begin{threeparttable}
  \caption{Impact of COVID-19 pressures on staff absence rates  \emph{without controlling for community COVID-19 transmission} \label{table:staff absences-nocomm}}
\begin{tabular}{lccccccccc}
  \hline

%   & \multicolumn{6}{c}{\emph{DV:  }}  \\
% \cmidrule(lr){2-7}     
\addlinespace
 &\multicolumn{1}{c}{(1)}   &\multicolumn{1}{c}{(2)}   &\multicolumn{1}{c}{(3)}   &\multicolumn{1}{c}{(4)}   &\multicolumn{1}{c}{(5)}  &\multicolumn{1}{c}{(6)} \\ %&\multicolumn{1}{c}{(7)}   &\multicolumn{1}{c}{(8)}  &\multicolumn{1}{c}{(9)}  \\

%  & \multicolumn{6}{c}{\emph{NHS staff vaccination uptake measure}} \\
%     \cmidrule(lr){2-7} 
  & \multicolumn{3}{c}{\emph{All staff groups}} & \multicolumn{3}{c}{\emph{By staff group}}  \\% & \multicolumn{3}{c}{\emph{Above 75\% percentile}}      \\
    \cmidrule(lr){2-4}        \cmidrule(lr){5-7}  % \cmidrule(lr){8-10} 

%   &\multicolumn{1}{c}{ }   &\multicolumn{2}{c}{treatment}     &\multicolumn{1}{c}{referral}   & \multicolumn{2}{c}{treatment}   \\ 
   
   &\multicolumn{1}{c}{ }   &\multicolumn{1}{c}{ }   &\multicolumn{1}{c}{}   &\multicolumn{1}{c}{Nurses}   &\multicolumn{1}{c}{Doctors} &\multicolumn{1}{c}{Managers}  \\
 
  \addlinespace
Panel A: \\
\input{fragment-absence-new_admissions.tex} \\
\addlinespace
\addlinespace
\addlinespace

Panel B:  \\
\input{fragment-absence-hospital_cases.tex} \\
\addlinespace
\addlinespace

Panel C: \\
\input{fragment-absence-occupied_mv_beds.tex} \\
\addlinespace
\addlinespace

Provider  FE &X & X & X &X & X  & X  \\
Time FE &X & X & X &X & X  & X\\ 
 \% Population vaccinated      & & X & X &X & X  & X \\
Provider specific linear trends  &  &   & X &X & X  & X\\
   \hline
   \end{tabular}

    \begin{tablenotes} {\footnotesize
    \item Notes:  Regressions capture the changing effect of NHS trust hospital admissions on provider specific excess mortality documenting how the vaccination roll out across the NHS is moderating this relationship. The excess death measure captures month-on-month changes in excess death constructed from the twelve month cumulative windows.  Standard errors are clustered at the provider level with stars indicating *** p$<$ 0.01, ** p$<$ 0.05, * p$<$ 0.1.} \end{tablenotes}
      \end{threeparttable}
      
 }
  }    
 
\end{table}
 %%%%%%%%%%%%%%%%%%%%%%%%%%%%%%

%%%%%%%%%%%%%%%%%%%%%%%%%%%%%%
 \begin{table}[h!]
\centering{
\scalebox{0.7}{
  \begin{threeparttable}
  \caption{Impact of COVID-19 pressures on staff absence rates: the effect of NHS vaccination uptake   \emph{without controlling for community COVID-19 transmission}  \label{table:staff absences-vaccination-nocomm}}
\begin{tabular}{lccccccccc}
  \hline

%   & \multicolumn{6}{c}{\emph{DV:  }}  \\
% \cmidrule(lr){2-7}     
\addlinespace
 &\multicolumn{1}{c}{(1)}   &\multicolumn{1}{c}{(2)}   &\multicolumn{1}{c}{(3)}   &\multicolumn{1}{c}{(4)}   &\multicolumn{1}{c}{(5)}  &\multicolumn{1}{c}{(6)} \\ %&\multicolumn{1}{c}{(7)}   &\multicolumn{1}{c}{(8)}  &\multicolumn{1}{c}{(9)}  \\

%  & \multicolumn{6}{c}{\emph{NHS staff vaccination uptake measure}} \\
%     \cmidrule(lr){2-7} 
  \emph{DV: staff absence rates}    & \multicolumn{6}{c}{\emph{NHS staff vaccination uptake measure}} \\
     \cmidrule(lr){2-7} 
  & \multicolumn{3}{c}{\emph{\% of NHS staff with with 2 doses}} & \multicolumn{3}{c}{\emph{\% of NHS staff with at least 1 dose}}  \\% & \multicolumn{3}{c}{\emph{Above 75\% percentile}}      \\
    \cmidrule(lr){2-4}        \cmidrule(lr){5-7}  % \cmidrule(lr){8-10}  
  \addlinespace
Panel A: \\
\input{fragment-absence-vaccination-new_admissions.tex} \\
\addlinespace
\addlinespace
\addlinespace

Panel B:  \\
\input{fragment-absence-vaccination-hospital_cases.tex} \\
\addlinespace
\addlinespace

Panel C: \\
\input{fragment-absence-vaccination-occupied_mv_beds.tex} \\
\addlinespace
\addlinespace

Provider  FE &X & X & X &X & X  & X  \\
Time FE &X & X & X &X & X  & X\\ 
 \% Population vaccinated      &   & X & X &  & X  & X \\
Provider specific linear trends  &  &   & X & &   & X\\
   \hline
   \end{tabular}

    \begin{tablenotes} {\footnotesize
    \item Notes:  Regressions capture the changing effect of different measures of COVID-19 pressures on staff absence rates depending on the vaccination uptake of NHS staff.   Standard errors are clustered at the provider level with stars indicating *** p$<$ 0.01, ** p$<$ 0.05, * p$<$ 0.1.} \end{tablenotes}
      \end{threeparttable}
      
 }
  }    
 
\end{table}
 %%%%%%%%%%%%%%%%%%%%%%%%%%%%%%

%%%%%%%%%%%%%%%%%%%%%%%%%%%%%%
 \begin{table}[h!]
\centering{
\scalebox{0.725}{
  \begin{threeparttable}
  \caption{Impact of COVID-19 pressures on non-COVID-19 excess mortality: the moderating effect of NHS vaccination uptake \emph{without controlling for community COVID-19 transmission}  \label{table:excess-deaths-nhsvaxx-nocomm}}
\begin{tabular}{lccccccccc}
  \hline

 & \multicolumn{6}{c}{}  \\
% \cmidrule(lr){2-7}     
\addlinespace
 &\multicolumn{1}{c}{(1)}   &\multicolumn{1}{c}{(2)}   &\multicolumn{1}{c}{(3)}   &\multicolumn{1}{c}{(4)}   &\multicolumn{1}{c}{(5)}  &\multicolumn{1}{c}{(6)} \\ %&\multicolumn{1}{c}{(7)}   &\multicolumn{1}{c}{(8)}  &\multicolumn{1}{c}{(9)}  \\

  \emph{DV: Non-COVID-19 excess death}    & \multicolumn{6}{c}{\emph{NHS staff vaccination uptake measure}} \\
     \cmidrule(lr){2-7} 
  & \multicolumn{3}{c}{\emph{\% of NHS staff with with 2 doses}} & \multicolumn{3}{c}{\emph{\% of NHS staff with at least 1 dose}}  \\% & \multicolumn{3}{c}{\emph{Above 75\% percentile}}      \\
    \cmidrule(lr){2-4}        \cmidrule(lr){5-7}  % \cmidrule(lr){8-10} 

%   &\multicolumn{1}{c}{ }   &\multicolumn{2}{c}{treatment}     &\multicolumn{1}{c}{referral}   & \multicolumn{2}{c}{treatment}   \\ 
   
 %  &\multicolumn{1}{c}{referrals}   &\multicolumn{1}{c}{decision}   &\multicolumn{1}{c}{start}   &\multicolumn{1}{c}{seen $>$ 14 days}   &\multicolumn{1}{c}{decision $>$ 31 days} &\multicolumn{1}{c}{start $>$ 62 days}  \\
 
  \addlinespace
\input{interaction-nhsvax-shmi-dchexcess.tex} \\
\addlinespace
\addlinespace
\addlinespace

Provider  FE &X & X & X &X & X  & X  \\
Time FE &X & X & X &X & X  & X\\ 
$ \text{Excess deaths}_{p,t-13} $  &X & X & X &X & X  & X\\
$\Delta \text{Spells}_{p,t}$     &X & X & X &X & X  & X \\
\% Population vaccination      &X & X & X &X & X  & X \\

   \hline
   \end{tabular}

    \begin{tablenotes} {\footnotesize
    \item Notes:  Regressions capture the changing effect of NHS trust hospital admissions on provider specific excess mortality documenting how the vaccination roll out across the NHS is moderating this relationship. The excess death measure captures month-on-month changes in excess death constructed from the twelve month cumulative windows.  Standard errors are clustered at the provider level with stars indicating *** p$<$ 0.01, ** p$<$ 0.05, * p$<$ 0.1.} \end{tablenotes}
      \end{threeparttable}
      
 }
  }    
 
\end{table}
 %%%%%%%%%%%%%%%%%%%%%%%%%%%%%%

    %%%%%%%%%%%%%%%%%%%%%%%%%%%%%%
 \begin{table}[h!]
\centering{
\scalebox{0.825}{
  \begin{threeparttable}
  \caption{Impact of COVID-19 health care system pressures and non-COVID-19 excess mortality \label{table:did-excess-mortality}}
\begin{tabular}{lcccccc}
  \hline

%   & \multicolumn{6}{c}{\emph{DV:  }}  \\
% \cmidrule(lr){2-7}     
\addlinespace
  & \multicolumn{5}{c}{\emph{COVID-19 pressures measured in the last ... months }}     \\
   & \multicolumn{1}{c}{0}  &\multicolumn{1}{c}{ 1} &\multicolumn{1}{c}{ 2} &\multicolumn{1}{c}{ 3} &\multicolumn{1}{c}{ 6}  &\multicolumn{1}{c}{ 9} \\
 \cmidrule(lr){2-4}        \cmidrule(lr){5-7}

 &\multicolumn{1}{c}{(1)}   &\multicolumn{1}{c}{(2)}   &\multicolumn{1}{c}{(3)}   &\multicolumn{1}{c}{(4)}   &\multicolumn{1}{c}{(5)} &\multicolumn{1}{c}{(5)} \\

   \hline
  \addlinespace
\multicolumn{5}{l}{\emph{Panel A}:  }\\
\input{fragment-shmi-dlogexcess-new_admissions.tex} \\
\addlinespace
\addlinespace
\multicolumn{5}{l}{\emph{Panel B}:   } \\
\input{fragment-shmi-dlogexcess-hospital_cases.tex} \\
\addlinespace
\addlinespace
\multicolumn{5}{l}{\emph{Panel C}:  } \\
\input{fragment-shmi-dlogexcess-occupied_mv_beds.tex} \\
\addlinespace
\addlinespace

%\multicolumn{5}{l}{\textbf{Non-linear time trends in COVID-19 intensity weeks 36-38}  } \\
Provider FE &X & X & X &X & X & X\\
Time FE &  X& X &  X &  X & X & X \\ 
Spells &  X& X &  X &  X & X & X \\ 

%"date"  "rgndate " "rgndate i.week#c.preshockcases"  "rgndate i.week#i.nqpreshock" "rgndate i.week#i.nqpreshock i.week#i.nqpopdens" {

   \hline
   \end{tabular}

    \begin{tablenotes} {\footnotesize
    \item Notes:  Regressions present results at the NHS provider level documenting a positive relationship between COVID-19 pressures measured in different ways across Panels A - D and diagnostic-specific excess mortality for non COVID-19 patients.  The dependent variable measures the log difference in observed versus expected number of deaths. The expected number of deaths is constructed by \cite{SHMI2021} based on case-level data. The right hand-side measures across columns are in logs measuring the COVID-19 pressures cumulative over the number of months indicated in the column head.  That is, column (3) studies how COVID-19 pressures measured at the provider level affects excess deaths over the last 12 months. Standard errors are clustered at the provider level with stars indicating *** p$<$ 0.01, ** p$<$ 0.05, * p$<$ 0.1.} \end{tablenotes}
      
      \end{threeparttable}
      
 }
  }    
  
\end{table}
 %%%%%%%%%%%%%%%%%%%%%%%%%%%%%%

    %%%%%%%%%%%%%%%%%%%%%%%%%%%%%%
 \begin{table}[h!]
\centering{
\scalebox{0.825}{
  \begin{threeparttable}
  \caption{Impact of COVID-19 health care system pressures and non-COVID-19 mortality -- controlling for expected mortality \label{table:did-logobserved-mortality}}
\begin{tabular}{lcccccc}
  \hline

%   & \multicolumn{6}{c}{\emph{DV:  }}  \\
% \cmidrule(lr){2-7}     
\addlinespace
  & \multicolumn{5}{c}{\emph{COVID-19 pressures measured in the last ... months }}     \\
   & \multicolumn{1}{c}{0}  &\multicolumn{1}{c}{ 1} &\multicolumn{1}{c}{ 2} &\multicolumn{1}{c}{ 3} &\multicolumn{1}{c}{ 6}  &\multicolumn{1}{c}{ 9} \\
 \cmidrule(lr){2-4}        \cmidrule(lr){5-7}

 &\multicolumn{1}{c}{(1)}   &\multicolumn{1}{c}{(2)}   &\multicolumn{1}{c}{(3)}   &\multicolumn{1}{c}{(4)}   &\multicolumn{1}{c}{(5)} &\multicolumn{1}{c}{(5)} \\

   \hline
  \addlinespace
\multicolumn{5}{l}{\emph{Panel A}:    } \\
\input{fragment-shmi-logobserved-new_admissions.tex} \\
\addlinespace
\addlinespace
\multicolumn{5}{l}{\emph{Panel B}:  }\\
\input{fragment-shmi-logobserved-hospital_cases.tex} \\
\addlinespace
\addlinespace
\multicolumn{5}{l}{\emph{Panel C}:  } \\
\input{fragment-shmi-logobserved-occupied_mv_beds.tex} \\
\addlinespace
\addlinespace

%\multicolumn{5}{l}{\textbf{Non-linear time trends in COVID-19 intensity weeks 36-38}  } \\
Provider FE &X & X & X &X & X & X\\
Time FE &  X& X &  X &  X & X & X \\ 
Spells &  X& X &  X &  X & X & X \\ 

%"date"  "rgndate " "rgndate i.week#c.preshockcases"  "rgndate i.week#i.nqpreshock" "rgndate i.week#i.nqpreshock i.week#i.nqpopdens" {

   \hline
   \end{tabular}

    \begin{tablenotes} {\footnotesize
    \item Notes:  Regressions present results at the NHS provider level documenting a positive relationship between COVID-19 pressures measured in different ways across Panels A - D and mortality for non COVID-19 patients.  The dependent variable measures the log in observed deaths.  The expected number of deaths is added as a control variable. The expected number of deaths is constructed by \cite{SHMI2021} based on case-level data. The measures across columns are in logs measuring the COVID-19 pressures cumulative over the number of months indicated in the column head. That is, column (3) studies how COVID-19 pressures measured at the provider level affects excess deaths over the last 12 months. Standard errors are clustered at the provider level with stars indicating *** p$<$ 0.01, ** p$<$ 0.05, * p$<$ 0.1.} \end{tablenotes}
      
      \end{threeparttable}
      
 }
  }    
\end{table}
 %%%%%%%%%%%%%%%%%%%%%%%%%%%%%%

    %%%%%%%%%%%%%%%%%%%%%%%%%%%%%%
 \begin{table}[h!]
\centering{
\scalebox{0.825}{
  \begin{threeparttable}
  \caption{Impact of COVID-19 health care system pressures and non-COVID-19 death rates  \label{table:did-deathrate-mortality}}
\begin{tabular}{lcccccc}
  \hline

%   & \multicolumn{6}{c}{\emph{DV:  }}  \\
% \cmidrule(lr){2-7}     
\addlinespace
  & \multicolumn{5}{c}{\emph{COVID-19 pressures measured in the last ... months }}     \\
   & \multicolumn{1}{c}{0}  &\multicolumn{1}{c}{ 1} &\multicolumn{1}{c}{ 2} &\multicolumn{1}{c}{ 3} &\multicolumn{1}{c}{ 6}  &\multicolumn{1}{c}{ 9} \\
 \cmidrule(lr){2-4}        \cmidrule(lr){5-7}

 &\multicolumn{1}{c}{(1)}   &\multicolumn{1}{c}{(2)}   &\multicolumn{1}{c}{(3)}   &\multicolumn{1}{c}{(4)}   &\multicolumn{1}{c}{(5)} &\multicolumn{1}{c}{(5)} \\

   \hline
  \addlinespace
\multicolumn{5}{l}{\emph{Panel A}:    } \\
\input{fragment-shmi-deathrate-new_admissions.tex} \\
\addlinespace
\addlinespace
\multicolumn{5}{l}{\emph{Panel B}:  }\\
\input{fragment-shmi-deathrate-hospital_cases.tex} \\
\addlinespace
\addlinespace
\multicolumn{5}{l}{\emph{Panel C}:  } \\
\input{fragment-shmi-deathrate-occupied_mv_beds.tex} \\
\addlinespace
\addlinespace

%\multicolumn{5}{l}{\textbf{Non-linear time trends in COVID-19 intensity weeks 36-38}  } \\
Provider FE &X & X & X &X & X & X\\
Time FE &  X& X &  X &  X & X & X \\ 

%"date"  "rgndate " "rgndate i.week#c.preshockcases"  "rgndate i.week#i.nqpreshock" "rgndate i.week#i.nqpreshock i.week#i.nqpopdens" {

   \hline
   \end{tabular}

    \begin{tablenotes} {\footnotesize
    \item Notes:  Regressions present results at the NHS provider level documenting a positive relationship between COVID-19 pressures measured in different ways across Panels A - D and mortality for non COVID-19 patients.  The dependent variable measures the share of hospital admissions that result in a death.  The expected share of deaths per admission is added as a control variable. The expected number of deaths is constructed by \cite{SHMI2021} based on case-level data. The measures across columns are in logs measuring the COVID-19 pressures cumulative over the number of months indicated in the column head. That is, column (3) studies how COVID-19 pressures measured at the provider level affects excess deaths over the last 12 months. Standard errors are clustered at the provider level with stars indicating *** p$<$ 0.01, ** p$<$ 0.05, * p$<$ 0.1.} \end{tablenotes}
      
      \end{threeparttable}
      
 }
  }    
\end{table}
 %%%%%%%%%%%%%%%%%%%%%%%%%%%%%%

\end{document}

%% file: quantity_outbreak.tex
A\&E attendance&1325115&910772&1193893&-414343&-131222 \\ 
Diagnostic waiting list &941397&894607&1194390&-46789&252992 \\ 
Referral to treatment incomplete&4591780&4458938&5440273&-132842&848492 \\ 
Referral to treatment complete (admitted)&249821&87764&178768&-162057&-71053 \\ 
Referral to treatment complete (non-admitted)&958777&604443&821883&-354334&-136894 \\ 
Referral to treatment complete&1208599&692207&1000651&-516391&-207947 \\ 
Cancer referral to specialist consultation&189983&129211&204615&-60772&14632 \\ 
Cancer from referral to treatment &13326&10974&13228&-2352&-97 \\ 
Cancer from decision to treat to treatment   &24941&20382&24120&-4558&-821 \\ 

%% file: quality_outbreak.tex
A\&E attendance&.798&.876&.759&.079&-.038 \\ 
Diagnostic waiting list&.967&.556&.712&-.411&-.255 \\ 
Referral to treatment incomplete&.829&.619&.613&-.209&-.216 \\ 
Referral to treatment complete (admitted)&.7&.748&.634&.048&-.066 \\ 
Referral to treatment complete (non-admitted)&.861&.812&.776&-.049&-.085 \\ 
Referral to treatment complete&.828&.804&.751&-.024&-.077 \\ 
Cancer referral to specialist consultation&.914&.919&.857&.006&-.056 \\ 
Cancer from referral to treatment&.781&.751&.725&-.03&-.057 \\ 
Cancer from decision to treat to treatment&.964&.954&.943&-.01&-.021 \\ 

%% file: fragment-ae_statisticslognew_cases-new_admissions.tex
log(New COVID-19 admissions$_t$)&       0.019   &       0.013   &       2.661***&       1.133***&       0.096** \\
                    &     (0.014)   &     (0.010)   &     (0.546)   &     (0.222)   &     (0.040)   \\
\addlinespace Mean of DV&     9024.57   &     2911.65   &       21.63   &        5.54   &        0.23   \\
Observations        &        2773   &        2652   &        2652   &        2652   &        2652   \\
Clusters            &         127   &         123   &         123   &         123   &         123   \\

%% file: fragment-ae_statisticslognew_cases-hospital_cases.tex
log(COVID-19 cases in hospital$_t$)&       0.004   &      -0.007   &       1.316***&       0.456***&       0.057*  \\
                    &     (0.008)   &     (0.007)   &     (0.375)   &     (0.150)   &     (0.033)   \\
\addlinespace Mean of DV&     9024.57   &     2911.65   &       21.63   &        5.54   &        0.23   \\
Observations        &        2773   &        2652   &        2652   &        2652   &        2652   \\
Clusters            &         127   &         123   &         123   &         123   &         123   \\

%% file: fragment-ae_statisticslognew_cases-occupied_mv_beds.tex
log(COVID-19 cases on ventilators$_t$)&      -0.012   &      -0.008   &       0.996***&       0.522***&       0.066** \\
                    &     (0.013)   &     (0.009)   &     (0.361)   &     (0.136)   &     (0.026)   \\
\addlinespace Mean of DV&     9024.57   &     2911.65   &       21.63   &        5.54   &        0.23   \\
Observations        &        2773   &        2652   &        2652   &        2652   &        2652   \\
Clusters            &         127   &         123   &         123   &         123   &         123   \\

%% file: fragment-rttlognew_cases-new_admissions.tex
log(New COVID-19 admissions$_t$)&      -0.019   &      -0.077***&      -0.015   &               &               &               &       0.418*  &               &               \\
                    &     (0.015)   &     (0.020)   &     (0.018)   &               &               &               &     (0.236)   &               &               \\
log($\sum_{t-1}^t$ New COVID-19 admissions$_t$ )&               &               &               &       0.018   &       0.033*  &       0.015*  &               &       0.748** &               \\
                    &               &               &               &     (0.012)   &     (0.018)   &     (0.009)   &               &     (0.360)   &               \\
log($\sum_{t-2}^t$ New COVID-19 admissions$_t$ )&               &               &               &               &               &               &               &               &       0.787** \\
                    &               &               &               &               &               &               &               &               &     (0.386)   \\
\addlinespace Mean of DV&      587.13   &       99.50   &      393.98   &     2595.89   &    46977.22   &       15.49   &       76.52   &       59.50   &       46.04   \\
Observations        &       40874   &       33026   &       40456   &       41337   &       41336   &       41336   &       41337   &       41337   &       41337   \\
Clusters            &         123   &         123   &         123   &         123   &         123   &         123   &         123   &         123   &         123   \\

%% file: fragment-rttlognew_cases-hospital_cases.tex
log(COVID-19 cases in hospital$_t$)&      -0.013   &      -0.057***&      -0.006   &               &               &               &       0.455** &               &               \\
                    &     (0.012)   &     (0.018)   &     (0.011)   &               &               &               &     (0.187)   &               &               \\
log($\sum_{t-1}^t$ COVID-19 cases in hospital$_t$ )&               &               &               &       0.021** &       0.039***&       0.017** &               &       0.817***&               \\
                    &               &               &               &     (0.009)   &     (0.013)   &     (0.007)   &               &     (0.297)   &               \\
log($\sum_{t-2}^t$ COVID-19 cases in hospital$_t$ )&               &               &               &               &               &               &               &               &       0.973***\\
                    &               &               &               &               &               &               &               &               &     (0.317)   \\
\addlinespace Mean of DV&      587.13   &       99.50   &      393.98   &     2595.89   &    46977.22   &       15.49   &       76.52   &       59.50   &       46.04   \\
Observations        &       40874   &       33026   &       40456   &       41337   &       41336   &       41336   &       41337   &       41337   &       41337   \\
Clusters            &         123   &         123   &         123   &         123   &         123   &         123   &         123   &         123   &         123   \\

%% file: fragment-rttlognew_cases-occupied_mv_beds.tex
log(COVID-19 cases on ventilators$_t$)&      -0.005   &      -0.050***&      -0.007   &               &               &               &       0.325** &               &               \\
                    &     (0.010)   &     (0.017)   &     (0.011)   &               &               &               &     (0.159)   &               &               \\
log($\sum_{t-1}^t$ COVID-19 cases on ventilators$_t$ )&               &               &               &       0.016*  &       0.031***&       0.014** &               &       0.565** &               \\
                    &               &               &               &     (0.009)   &     (0.012)   &     (0.006)   &               &     (0.225)   &               \\
log($\sum_{t-2}^t$ COVID-19 cases on ventilators$_t$ )&               &               &               &               &               &               &               &               &       0.662***\\
                    &               &               &               &               &               &               &               &               &     (0.235)   \\
\addlinespace Mean of DV&      587.13   &       99.50   &      393.98   &     2595.89   &    46977.22   &       15.49   &       76.52   &       59.50   &       46.04   \\
Observations        &       40874   &       33026   &       40456   &       41337   &       41336   &       41336   &       41337   &       41337   &       41337   \\
Clusters            &         123   &         123   &         123   &         123   &         123   &         123   &         123   &         123   &         123   \\

%% file: fragment-diagnosticlognew_cases-new_admissions.tex
log(New COVID-19 admissions$_t$)&      -0.023   &               &               &               &               \\
                    &     (0.031)   &               &               &               &               \\
log($\sum_{t-1}^t$ New COVID-19 admissions$_t$ )&               &       0.010   &       0.073** &       0.382   &       3.319***\\
                    &               &     (0.016)   &     (0.028)   &     (0.763)   &     (1.108)   \\
\addlinespace Mean of DV&      927.23   &        5.46   &        3.87   &       30.23   &       16.55   \\
Observations        &       33918   &       31238   &        2518   &       31238   &        2518   \\
Clusters            &         123   &         123   &         123   &         123   &         123   \\

%% file: fragment-diagnosticlognew_cases-hospital_cases.tex
log(COVID-19 cases in hospital$_t$)&       0.020   &               &               &               &               \\
                    &     (0.025)   &               &               &               &               \\
log($\sum_{t-1}^t$ COVID-19 cases in hospital$_t$ )&               &      -0.000   &       0.014   &      -0.085   &       0.817   \\
                    &               &     (0.010)   &     (0.023)   &     (0.524)   &     (0.877)   \\
\addlinespace Mean of DV&      927.23   &        5.46   &        3.87   &       30.23   &       16.55   \\
Observations        &       33918   &       31238   &        2518   &       31238   &        2518   \\
Clusters            &         123   &         123   &         123   &         123   &         123   \\

%% file: fragment-diagnosticlognew_cases-occupied_mv_beds.tex
log(COVID-19 cases on ventilators$_t$)&      -0.027   &               &               &               &               \\
                    &     (0.022)   &               &               &               &               \\
log($\sum_{t-1}^t$ COVID-19 cases on ventilators$_t$ )&               &       0.007   &       0.010   &       0.351   &       0.599   \\
                    &               &     (0.010)   &     (0.020)   &     (0.522)   &     (0.807)   \\
\addlinespace Mean of DV&      927.23   &        5.46   &        3.87   &       30.23   &       16.55   \\
Observations        &       33918   &       31238   &        2518   &       31238   &        2518   \\
Clusters            &         123   &         123   &         123   &         123   &         123   \\

%% file: fragment-cancerlognew_cases-new_admissions.tex
log($\sum_{t-1}^t$ New COVID-19 admissions$_t$ )&      -0.012   &      -0.015   &      -0.013   &       0.375   &       0.393** &       1.920***\\
                    &     (0.014)   &     (0.013)   &     (0.011)   &     (0.423)   &     (0.175)   &     (0.579)   \\
\addlinespace Mean of DV&      117.45   &       16.12   &        8.83   &        9.55   &        4.22   &       29.84   \\
Observations        &       27404   &       29779   &       29773   &       29468   &       28187   &       27689   \\
Clusters            &         123   &         123   &         123   &         123   &         123   &         123   \\

%% file: fragment-cancerlognew_cases-hospital_cases.tex
log($\sum_{t-1}^t$ COVID-19 cases in hospital$_t$ )&      -0.008   &       0.005   &      -0.004   &       0.203   &       0.363** &       1.300***\\
                    &     (0.010)   &     (0.014)   &     (0.011)   &     (0.348)   &     (0.171)   &     (0.406)   \\
\addlinespace Mean of DV&      117.45   &       16.12   &        8.83   &        9.55   &        4.22   &       29.84   \\
Observations        &       27404   &       29779   &       29773   &       29468   &       28187   &       27689   \\
Clusters            &         123   &         123   &         123   &         123   &         123   &         123   \\

%% file: fragment-cancerlognew_cases-occupied_mv_beds.tex
log($\sum_{t-1}^t$ COVID-19 cases on ventilators$_t$ )&      -0.016   &      -0.005   &      -0.008   &       0.059   &       0.174   &       0.703** \\
                    &     (0.010)   &     (0.009)   &     (0.008)   &     (0.267)   &     (0.147)   &     (0.348)   \\
\addlinespace Mean of DV&      117.45   &       16.12   &        8.83   &        9.55   &        4.22   &       29.84   \\
Observations        &       27404   &       29779   &       29773   &       29468   &       28187   &       27689   \\
Clusters            &         123   &         123   &         123   &         123   &         123   &         123   \\

%% file: fragment-shmilognew_cases-dchexcess-new_admissions.tex
log(New COVID-19 admissions$_t$)&       2.933*  &       4.494***&       4.481***&       4.275***&       5.012***\\
                    &     (1.490)   &     (1.550)   &     (1.576)   &     (1.549)   &     (1.601)   \\
\addlinespace Observations&        2163   &        2163   &        2145   &        2145   &        2145   \\
Clusters            &         123   &         123   &         122   &         122   &         122   \\

%% file: fragment-shmilognew_cases-dchexcess-hospital_cases.tex
log(COVID-19 cases in hospital$_t$)&      -0.637   &       0.303   &       0.365   &       0.332   &       1.066   \\
                    &     (1.310)   &     (1.115)   &     (1.129)   &     (1.136)   &     (1.162)   \\
\addlinespace Observations&        2163   &        2163   &        2145   &        2145   &        2145   \\
Clusters            &         123   &         123   &         122   &         122   &         122   \\

%% file: fragment-shmilognew_cases-dchexcess-occupied_mv_beds.tex
log(COVID-19 cases on ventilators$_t$)&       1.981   &       3.043** &       3.208** &       3.053** &       4.058***\\
                    &     (1.558)   &     (1.372)   &     (1.379)   &     (1.430)   &     (1.525)   \\
\addlinespace Observations&        2163   &        2163   &        2145   &        2145   &        2145   \\
Clusters            &         123   &         123   &         122   &         122   &         122   \\

%% file: fragment-absencelognew_cases-new_admissions.tex
log(New COVID-19 admissions$_t$)&       0.503***&       0.417***&       0.445***&       0.646***&       0.226***&       0.106   \\
                    &     (0.061)   &     (0.073)   &     (0.059)   &     (0.089)   &     (0.047)   &     (0.096)   \\
\addlinespace Observations&        2259   &        2154   &        2154   &        2154   &        2154   &        2154   \\
Clusters            &         127   &         123   &         123   &         123   &         123   &         123   \\

%% file: fragment-absencelognew_cases-hospital_cases.tex
log(COVID-19 cases in hospital$_t$)&       0.347***&       0.314***&       0.274***&       0.440***&       0.097***&      -0.012   \\
                    &     (0.031)   &     (0.032)   &     (0.034)   &     (0.052)   &     (0.036)   &     (0.071)   \\
\addlinespace Observations&        2259   &        2154   &        2154   &        2154   &        2154   &        2154   \\
Clusters            &         127   &         123   &         123   &         123   &         123   &         123   \\

%% file: fragment-absencelognew_cases-occupied_mv_beds.tex
log(COVID-19 cases on ventilators$_t$)&       0.306***&       0.255***&       0.233***&       0.405***&       0.017   &       0.052   \\
                    &     (0.038)   &     (0.037)   &     (0.040)   &     (0.063)   &     (0.037)   &     (0.073)   \\
\addlinespace Observations&        2259   &        2154   &        2154   &        2154   &        2154   &        2154   \\
Clusters            &         127   &         123   &         123   &         123   &         123   &         123   \\

%% file: fragment-absencelognew_cases-vaccination-new_admissions.tex
log(New COVID-19 admissions$_t$)&       0.478***&       0.395***&       0.412***&       0.480***&       0.397***&       0.415***\\
                    &     (0.058)   &     (0.072)   &     (0.060)   &     (0.059)   &     (0.072)   &     (0.060)   \\
NHS vaccination uptake $\times$ log(New COVID-19 admissions$_t$)&      -0.051** &      -0.039   &      -0.059** &      -0.049*  &      -0.037   &      -0.058** \\
                    &     (0.026)   &     (0.025)   &     (0.027)   &     (0.025)   &     (0.025)   &     (0.027)   \\
\addlinespace Observations&        2259   &        2154   &        2154   &        2259   &        2154   &        2154   \\
Clusters            &         127   &         123   &         123   &         127   &         123   &         123   \\

%% file: fragment-absencelognew_cases-vaccination-hospital_cases.tex
log(COVID-19 cases in hospital$_t$)&       0.353***&       0.312***&       0.269***&       0.353***&       0.312***&       0.269***\\
                    &     (0.032)   &     (0.032)   &     (0.034)   &     (0.032)   &     (0.032)   &     (0.034)   \\
NHS vaccination uptake $\times$ log(COVID-19 cases in hospital$_t$)&      -0.061***&      -0.047** &      -0.063***&      -0.058***&      -0.044** &      -0.061***\\
                    &     (0.019)   &     (0.019)   &     (0.020)   &     (0.019)   &     (0.019)   &     (0.019)   \\
\addlinespace Observations&        2259   &        2154   &        2154   &        2259   &        2154   &        2154   \\
Clusters            &         127   &         123   &         123   &         127   &         123   &         123   \\

%% file: fragment-absencelognew_cases-vaccination-occupied_mv_beds.tex
log(COVID-19 cases on ventilators$_t$)&       0.301***&       0.240***&       0.188***&       0.302***&       0.241***&       0.191***\\
                    &     (0.039)   &     (0.040)   &     (0.043)   &     (0.039)   &     (0.040)   &     (0.043)   \\
NHS vaccination uptake $\times$ log(COVID-19 cases on ventilators$_t$)&      -0.009   &      -0.022   &      -0.065***&      -0.007   &      -0.020   &      -0.063***\\
                    &     (0.017)   &     (0.017)   &     (0.021)   &     (0.018)   &     (0.018)   &     (0.022)   \\
\addlinespace Observations&        2259   &        2154   &        2154   &        2259   &        2154   &        2154   \\
Clusters            &         127   &         123   &         123   &         127   &         123   &         123   \\

%% file: interaction-nhsvaxlognew_cases-shmi-dchexcess.tex
log(New COVID-19 admissions$_t$)&       3.567** &               &               &       3.615** &               &               \\
                    &     (1.502)   &               &               &     (1.503)   &               &               \\
NHS vaccination uptake $\times$ log(New COVID-19 admissions$_t$)&      -1.371** &               &               &      -1.329** &               &               \\
                    &     (0.561)   &               &               &     (0.558)   &               &               \\
log(COVID-19 cases in hospital$_t$)&               &       1.338   &               &               &       1.324   &               \\
                    &               &     (1.161)   &               &               &     (1.161)   &               \\
NHS vaccination uptake $\times$ log(COVID-19 cases in hospital$_t$)&               &      -0.945** &               &               &      -0.915** &               \\
                    &               &     (0.421)   &               &               &     (0.411)   &               \\
log(COVID-19 cases on ventilators$_t$)&               &               &       3.299** &               &               &       3.319** \\
                    &               &               &     (1.402)   &               &               &     (1.397)   \\
NHS vaccination uptake $\times$ log(COVID-19 cases on ventilators$_t$)&               &               &      -0.536   &               &               &      -0.519   \\
                    &               &               &     (0.533)   &               &               &     (0.514)   \\
\emph{Joint Test:}  &               &               &               &               &               &               \\
COVID-19 + Vaccination x COVID-19 = 0&       2.196   &        .393   &      2.763*   &       2.287   &        .408   &      2.799*   \\
                    &      (1.63)   &      (1.27)   &      (1.57)   &      (1.63)   &      (1.27)   &      (1.55)   \\
                    &               &               &               &               &               &               \\
Observations        &        2145   &        2145   &        2145   &        2145   &        2145   &        2145   \\
Clusters            &         122   &         122   &         122   &         122   &         122   &         122   \\

%% file: fragment-ae_statistics-new_admissions.tex
log(New COVID-19 admissions$_t$)&       0.019   &       0.022*  &       2.903***&       1.163***&       0.135***\\
                    &     (0.014)   &     (0.013)   &     (0.498)   &     (0.186)   &     (0.049)   \\
\addlinespace Mean of DV&     9024.57   &     2836.28   &       21.24   &        5.35   &        0.22   \\
Observations        &        2773   &        2773   &        2772   &        2772   &        2772   \\
Clusters            &         127   &         127   &         127   &         127   &         127   \\

%% file: fragment-ae_statistics-hospital_cases.tex
log(COVID-19 cases in hospital$_t$)&       0.004   &       0.002   &       1.640***&       0.521***&       0.086** \\
                    &     (0.008)   &     (0.008)   &     (0.338)   &     (0.120)   &     (0.039)   \\
\addlinespace Mean of DV&     9024.57   &     2836.28   &       21.24   &        5.35   &        0.22   \\
Observations        &        2773   &        2773   &        2772   &        2772   &        2772   \\
Clusters            &         127   &         127   &         127   &         127   &         127   \\

%% file: fragment-ae_statistics-occupied_mv_beds.tex
log(COVID-19 cases on ventilators$_t$)&      -0.012   &      -0.018   &       1.366***&       0.655***&       0.086***\\
                    &     (0.013)   &     (0.012)   &     (0.367)   &     (0.134)   &     (0.027)   \\
\addlinespace Mean of DV&     9024.57   &     2836.28   &       21.24   &        5.35   &        0.22   \\
Observations        &        2773   &        2773   &        2772   &        2772   &        2772   \\
Clusters            &         127   &         127   &         127   &         127   &         127   \\

%% file: fragment-rtt-new_admissions.tex
log(New COVID-19 admissions$_t$)&      -0.027** &      -0.091***&      -0.021   &               &               &               &       0.529** &               &               \\
                    &     (0.012)   &     (0.017)   &     (0.014)   &               &               &               &     (0.203)   &               &               \\
log($\sum_{t-1}^t$ New COVID-19 admissions$_t$ )&               &               &               &       0.015   &       0.032** &       0.017** &               &       0.998***&               \\
                    &               &               &               &     (0.010)   &     (0.015)   &     (0.008)   &               &     (0.306)   &               \\
log($\sum_{t-2}^t$ New COVID-19 admissions$_t$ )&               &               &               &               &               &               &               &               &       0.991***\\
                    &               &               &               &               &               &               &               &               &     (0.328)   \\
\addlinespace Mean of DV&      586.42   &       99.51   &      394.26   &     2593.53   &    46966.38   &       15.52   &       76.54   &       59.56   &       46.16   \\
Observations        &       41883   &       33827   &       41431   &       42351   &       42350   &       42350   &       42351   &       42351   &       42351   \\
Clusters            &         127   &         127   &         127   &         127   &         127   &         127   &         127   &         127   &         127   \\

%% file: fragment-rtt-hospital_cases.tex
log(COVID-19 cases in hospital$_t$)&      -0.021** &      -0.069***&      -0.014   &               &               &               &       0.511***&               &               \\
                    &     (0.009)   &     (0.014)   &     (0.009)   &               &               &               &     (0.166)   &               &               \\
log($\sum_{t-1}^t$ COVID-19 cases in hospital$_t$ )&               &               &               &       0.015*  &       0.033***&       0.018***&               &       0.945***&               \\
                    &               &               &               &     (0.008)   &     (0.012)   &     (0.006)   &               &     (0.258)   &               \\
log($\sum_{t-2}^t$ COVID-19 cases in hospital$_t$ )&               &               &               &               &               &               &               &               &       1.039***\\
                    &               &               &               &               &               &               &               &               &     (0.276)   \\
\addlinespace Mean of DV&      586.42   &       99.51   &      394.26   &     2593.53   &    46966.38   &       15.52   &       76.54   &       59.56   &       46.16   \\
Observations        &       41883   &       33827   &       41431   &       42351   &       42350   &       42350   &       42351   &       42351   &       42351   \\
Clusters            &         127   &         127   &         127   &         127   &         127   &         127   &         127   &         127   &         127   \\

%% file: fragment-rtt-occupied_mv_beds.tex
log(COVID-19 cases on ventilators$_t$)&      -0.010   &      -0.061***&      -0.011   &               &               &               &       0.409** &               &               \\
                    &     (0.010)   &     (0.016)   &     (0.011)   &               &               &               &     (0.157)   &               &               \\
log($\sum_{t-1}^t$ COVID-19 cases on ventilators$_t$ )&               &               &               &       0.015*  &       0.031***&       0.015***&               &       0.724***&               \\
                    &               &               &               &     (0.008)   &     (0.011)   &     (0.005)   &               &     (0.227)   &               \\
log($\sum_{t-2}^t$ COVID-19 cases on ventilators$_t$ )&               &               &               &               &               &               &               &               &       0.786***\\
                    &               &               &               &               &               &               &               &               &     (0.231)   \\
\addlinespace Mean of DV&      586.42   &       99.51   &      394.26   &     2593.53   &    46966.38   &       15.52   &       76.54   &       59.56   &       46.16   \\
Observations        &       41883   &       33827   &       41431   &       42351   &       42350   &       42350   &       42351   &       42351   &       42351   \\
Clusters            &         127   &         127   &         127   &         127   &         127   &         127   &         127   &         127   &         127   \\

%% file: fragment-diagnostic-new_admissions.tex
log(New COVID-19 admissions$_t$)&      -0.037   &               &               &               &               \\
                    &     (0.025)   &               &               &               &               \\
log($\sum_{t-1}^t$ New COVID-19 admissions$_t$ )&               &       0.020   &       0.060***&       0.974   &       2.732***\\
                    &               &     (0.012)   &     (0.021)   &     (0.621)   &     (0.879)   \\
\addlinespace Mean of DV&      902.78   &        5.47   &        3.89   &       30.31   &       16.75   \\
Observations        &       35504   &       32502   &        2634   &       32502   &        2634   \\
Clusters            &         127   &         127   &         127   &         127   &         127   \\

%% file: fragment-diagnostic-hospital_cases.tex
log(COVID-19 cases in hospital$_t$)&      -0.009   &               &               &               &               \\
                    &     (0.019)   &               &               &               &               \\
log($\sum_{t-1}^t$ COVID-19 cases in hospital$_t$ )&               &       0.012   &       0.023   &       0.542   &       1.112*  \\
                    &               &     (0.009)   &     (0.016)   &     (0.453)   &     (0.662)   \\
\addlinespace Mean of DV&      902.78   &        5.47   &        3.89   &       30.31   &       16.75   \\
Observations        &       35504   &       32502   &        2634   &       32502   &        2634   \\
Clusters            &         127   &         127   &         127   &         127   &         127   \\

%% file: fragment-diagnostic-occupied_mv_beds.tex
log(COVID-19 cases on ventilators$_t$)&      -0.044** &               &               &               &               \\
                    &     (0.021)   &               &               &               &               \\
log($\sum_{t-1}^t$ COVID-19 cases on ventilators$_t$ )&               &       0.010   &       0.016   &       0.498   &       0.903   \\
                    &               &     (0.010)   &     (0.018)   &     (0.517)   &     (0.750)   \\
\addlinespace Mean of DV&      902.78   &        5.47   &        3.89   &       30.31   &       16.75   \\
Observations        &       35504   &       32502   &        2634   &       32502   &        2634   \\
Clusters            &         127   &         127   &         127   &         127   &         127   \\

%% file: fragment-cancer-new_admissions.tex
log($\sum_{t-1}^t$ New COVID-19 admissions$_t$ )&      -0.017   &      -0.027** &      -0.020*  &       0.414   &       0.521***&       1.716***\\
                    &     (0.015)   &     (0.012)   &     (0.010)   &     (0.358)   &     (0.159)   &     (0.484)   \\
\addlinespace Mean of DV&      116.31   &       16.01   &        8.76   &        9.49   &        4.21   &       29.81   \\
Observations        &       28102   &       30525   &       30503   &       30155   &       28869   &       28304   \\
Clusters            &         127   &         127   &         126   &         127   &         127   &         125   \\

%% file: fragment-cancer-hospital_cases.tex
log($\sum_{t-1}^t$ COVID-19 cases in hospital$_t$ )&      -0.012   &      -0.010   &      -0.012   &       0.291   &       0.433***&       1.232***\\
                    &     (0.010)   &     (0.011)   &     (0.009)   &     (0.289)   &     (0.147)   &     (0.331)   \\
\addlinespace Mean of DV&      116.31   &       16.01   &        8.76   &        9.49   &        4.21   &       29.81   \\
Observations        &       28102   &       30525   &       30503   &       30155   &       28869   &       28304   \\
Clusters            &         127   &         127   &         126   &         127   &         127   &         125   \\

%% file: fragment-cancer-occupied_mv_beds.tex
log($\sum_{t-1}^t$ COVID-19 cases on ventilators$_t$ )&      -0.019*  &      -0.013   &      -0.010   &       0.151   &       0.290** &       0.877***\\
                    &     (0.010)   &     (0.009)   &     (0.008)   &     (0.276)   &     (0.142)   &     (0.324)   \\
\addlinespace Mean of DV&      116.31   &       16.01   &        8.76   &        9.49   &        4.21   &       29.81   \\
Observations        &       28102   &       30525   &       30503   &       30155   &       28869   &       28304   \\
Clusters            &         127   &         127   &         126   &         127   &         127   &         125   \\

%% file: fragment-shmi-dchexcess-new_admissions.tex
log(New COVID-19 admissions$_t$)&       4.146***&       5.428***&       5.302***&       5.107***&       5.131***\\
                    &     (1.237)   &     (1.298)   &     (1.308)   &     (1.264)   &     (1.300)   \\
\addlinespace Observations&        2215   &        2215   &        2185   &        2185   &        2185   \\
Clusters            &         124   &         124   &         123   &         123   &         123   \\

%% file: fragment-shmi-dchexcess-hospital_cases.tex
log(COVID-19 cases in hospital$_t$)&       0.887   &       1.644   &       1.616   &       1.551   &       1.660   \\
                    &     (1.081)   &     (1.006)   &     (1.011)   &     (1.011)   &     (1.034)   \\
\addlinespace Observations&        2215   &        2215   &        2185   &        2185   &        2185   \\
Clusters            &         124   &         124   &         123   &         123   &         123   \\

%% file: fragment-shmi-dchexcess-occupied_mv_beds.tex
log(COVID-19 cases on ventilators$_t$)&       2.394   &       3.446***&       3.515***&       3.362** &       4.040***\\
                    &     (1.469)   &     (1.311)   &     (1.324)   &     (1.374)   &     (1.456)   \\
\addlinespace Observations&        2215   &        2215   &        2185   &        2185   &        2185   \\
Clusters            &         124   &         124   &         123   &         123   &         123   \\

%% file: fragment-absence-new_admissions.tex
log(New COVID-19 admissions$_t$)&       0.503***&       0.491***&       0.507***&       0.730***&       0.269***&       0.156** \\
                    &     (0.061)   &     (0.061)   &     (0.050)   &     (0.075)   &     (0.045)   &     (0.077)   \\
\addlinespace Observations&        2259   &        2235   &        2235   &        2235   &        2235   &        2235   \\
Clusters            &         127   &         127   &         127   &         127   &         127   &         127   \\

%% file: fragment-absence-hospital_cases.tex
log(COVID-19 cases in hospital$_t$)&       0.347***&       0.365***&       0.329***&       0.500***&       0.149***&       0.051   \\
                    &     (0.031)   &     (0.030)   &     (0.030)   &     (0.047)   &     (0.035)   &     (0.057)   \\
\addlinespace Observations&        2259   &        2235   &        2235   &        2235   &        2235   &        2235   \\
Clusters            &         127   &         127   &         127   &         127   &         127   &         127   \\

%% file: fragment-absence-occupied_mv_beds.tex
log(COVID-19 cases on ventilators$_t$)&       0.306***&       0.314***&       0.303***&       0.496***&       0.064*  &       0.094   \\
                    &     (0.038)   &     (0.037)   &     (0.039)   &     (0.060)   &     (0.037)   &     (0.067)   \\
\addlinespace Observations&        2259   &        2235   &        2235   &        2235   &        2235   &        2235   \\
Clusters            &         127   &         127   &         127   &         127   &         127   &         127   \\

%% file: fragment-absence-vaccination-new_admissions.tex
log(New COVID-19 admissions$_t$)&       0.478***&       0.472***&       0.476***&       0.480***&       0.474***&       0.478***\\
                    &     (0.058)   &     (0.059)   &     (0.048)   &     (0.059)   &     (0.059)   &     (0.048)   \\
NHS vaccination uptake $\times$ log(New COVID-19 admissions$_t$)&      -0.051** &      -0.040   &      -0.060** &      -0.049*  &      -0.037   &      -0.059** \\
                    &     (0.026)   &     (0.026)   &     (0.027)   &     (0.025)   &     (0.025)   &     (0.027)   \\
\addlinespace Observations&        2259   &        2235   &        2235   &        2259   &        2235   &        2235   \\
Clusters            &         127   &         127   &         127   &         127   &         127   &         127   \\

%% file: fragment-absence-vaccination-hospital_cases.tex
log(COVID-19 cases in hospital$_t$)&       0.353***&       0.367***&       0.329***&       0.353***&       0.367***&       0.329***\\
                    &     (0.032)   &     (0.031)   &     (0.030)   &     (0.032)   &     (0.031)   &     (0.030)   \\
NHS vaccination uptake $\times$ log(COVID-19 cases in hospital$_t$)&      -0.061***&      -0.049***&      -0.064***&      -0.058***&      -0.046** &      -0.062***\\
                    &     (0.019)   &     (0.019)   &     (0.020)   &     (0.019)   &     (0.019)   &     (0.020)   \\
\addlinespace Observations&        2259   &        2235   &        2235   &        2259   &        2235   &        2235   \\
Clusters            &         127   &         127   &         127   &         127   &         127   &         127   \\

%% file: fragment-absence-vaccination-occupied_mv_beds.tex
log(COVID-19 cases on ventilators$_t$)&       0.301***&       0.313***&       0.278***&       0.302***&       0.314***&       0.280***\\
                    &     (0.039)   &     (0.038)   &     (0.041)   &     (0.039)   &     (0.038)   &     (0.041)   \\
NHS vaccination uptake $\times$ log(COVID-19 cases on ventilators$_t$)&      -0.009   &      -0.002   &      -0.044** &      -0.007   &       0.000   &      -0.042*  \\
                    &     (0.017)   &     (0.017)   &     (0.021)   &     (0.018)   &     (0.017)   &     (0.021)   \\
\addlinespace Observations&        2259   &        2235   &        2235   &        2259   &        2235   &        2235   \\
Clusters            &         127   &         127   &         127   &         127   &         127   &         127   \\

%% file: interaction-nhsvax-shmi-dchexcess.tex
log(New COVID-19 admissions$_t$)&       4.411***&               &               &       4.460***&               &               \\
                    &     (1.311)   &               &               &     (1.306)   &               &               \\
NHS vaccination uptake $\times$ log(New COVID-19 admissions$_t$)&      -1.375** &               &               &      -1.329** &               &               \\
                    &     (0.554)   &               &               &     (0.552)   &               &               \\
log(COVID-19 cases in hospital$_t$)&               &       2.358** &               &               &       2.347** &               \\
                    &               &     (1.006)   &               &               &     (1.006)   &               \\
NHS vaccination uptake $\times$ log(COVID-19 cases in hospital$_t$)&               &      -0.974** &               &               &      -0.941** &               \\
                    &               &     (0.414)   &               &               &     (0.406)   &               \\
log(COVID-19 cases on ventilators$_t$)&               &               &       3.742***&               &               &       3.757***\\
                    &               &               &     (1.331)   &               &               &     (1.327)   \\
NHS vaccination uptake $\times$ log(COVID-19 cases on ventilators$_t$)&               &               &      -0.359   &               &               &      -0.344   \\
                    &               &               &     (0.532)   &               &               &     (0.511)   \\
\emph{Joint Test:}  &               &               &               &               &               &               \\
COVID-19 + Vaccination x COVID-19 = 0&     3.036**   &       1.384   &     3.384**   &      3.13**   &       1.406   &     3.413**   \\
                    &      (1.51)   &      (1.13)   &      (1.48)   &       (1.5)   &      (1.12)   &      (1.45)   \\
                    &               &               &               &               &               &               \\
Observations        &        2180   &        2180   &        2180   &        2180   &        2180   &        2180   \\
Clusters            &         122   &         122   &         122   &         122   &         122   &         122   \\

%% file: fragment-shmi-dlogexcess-new_admissions.tex
log(New COVID-19 hospital admissions)&       0.006***&       0.008***&       0.010***&       0.012** &       0.014   &       0.013   \\
                    &     (0.002)   &     (0.003)   &     (0.004)   &     (0.005)   &     (0.009)   &     (0.014)   \\
\addlinespace Observations&        2225   &        2225   &        2225   &        2225   &        2225   &        2225   \\
Clusters            &         126   &         126   &         126   &         126   &         126   &         126   \\

%% file: fragment-shmi-dlogexcess-hospital_cases.tex
log(confirmed COVID-19 patients in hospital)&       0.007** &       0.008** &       0.011** &       0.014** &       0.018   &       0.016   \\
                    &     (0.003)   &     (0.004)   &     (0.004)   &     (0.006)   &     (0.012)   &     (0.017)   \\
\addlinespace Observations&        2225   &        2225   &        2225   &        2225   &        2225   &        2225   \\
Clusters            &         126   &         126   &         126   &         126   &         126   &         126   \\

%% file: fragment-shmi-dlogexcess-occupied_mv_beds.tex
log(\# of COVID-19 cases in ventilator beds)&       0.004** &       0.005***&       0.006** &       0.008***&       0.014** &       0.018** \\
                    &     (0.002)   &     (0.002)   &     (0.003)   &     (0.003)   &     (0.006)   &     (0.008)   \\
\addlinespace Observations&        2225   &        2225   &        2225   &        2225   &        2225   &        2225   \\
Clusters            &         126   &         126   &         126   &         126   &         126   &         126   \\

%% file: fragment-shmi-logobserved-new_admissions.tex
log(New COVID-19 admissions)&       0.006***&       0.008***&       0.010***&       0.012** &       0.013   &       0.012   \\
                    &     (0.002)   &     (0.003)   &     (0.004)   &     (0.005)   &     (0.009)   &     (0.013)   \\
log(expected deaths)&       0.886***&       0.887***&       0.887***&       0.887***&       0.889***&       0.889***\\
                    &     (0.082)   &     (0.083)   &     (0.083)   &     (0.083)   &     (0.083)   &     (0.083)   \\
\addlinespace Observations&        2225   &        2225   &        2225   &        2225   &        2225   &        2225   \\
Clusters            &         126   &         126   &         126   &         126   &         126   &         126   \\

%% file: fragment-shmi-logobserved-hospital_cases.tex
log(COVID-19 cases in hospital)&       0.007** &       0.009** &       0.011** &       0.015***&       0.018   &       0.015   \\
                    &     (0.003)   &     (0.004)   &     (0.004)   &     (0.006)   &     (0.012)   &     (0.017)   \\
log(expected deaths)&       0.885***&       0.885***&       0.885***&       0.884***&       0.887***&       0.888***\\
                    &     (0.083)   &     (0.083)   &     (0.083)   &     (0.083)   &     (0.083)   &     (0.083)   \\
\addlinespace Observations&        2225   &        2225   &        2225   &        2225   &        2225   &        2225   \\
Clusters            &         126   &         126   &         126   &         126   &         126   &         126   \\

%% file: fragment-shmi-logobserved-occupied_mv_beds.tex
log(COVID-19 cases on ventilators)&       0.004***&       0.006***&       0.007***&       0.009***&       0.015** &       0.018** \\
                    &     (0.002)   &     (0.002)   &     (0.003)   &     (0.003)   &     (0.006)   &     (0.008)   \\
log(expected deaths)&       0.886***&       0.886***&       0.886***&       0.885***&       0.886***&       0.887***\\
                    &     (0.082)   &     (0.082)   &     (0.082)   &     (0.082)   &     (0.081)   &     (0.081)   \\
\addlinespace Observations&        2225   &        2225   &        2225   &        2225   &        2225   &        2225   \\
Clusters            &         126   &         126   &         126   &         126   &         126   &         126   \\

%% file: fragment-shmi-deathrate-new_admissions.tex
log( New COVID-19 admissions )&       0.018***&       0.024** &       0.030** &       0.034** &       0.034   &       0.018   \\
                    &     (0.007)   &     (0.009)   &     (0.012)   &     (0.015)   &     (0.030)   &     (0.045)   \\
Expected deaths / \# of spells&       0.930***&       0.930***&       0.930***&       0.930***&       0.931***&       0.930***\\
                    &     (0.105)   &     (0.105)   &     (0.105)   &     (0.105)   &     (0.105)   &     (0.105)   \\
\addlinespace Observations&        2225   &        2225   &        2225   &        2225   &        2225   &        2225   \\
Clusters            &         126   &         126   &         126   &         126   &         126   &         126   \\

%% file: fragment-shmi-deathrate-hospital_cases.tex
log( COVID-19 cases in hospital )&       0.028** &       0.035** &       0.044** &       0.056** &       0.064   &       0.046   \\
                    &     (0.011)   &     (0.014)   &     (0.018)   &     (0.022)   &     (0.044)   &     (0.061)   \\
Expected deaths / \# of spells&       0.926***&       0.925***&       0.925***&       0.925***&       0.930***&       0.930***\\
                    &     (0.105)   &     (0.105)   &     (0.105)   &     (0.105)   &     (0.105)   &     (0.106)   \\
\addlinespace Observations&        2225   &        2225   &        2225   &        2225   &        2225   &        2225   \\
Clusters            &         126   &         126   &         126   &         126   &         126   &         126   \\

%% file: fragment-shmi-deathrate-occupied_mv_beds.tex
log( COVID-19 cases on ventilators )&       0.014***&       0.019***&       0.023***&       0.029***&       0.049** &       0.056** \\
                    &     (0.005)   &     (0.007)   &     (0.009)   &     (0.011)   &     (0.022)   &     (0.028)   \\
Expected deaths / \# of spells&       0.929***&       0.928***&       0.928***&       0.928***&       0.929***&       0.930***\\
                    &     (0.104)   &     (0.104)   &     (0.104)   &     (0.104)   &     (0.104)   &     (0.103)   \\
\addlinespace Observations&        2225   &        2225   &        2225   &        2225   &        2225   &        2225   \\
Clusters            &         126   &         126   &         126   &         126   &         126   &         126   \\

%% file: coronabib.bib
@article{proto2021covid,
	author = {Proto, Eugenio and Quintana-Domeque, Climent},
	journal = {PloS one},
	number = {1},
	pages = {e0244419},
	publisher = {Public Library of Science San Francisco, CA USA},
	title = {COVID-19 and mental health deterioration by ethnicity and gender in the UK},
	volume = {16},
	year = {2021}}

@techreport{etheridge2020gender,
	author = {Etheridge, Ben and Spantig, Lisa},
	institution = {ISER Working paper series},
	title = {The gender gap in mental well-being during the Covid-19 outbreak: evidence from the UK},
	year = {2020}}

@article{adams2020impact,
	author = {Adams-Prassl, Abi and Boneva, Teodora and Golin, Marta and Rauh, Christopher},
	journal = {Economic Policy},
	title = {The impact of the {C}oronavirus lockdown on mental health: evidence from the {US}},
	year = {2022}}

@article{lai2020estimated,
	author = {Lai, Alvina G and Pasea, Laura and Banerjee, Amitava and Hall, Geoff and Denaxas, Spiros and Chang, Wai Hoong and Katsoulis, Michail and Williams, Bryan and Pillay, Deenan and Noursadeghi, Mahdad and others},
	journal = {BMJ open},
	number = {11},
	pages = {e043828},
	publisher = {British Medical Journal Publishing Group},
	title = {Estimated impact of the COVID-19 pandemic on cancer services and excess 1-year mortality in people with cancer and multimorbidity: near real-time data on cancer care, cancer deaths and a population-based cohort study},
	volume = {10},
	year = {2020}}

@techreport{bethune2020covid,
	author = {Bethune, Zachary A and Korinek, Anton},
	institution = {National Bureau of Economic Research},
	title = {Covid-19 infection externalities: Trading off lives vs. livelihoods},
	year = {2020}}

@techreport{macmillan2020,
	author = {Macmillan},
	institution = {Macmillan Cancer Support},
	title = {The forgotten C? The impact of Covid-19 on cancer care},
	year = {2020}}

@misc{kutikov2020war,
	author = {Kutikov, Alexander and Weinberg, David S and Edelman, Martin J and Horwitz, Eric M and Uzzo, Robert G and Fisher, Richard I},
	journal = {Annals of internal medicine},
	number = {11},
	pages = {756--758},
	publisher = {American College of Physicians},
	title = {A war on two fronts: cancer care in the time of COVID-19},
	volume = {172},
	year = {2020}}

@article{richards2020impact,
	author = {Richards, Mike and Anderson, Michael and Carter, Paul and Ebert, Benjamin L and Mossialos, Elias},
	journal = {Nature Cancer},
	number = {6},
	pages = {565--567},
	publisher = {Nature Publishing Group},
	title = {The impact of the COVID-19 pandemic on cancer care},
	volume = {1},
	year = {2020}}

@article{Clements2008,
	author = {Clements, Archie and Halton, Kate and Graves, Nicholas and Pettitt, Anthony and Morton, Anthony and Looke, David and Whitby, Michael},
	doi = {10.1016/S1473-3099(08)70151-8},
	issn = {14733099},
	journal = {The Lancet Infectious Diseases},
	number = {7},
	pages = {427--434},
	pmid = {18582835},
	title = {{Overcrowding and understaffing in modern health-care systems: key determinants in meticillin-resistant Staphylococcus aureus transmission}},
	volume = {8},
	year = {2008}}

@article{Lasater2021,
	author = {Lasater, Karen B. and Aiken, Linda H. and Sloane, Douglas M. and French, Rachel and Martin, Brendan and Reneau, Kyrani and Alexander, Maryann and McHugh, Matthew D.},
	doi = {10.1136/bmjqs-2020-011512},
	issn = {20445415},
	journal = {BMJ Quality and Safety},
	number = {8},
	pages = {639--647},
	pmid = {32817399},
	title = {{Chronic hospital nurse understaffing meets COVID-19: An observational study}},
	volume = {30},
	year = {2021}}

@article{Bandyopadhyay2020,
	author = {Bandyopadhyay, Soham and Baticulon, Ronnie E. and Kadhum, Murtaza and Alser, Muath and Ojuka, Daniel K. and Badereddin, Yara and Kamath, Archith and Parepalli, Sai Arathi and Brown, Grace and Iharchane, Sara and Gandino, Sofia and Markovic-Obiago, Zara and Scott, Samuel and Manirambona, Emery and Machhada, Asif and Aggarwal, Aditi and Benazaize, Lydia and Ibrahim, Mina and Kim, David and Tol, Isabel and Taylor, Elliott H. and Knighton, Alexandra and Bbaale, Dorothy and Jasim, Duha and Alghoul, Heba and Reddy, Henna and Abuelgasim, Hibatullah and Saini, Kirandeep and Sigler, Alicia and Abuelgasim, Leenah and Moran-Romero, Mario and Kumarendran, Mary and Jamie, Najlaa Abu and Ali, Omaima and Sudarshan, Raghav and Dean, Riley and Kissyova, Rumi and Kelzang, Sonam and Roche, Sophie and Ahsan, Tazin and Mohamed, Yethrib and Dube, Andile Maqhawe and Gwini, Grace Paida and Gwokyala, Rashidah and Brown, Robin and Papon, Mohammad Rabiul Karim Khan and Li, Zoe and Ruzats, Salvador Sun and Charuvila, Somy and Peter, Noel and Khalidy, Khalil and Moyo, Nkosikhona and Alser, Osaid and Solano, Arielis and Robles-Perez, Eduardo and Tariq, Aiman and Gaddah, Mariam and Kolovos, Spyros and Muchemwa, Faith C. and Saleh, Abdullah and Gosman, Amanda and Pinedo-Villanueva, Rafael and Jani, Anant and Khundkar, Roba},
	doi = {10.1136/bmjgh-2020-003097},
	issn = {20597908},
	journal = {BMJ Global Health},
	number = {12},
	title = {{Infection and mortality of healthcare workers worldwide from COVID-19: A systematic review}},
	volume = {5},
	year = {2020}}

@article{Vandoros2021,
	author = {Vandoros, Sotiris},
	doi = {10.1136/injuryprev-2020-044139},
	issn = {14755785},
	journal = {Injury Prevention},
	pages = {1--5},
	title = {{COVID-19, lockdowns and motor vehicle collisions: Empirical evidence from Greece}},
	year = {2021}}

@article{Hanna2020,
	author = {Hanna, Timothy P. and King, Will D. and Thibodeau, Stephane and Jalink, Matthew and Paulin, Gregory A. and Harvey-Jones, Elizabeth and O'Sullivan, Dylan E. and Booth, Christopher M. and Sullivan, Richard and Aggarwal, Ajay},
	doi = {10.1136/bmj.m4087},
	isbn = {0000000332286},
	issn = {17561833},
	journal = {BMJ (Clinical research ed.)},
	pages = {m4087},
	pmid = {33148535},
	title = {{Mortality due to cancer treatment delay: systematic review and meta-analysis}},
	volume = {371},
	year = {2020}}

@article{Fajgelbaum2020,
author = {Fajgelbaum, Pablo and Khandelwal, Amit and Kim, Wookun and Mantovani, Cristiano and Schaal, Edouard},
journal = {NBER Working Paper No. 27441},
title = {{Optimal Lockdown in a Commuting Network}},
url = {http://www.nber.org/papers/w27441},
year = {2020}
}

@article{Banerjee2021,
	author = {Banerjee, Amitava and Chen, Suliang and Pasea, Laura and Lai, Alvina G and Katsoulis, Michail and Denaxas, Spiros and Nafilyan, Vahe and Williams, Bryan and Wong, Wai Keong and Bakhai, Ameet and Khunti, Kamlesh and Pillay, Deenan and Noursadeghi, Mahdad and Wu, Honghan and Pareek, Nilesh and Bromage, Daniel and McDonagh, Theresa A and Byrne, Jonathan and Teo, James T H and Shah, Ajay M and Humberstone, Ben and Tang, Liang V and Shah, Anoop S V and Rubboli, Andrea and Guo, Yutao and Hu, Yu and Sudlow, Cathie L M and Lip, Gregory Y H and Hemingway, Harry},
	doi = {10.1093/eurjpc/zwaa155},
	issn = {2047-4873},
	journal = {European Journal of Preventive Cardiology},
	number = {14},
	pages = {1599--1609},
	title = {{Excess deaths in people with cardiovascular diseases during the COVID-19 pandemic}},
	volume = {28},
	year = {2021}}

@article{Mitze2020,
	author = {Mitze, Timo and Kosfeld, Reinhold and Rode, Johannes and W{\"{a}}lde, Klaus},
	doi = {10.1073/pnas.2015954117},
	issn = {0027-8424},
	journal = {Proceedings of the National Academy of Sciences},
	month = {dec},
	number = {13319},
	pages = {202015954},
	title = {{Face masks considerably reduce COVID-19 cases in Germany}},
	url = {http://www.pnas.org/lookup/doi/10.1073/pnas.2015954117},
	year = {2020}}

@article{Sun2020a,
	author = {Sun, Niuniu and Wei, Luoqun and Shi, Suling and Jiao, Dandan and Song, Runluo and Ma, Lili and Wang, Hongwei and Wang, Chao and Wang, Zhaoguo and You, Yanli and Liu, Shuhua and Wang, Hongyun},
	doi = {10.1016/j.ajic.2020.03.018},
	issn = {15273296},
	journal = {American Journal of Infection Control},
	number = {6},
	pages = {592--598},
	pmid = {32334904},
	publisher = {Elsevier Inc.},
	title = {{A qualitative study on the psychological experience of caregivers of COVID-19 patients}},
	url = {https://doi.org/10.1016/j.ajic.2020.03.018},
	volume = {48},
	year = {2020}}

@article{Quintana-Domeque2021,
	author = {Quintana-Domeque, Climent and Lee, Ines and Zhang, Anwen and Proto, Eugenio and Battisti, Michele and Ho, Antonia},
	doi = {10.1371/journal.pone.0259213},
	isbn = {1111111111},
	issn = {19326203},
	journal = {PLoS ONE},
	number = {November},
	pages = {1--14},
	pmid = {34727110},
	title = {{Anxiety and depression among medical doctors in Catalonia, Italy, and the UK during the COVID-19 pandemic}},
	url = {http://dx.doi.org/10.1371/journal.pone.0259213},
	volume = {16},
	year = {2021}}

@article{Mahasen2664,
	author = {Mahase, Elisabeth},
	doi = {10.1136/bmj.n2664},
	journal = {BMJ},
	publisher = {BMJ Publishing Group Ltd},
	title = {{Under pressure: when does the NHS reach breaking point?}},
	url = {https://www.bmj.com/content/375/bmj.n2664},
	volume = {375},
	year = {2021}}

@article{deOliveiraAndradem3032,
	author = {{de Oliveira Andrade}, Rodrigo},
	doi = {10.1136/bmj.m3032},
	journal = {BMJ},
	publisher = {BMJ Publishing Group Ltd},
	title = {{Covid-19 is causing the collapse of Brazil's national health service}},
	url = {https://www.bmj.com/content/370/bmj.m3032},
	volume = {370},
	year = {2020}}

@article{Adam2022,
	author = {Adam, David},
	doi = {10.1038/d41586-022-00104-8},
	issn = {0028-0836},
	journal = {Nature},
	month = {jan},
	number = {7893},
	pages = {312--315},
	title = {{The pandemic's true death toll: millions more than official counts}},
	url = {https://www.nature.com/articles/d41586-022-00104-8},
	volume = {601},
	year = {2022}}

@article{Jha2022,
	author = {Jha, Prabhat and Deshmukh, Yashwant and Tumbe, Chinmay and Suraweera, Wilson and Bhowmick, Aditi and Sharma, Sankalp and Novosad, Paul and Fu, Sze Hang and Newcombe, Leslie and Gelband, Hellen and Brown, Patrick},
	doi = {10.1126/science.abm5154},
	issn = {0036-8075},
	journal = {Science},
	month = {jan},
	number = {March 2020},
	pages = {1--10},
	title = {{COVID mortality in India: National survey data and health facility deaths}},
	url = {https://www.science.org/doi/10.1126/science.abm5154},
	volume = {5154},
	year = {2022}}

@techreport{SHMI2021,
	author = {{NHS Digital}},
	number = {June 2021},
	title = {{Summary Hospital-level Mortality Indicator (SHMI) - Deaths associated with hospitalisation}},
	url = {http://www.hscic.gov.uk/catalogue/PUB20949/shmi-deat-hosp-eng-jan-15-dec-15-rep.pdf},
	year = {2021}}

@article{Wymant2021,
	author = {Wymant, Chris and Ferretti, Luca and Tsallis, Daphne and Charalambides, Marcos and Abeler-D{\"{o}}rner, Lucie and Bonsall, David and Hinch, Robert and Kendall, Michelle and Milsom, Luke and Ayres, Matthew and Holmes, Chris and Briers, Mark and Fraser, Christophe},
	doi = {10.1038/s41586-021-03606-z},
	issn = {14764687},
	journal = {Nature},
	number = {7863},
	pages = {408--412},
	pmid = {33979832},
	publisher = {Springer US},
	title = {{The epidemiological impact of the NHS COVID-19 app}},
	url = {http://dx.doi.org/10.1038/s41586-021-03606-z},
	volume = {594},
	year = {2021}}

@article{Fetzer2021,
	author = {Fetzer, Thiemo and Graeber, Thomas},
	doi = {10.1073/pnas.2100814118},
	issn = {0027-8424},
	journal = {Proceedings of the National Academy of Sciences},
	number = {33},
	pages = {e2100814118},
	title = {{Measuring the scientific effectiveness of contact tracing: Evidence from a natural experiment}},
	volume = {118},
	year = {2021}}

@article{Abaluck2021,
	author = {Abaluck, Jason and Kwong, Laura H and Styczynski, Ashley and Haque, Ashraful and Kabir, Md. Alamgir and Bates-Jefferys, Ellen and Crawford, Emily and Benjamin-Chung, Jade and Benhachmi, Salim and Raihan, Shabib and Rahman, Shadman and Zaman, Neeti and Winch, Peter J. and Hossain, Md. Maqsud and Re, Hasan Mahmud and Mobarak, Ahmed Mushfiq},
	journal = {NBER Working Paper},
	title = {{Normalizing Community Mask-Wearing: A Cluster Randomized Trial in Bangladesh}},
	year = {2021}}

@article{Fetzer2020c,
author = {Fetzer, Thiemo},
doi = {10.1093/ej/ueab074},
issn = {0013-0133},
journal = {The Economic Journal},
month = {oct},
title = {{Subsidising the spread of COVID-19: Evidence from the UK's Eat-Out-to-Help-Out Scheme}},
url = {https://academic.oup.com/ej/advance-article/doi/10.1093/ej/ueab074/6382847},
year = {2021}
}
